\documentclass[10pt,a4paper]{elsarticle}

\usepackage{amsmath,amssymb}
\usepackage{changepage}
\usepackage[utf8x]{inputenc}
\usepackage{textcomp,marvosym}
\usepackage{nameref,hyperref}
\usepackage[right]{lineno}
\usepackage{microtype}
\DisableLigatures[f]{encoding = *, family = * }
\usepackage[table,dvipsnames]{xcolor}
\usepackage{array}
\usepackage{multirow}
\usepackage{longtable}

\usepackage[labelfont=bf]{caption}
\usepackage{hyphenat}
\usepackage{lastpage,fancyhdr,graphicx}
\usepackage{epstopdf}
 \usepackage{siunitx}   
\usepackage{algorithm,algorithmicx,algpseudocode}
\usepackage{dcolumn}
\usepackage{listings}
\usepackage{xcolor}
\usepackage{fancyvrb}
\usepackage {tikz}
\usepackage [notextcomp]{stix}
\usepackage{makecell}
\newcolumntype{d}[1]{D{.}{.}{#1}}
\tikzstyle{rect} = [rectangle, draw, fill=white!20, node distance=2cm, text width=6em, text centered, sharp corners, minimum height=3em]
\usetikzlibrary{plotmarks,intersections,calc,decorations.pathreplacing,positioning,arrows,shapes.geometric,matrix,decorations.markings,,hobby,backgrounds}
\usetikzlibrary {positioning}

\usepackage[super]{nth} 
\usepackage[framemethod=tikz]{mdframed}

\usepackage{lscape}
\usepackage{rotating}

\usepackage{letltxmacro}
\newmdenv[bottomline=false, linecolor=black!60!black, innertopmargin=0.1cm,innerleftmargin=0.1cm,innerbottommargin=0.1cm]{bottomless}
\newfloat{lstfloat}{htbp}{lst}
\floatname{lstfloat}{Listing}

\usepackage{color}
\usepackage{tcolorbox}    

\usepackage{enumitem}
\usepackage{ascii}

\renewcommand\_{\textunderscore\allowbreak}   

\newcommand{\qnclass}[2]{{\scriptsize \texttt{#1}$:$\textnormal{#2}}}

\newcommand{\property}[2]{{\scriptsize \texttt{#1}$:$\texttt{#2}}}

\newcommand{\instance}[2]{{\textit{#1:#2}}}

\newcommand{\prop}[2]{{\textit{#1:#2}}}

\newcommand{\class}[2]{\textit{\nohyphens{#1}:\nohyphens{#2}}}

\newcommand{\topico}{\textit{Topico}}
\newcommand{\stopics}{\sbounti\ topics}

\newcommand{\spot}[1]{$[$#1$]$}
\newcommand{\lentity}[2]{\spot{#1} $\rightarrowtail$ $[$#2$]$}
\newcommand{\setid}[1]{$[\textsc{#1}]$}

\newcommand{\task}[1]{\textit{Task-$#1$}}

\newcommand{\tagmeP}{\mathtt{p}}

\newcommand{\covidnineteen}{\textsc{covid}-19}

\newcommand{\etal}{\textit{et al.}}

\newcommand{\wwwc}{\textsc{w3c}}

\newcommand{\timeo}{\wwwc\ \textsc{owl}-Time ontology}

\newcommand{\dbp}{{DBpedia}}

\newcommand{\wikidata}{{Wikidata}}
\newcommand{\foaf}{\textsc{foaf}}

\newcommand{\bounti}{\mbox{\textsc{b{oun}-ti}}}
\newcommand{\sparql}{\textsc{sparql}}
\newcommand{\lod}{\textsc{lod}}
\newcommand{\sbounti}{\mbox{\textsc{s-b{oun}-ti}}}
\newcommand{\lda}{\textsc{lda}}
\newcommand{\nmf}{\textsc{nmf}}
\newcommand{\lsa}{\textsc{lsa}}
\newcommand{\nlp}{\textsc{nlp}}
\newcommand{\api}{\textsc{api}}
\newcommand{\rdf}{\textsc{rdf}}
\newcommand{\rdfs}{\textsc{rdfs}}

\newcommand{\res}[1]{\url{#1}}

\newcommand{\tweettext}[1]{{{\em #1}}}

\newcommand{\tfidf}{\textit{tf-idf}}

\newcommand{\kwb}{\textsc{wlb}}
\newcommand{\ei}{\textsc{ei}}
\newcommand{\tr}{\textsc{tr}}
\newcommand{\ex}{\textsc{ex}}
\newcommand{\ti}{\textsc{ti}}

\newcommand{\rd}{\textsc{rd}}
\newcommand{\qo}{\textsc{qo}}
\newcommand{\sa}{\textsc{sa}}
\newcommand{\owl}{\textsc{owl}}
\newcommand{\uri}{\textsc{uri}}
\newcommand{\swrl}{\textsc{swrl}}
\newcommand{\messy}{untidy}

\def\kbbox[#1,#2,#3,#4,#5]#6{
        \draw[dashed] node[draw,color=gray!50,minimum height=#1,minimum width=#2] (#4) at #5 {}; 
        \node[anchor=#3,inner sep=2pt] at (#4.#3)  {#6};
        }
   
\newcommand{\gettikzxy}[3]{%
  \tikz@scan@one@point\pgfutil@firstofone#1\relax
  \edef#2{\the\pgf@x}%
  \edef#3{\the\pgf@y}%
}

\algnewcommand{\LineComment}[1]{\State \(\#\) #1}

\newcommand\lland{\scalebox{1}[1.3]{$\land$}}
\newcommand\ccup{\scalebox{1}[1.3]{$\cup$}}

\newcommand{\ie}{i.e.,}

\newcommand{\tagme}{{TagMe}}

\newcommand{\pdone}{\setid{pd$_1$}}
\newcommand{\pdtwo}{\setid{pd$_2$}}
\newcommand{\pdthree}{\setid{pd$_3$}}
\newcommand{\vp}{\setid{vp}}
\newcommand{\brangelina}{\setid{ba}}
\newcommand{\carriefisher}{\setid{cf}}
\newcommand{\concert}{\setid{co}}
\newcommand{\northdakota}{\setid{nd}}
\newcommand{\tonibraxton}{\setid{tb}}
\newcommand{\inauguration}{\setid{in}}
\newcommand{\public}{\setid{pub}}

\LetLtxMacro{\citeSupporting}{\nameref}
\renewcommand{\nameref}[1]{\textit{\citeSupporting{#1}}}

\tikzstyle{vertex}=[circle,fill=black!25,inner sep=0pt]
\tikzstyle{selected vertex} = [vertex, fill=red!24]
\tikzstyle{edge} = [draw,thick.--]
\tikzstyle{weight} = [font=\small]
\tikzstyle{selected edge} = [draw,line width=5pt,-,red!50]
\tikzstyle{ignored edge} = [draw,line width=5pt,-,black!20]

\newcommand\Algphase[1]{%
\vspace*{-.7\baselineskip}\Statex\hspace*{\dimexpr-\algorithmicindent-2pt\relax}\rule{\textwidth}{0.2pt}%
\Statex\hspace{-.5cm}\hspace*{\algorithmicindent}\textit{#1}%
\vspace*{-.5\baselineskip}\Statex\hspace*{\dimexpr-\algorithmicindent-2pt\relax}\rule{\textwidth}{0.2pt}%
}

\newcommand{\autour}[1]{\tikz[baseline=(X.base)]\node [draw=black,fill=white!40,line width=0.1mm,rectangle,inner sep=2pt, rounded corners=3pt] (X) {#1};}
\makeatletter
\newcounter{phase}[algorithm]
\newlength{\phaserulewidth}

\makeatother

\definecolor{LightGray}{rgb}{0.97,0.97,0.97}
\definecolor{gray}{rgb}{0.4,0.4,0.4}
\definecolor{darkblue}{rgb}{0.0,0.0,0.6}
\definecolor{cyan}{rgb}{0.0,0.6,0.6}

\hypersetup{
    colorlinks=true,
    linkcolor=blue,
    filecolor=magenta,      
    urlcolor=cyan,
}

\lstdefinelanguage{XML}
{
  basicstyle=\small\asciifamily,
  backgroundcolor=\color{LightGray},
  tabsize=2,
  showstringspaces=false,
  morecomment=[l][\color{gray}]{\#},       
  morecomment=[n][\color{blue}]{<http}{>}, 
  morestring=[b][\color{OliveGreen}]{\"},  
  commentstyle=\color{gray}\upshape
  columns=fullflexible,
  morestring=[b]",
  morestring=[s]{>}{<},
  morecomment=[s]{<?}{?>},
  identifierstyle=\color{Sepia},
  keywordstyle=\color{Sepia},
  breaklines=true,
  morekeywords={xmlns}
}

\lstdefinelanguage{SPARQL}{
  basicstyle=\asciifamily,
  backgroundcolor=\color{LightGray},
  columns=fullflexible,
  breaklines=false,
  sensitive=true,
  tabsize = 2,
  showstringspaces=false,
  morecomment=[l][\color{gray}]{\#},       
  morecomment=[n][\color{blue}]{<http}{>}, 
  morestring=[b][\color{OliveGreen}]{\"},  
  keywordsprefix=?,
  classoffset=0,
  keywordstyle=\color{Sepia},
  morekeywords={},
  classoffset=1,
  keywordstyle=\color{Purple},
  morekeywords={rdf,rdfs,owl,xsd,purl},
  classoffset=2,
  keywordstyle=\color{MidnightBlue},
  morekeywords={
    SELECT,CONSTRUCT,DESCRIBE,ASK,WHERE,FROM,NAMED,PREFIX,BASE,OPTIONAL,
    FILTER,GRAPH,LIMIT,OFFSET,SERVICE,UNION,EXISTS,NOT,BINDINGS,MINUS,a
  }
}

    \makeatletter
    \def\ps@pprintTitle{%
       \let\@oddhead\@empty
       \let\@evenhead\@empty
       \let\@oddfoot\@empty
       \let\@evenfoot\@oddfoot
    }
    \makeatother

\usetikzlibrary{automata, arrows}

\tikzstyle{vertex}=[circle,fill=black!25,inner sep=0pt]
\tikzstyle{selected vertex} = [vertex, fill=red!24]
\tikzstyle{edge} = [draw,thick.--]
\tikzstyle{weight} = [font=\small]
\tikzstyle{selected edge} = [draw,line width=5pt,-,red!50]
\tikzstyle{ignored edge} = [draw,line width=5pt,-,black!20]

\newlength\savedwidth

\newcommand\thickhline{\noalign{\global\savedwidth\arrayrulewidth\global\arrayrulewidth 2pt}%
\hline
\noalign{\global\arrayrulewidth\savedwidth}}

\begin{document}

\title{Microblog topic identification using Linked Open Data\tnoteref{t1}} 

\tnotetext[t1]{This is a more readable  \textsc{pdf}  version of the article published in PLOS ONE. Please cite as:
\begin{tcolorbox}[width=\textwidth]{Yıldırım A, Uskudarli S (2020) Microblog topic identification using Linked Open Data. PLOS ONE 15(8): e0236863. \url{https://doi.org/10.1371/journal.pone.0236863}.}\end{tcolorbox}}

\author{Yıldırım, A. \corref{cor}}
\ead{ahmet.yil@boun.edu.tr}
\cortext[cor]{Corresponding author}

\author{Uskudarli, S.}
\ead{suzan.uskudarli@boun.edu.tr}

\address{SosLab, Department of Computer Engineering, Bogazici University, Istanbul, Turkey}

\begin{abstract}
Much valuable information is embedded in social media posts (microposts) which are contributed by a great variety of persons about subjects that of interest to others. 
The automated utilization of this information is challenging due to the overwhelming quantity of posts and the distributed nature of the information related to subjects across several posts.
Numerous approaches have been proposed to detect topics from collections of microposts, where the topics are represented by lists of terms such as words, phrases, or word embeddings.
Such topics are used in tasks like classification  and recommendations.
The  interpretation of  topics  is considered a separate task in such methods, albeit they  are becoming increasingly human-interpretable.
This work proposes an approach for identifying machine-interpretable topics of collective interest.
We define topics as a set of related elements that are associated by having posted in the same contexts. 
To represent topics, we introduce an ontology specified according to the W3C recommended standards.
The elements of the topics are identified via linking entities to resources published on Linked Open Data (LOD). 
Such representation  enables processing topics to provide insights that go beyond what is explicitly expressed in the microposts. 
The feasibility of the proposed approach is examined by generating topics from more than one million tweets collected from Twitter during various events. 
The utility of these topics is demonstrated with a variety of topic-related tasks along with a comparison of the effort required to perform the same tasks with words-list-based representations. 
Manual evaluation of randomly selected 36 sets of topics  yielded 81.0\% and 93.3\% for the precision and F1 scores respectively.
\end{abstract}

\maketitle

\section{Introduction}
\label{sec:introduction}

Microblogging systems are widely used for sharing short messages (microposts) with online audiences. 
They are designed to support the creation of posts with minimal effort, which has resulted in a vast stream of posts relating to issues of current relevance such as  politics, product releases, entertainment, sports, conferences, and natural disasters. 
Twitter~\cite{TwitterCom}, the most popular microblogging platform, reports that over \num{500} million tweets are posted per day~\cite{Twitterstatistics}.
Such systems have become invaluable resources for learning what people are interested in and how they respond to events. 
However, making sense of such large volumes of posts is far from trivial, since  posts tend to be limited in context (due to their short length), informal,  \messy,  noisy, and cryptic \cite{eisenstein2013bad}.
Furthermore, content related to the same topic is typically distributed over many contributions posted by numerous users.

Various approaches have been developed to gain insight into the topics that emerge on microposts. 
Some of the most popular topic detection approaches are based on latent Dirichlet allocation (\lda)~\cite{yan2013biterm,Li:2017:ETM:3133943.3091108,Fang2019}, latent semantic analysis (\lsa)~\cite{AlexisPerrier2015,chodhary2016semantic}, and non-negative matrix factorization (\nmf)~\cite{0604371a2d4c430ea137e8d4086734b6,CHEN20191}.
These methods capture topical keywords from post sets to represent topics that can be utilized in classification, recommendation, and information retrieval tasks.
Topics are represented with bags-of-words, along with weights indicating the strength of their association with the microposts.
Alternative approaches are based on the change in the frequency of terms of interest (\ie\ words and hashtags)~\cite{ Alvanaki:2012:SWE:2247596.2247636,Cataldi:2010:ETD:1814245.1814249,kasiviswanathan2011emerging,marcus2011twitinfo,TwitterMonitor:Mathioudakis:2010:TTD:1807167.1807306,Sayyadi:2013:GAA:2542214.2542215} to capture trending topics.
More recently, word-embeddings have been used to capture  both the semantic and the syntactic features of words to improve the relevancy of the words representing the topics~\cite{Fang2019,Bicalho:2017:GFE:3062405.3062584,Li:2017:ETM:3133943.3091108}.

The determination of topics related to people or groups in social media facilitates content-specific user recommendation as opposed to the more familiar friend of friend recommendations obtained from follower networks~\cite{celebi2012content,degirmencioglu2010exploring}.
For this purpose, approaches that process the content and the user behavior to recommend users have been proposed that extracts content~\cite{celebi2012content} or social network analysis of co-occurring content~\cite{degirmencioglu2010exploring}.

Natural language processing (\nlp) techniques are utilized to yield more human-readable topics.
Sharifi~\etal\ propose a reinforcement-based method on consecutively co-occurring terms to summarize collections of microposts~\cite{sharifi2014summarization}.
\bounti~\cite{bounti} also produces human-readable topics using cosine similarity among collections of tweets and Wikipedia articles.
The up-to-date nature of Wikipedia pages successfully captures topics of current relevance with highly readable titles.
While such topics are easily human-interpretable, they are less suitable for automated processing.

Some approaches identify topics within single posts by linking meaningful fragments within them to external resources such as Wikipedia pages~\cite{8695381EntitiLinking1,sakor-etal-2019-oldEntityLinking2,ferragina2012tagme,Gattani2013entity}.
In the context of microposts, the detection of topics for single posts is not very effective due to their limited context. 
This work focuses on topics that have gained traction within the crowd by aggregating contributions relevant to topics from numerous posts. 

\begin{figure}[!h]
\begin{center}
\includegraphics[scale=1]{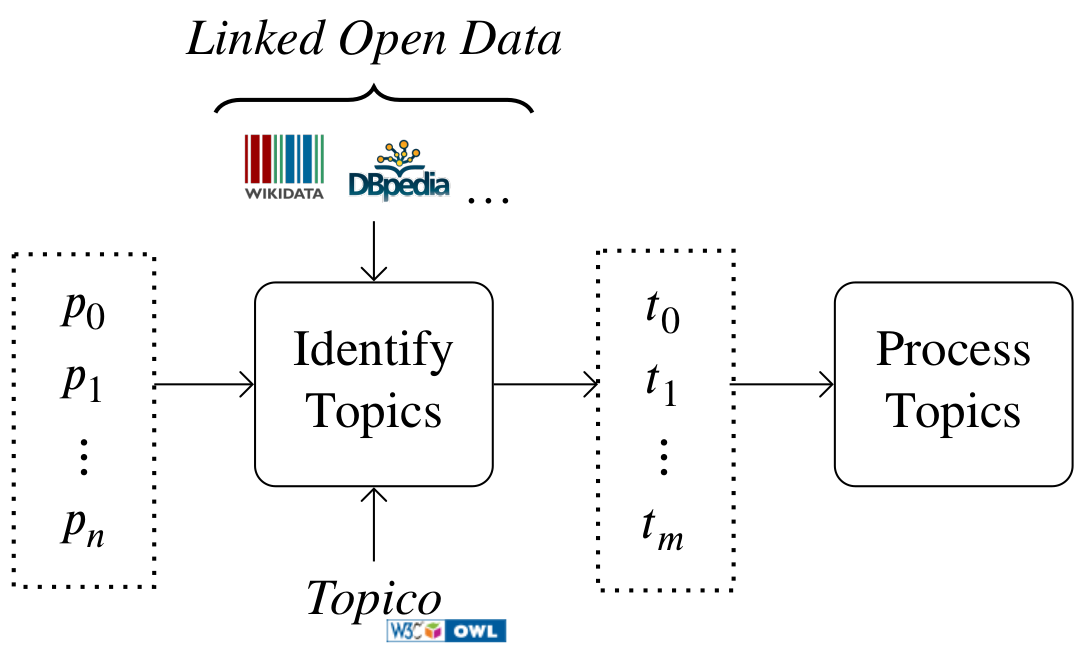}
\caption{{\bf Overview of identifying semantic topics from a set of microposts.}
Entities within microposts ($p_i$) are linked to semantic entities in \textit{Linked Open Data}, which are processed to yield semantic topics ($t_j$) expressed with the \topico\ ontology.
\label{fig:overview}}
\end{center}
\end{figure}

This work proposes an approach, \sbounti\ (Semantic-Boğaziçi University-Topic Identification) that produces machine-interpretable (actionable) semantic topics from collections of microposts (Fig~\ref{fig:overview}).
A topic is considered to be a collection of related \textit{topic elements}. The \textit{topic elements} are Linked Open Data (\lod)~\cite{bizer2011linked} resources that are linked to fragments of posts.
\lod\ is an ever-growing resource (\num{1,255} datasets with \num{16,174} links as of May 2020) of structured linked data regarding a wide range of human interests made accessible  with the semantic Web standards.
Such linking enables capturing the meaning of the content that is expressed in alternative manners such as the terms ``FBI'', ``feds''  and ``Federal Bureau of Investigation'' all link to the web resource \url{http://dbpedia.org/resource/Federal_Bureau_of_Investigation}, and ``guns n' roses'', and, ``gunsn'roses'' to the web resource \url{http://dbpedia.org/resource/Guns_N'_Roses} in \dbp.
These resources are further linked to other resources through their properties.
For example, the resource of the {\em FBI} provides information about \textit{law enforcement}, \textit{its director}, and the resource of {\em Guns N' Roses} provides information about their \textit{music genre} and \textit{group members}.
The linking of fragments within posts to such resources greatly expands the information at our disposal to make sense of  microposts. 
For example, the topics that are recently talked about regarding \textit{law enforcement} or \textit{rock concerts} could be easily retrieved.
To represent topics, an ontology (\topico) is specified under the W3C semantic Web~\cite{1637364,thesemanticweb} standards.

The main goal of this work is to explore the feasibility of linking informal conversational content in microposts to semantic resources in \lod\ to produce relevant machine-interpretable topics.
First, the potential elements of topics are determined by processing linked entities in a collection of posts.
Then, the elements are assigned to topics by processing a co-occurrence graph of the entities.
Finally, the topics are created by processing the elements and representing them with the \topico\ ontology. 
To the authors' knowledge, this is the first approach that utilizes semantic Web and \lod\ to identify topics within collections of microposts.

To assess the viability of the proposed approach, we developed a prototype to generate topics for \num{11} collections of tweets gathered during various events.
The utility of these topics (totaling \num{5248})  is demonstrated with a variety of topic-related tasks and a comparison of the effort required to perform the same tasks with words-list-based (\kwb) representations. 
An evaluation of randomly selected 36 sets of topics  yielded 81.0\% and 93.3\% for the precision and $F_1$ scores respectively.

The main contributions of this work are:
\begin{itemize}
\item{an approach for identifying semantic topics from collections of microposts using \lod,}
\item{the \topico\ ontology to represent semantic topics,}
\item{an analysis of semantic topics generated from \num{11} datasets with over one million tweets,} and
\item{a detailed evaluation of the utility of semantic topics through tasks of various complexities.}
\end{itemize}

To enable the reproducibility of our work and to support the research community, we contribute the following:
\begin{itemize}
\item{a prototype to generate semantic topics~\cite{EarlyPrototype}.}
\item{the semantic topics generated from the datasets and the identifiers of the tweets of \num{11} datasets~\cite{TopicExplorer,TopicDownload},}
\item{a demonstration endpoint for performing semantic queries over the generated topics (\url{http://soslab.cmpe.boun.edu.tr/sbounti}),} and
\item{the manual relevancy-annotations of \sbounti\ topics from \num{36} sets corresponding to approximately \num{5760} tweets~\cite{TopicDownload}.}
\end{itemize}

The remainder of this paper is organized as follows: The \nameref{sec:relatedwork} section provides an overview of topic identification approaches.
The key concepts and resources utilized in our work are presented in the \nameref{sec:background} section.
The proposed approach is described in the \nameref{sec:approach} section.
An analysis of the topics generated from various datasets and their utility is detailed in the \nameref{sec:evaluation} section.
Our observations related to the proposed approach and the resulting topics are presented along with future directions in the \nameref{sec:discussion} section.
Finally, in the \nameref{sec:conclusion} section,  we remark on our overall takeaways from this work.   

\section{Related work}
\label{sec:relatedwork}
The approaches to making sense of text can be characterized in terms of their input (\ie\ sets of short, long, structured, semi-structured text), their processing methods, the utilized resources (\ie\ Wikipedia, \dbp), and how the results are represented (\ie\ summaries, words, hashtags, word-embeddings, topics). 

Various statistical topic models, such as latent semantic analysis (\lsa)~\cite{AlexisPerrier2015,chodhary2016semantic}, non-negative matrix factorization (\nmf)~\cite{0604371a2d4c430ea137e8d4086734b6,CHEN20191}, and latent Dirichlet allocation (\lda)~\cite{blei2003latent},  aim to discover topics  in collections of documents.
They represent  documents  with topics and topics with words.
The topics are derived from a term-document matrix from which a  document-topic and a topic-term matrix are produced.
\lsa\ and \nmf\ methods achieve this with matrix factorization techniques.
\lda\ learns these matrices with a generative probabilistic approach that assumes that documents are represented with a mixture of a fixed number of topics, where each topic is characterized by a distribution over words.
The determination of the predefined number of topics can be difficult and is typically predetermined based on domain knowledge or experimentation.
The sparseness of the term-document matrix stemming from the shortness of the posts presents challenges to these approaches~\cite{abs-1904-07695,6778764,Lin:2014:DTM:2566486.2567980}.
The topics produced by these approaches are represented as lists of words and will be referred as words-lists-based approaches (\kwb).
The interpretation of the topics is considered a separate task.

\begin{figure}[!h]
\begin{center}
\includegraphics[scale=0.8]{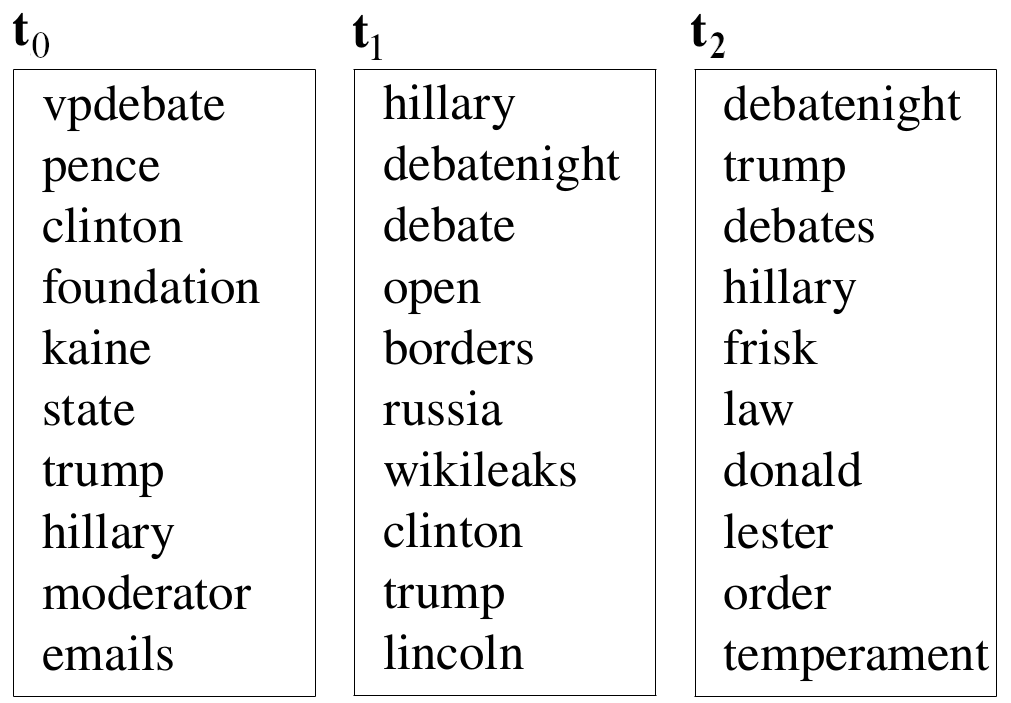}
\end{center}
\vspace{-0.5cm}\caption{{\bf The top 10 words of topics generated (using \textsc{t}witter-\lda) from a collection of tweets gathered during the 2016 U.S. presidential debate. }
\label{fig:lda-topics}
}
\end{figure}

\lda\ has been widely utilized for detecting topics in microposts.
Some of these approaches associate a single topic to an individual post~\cite{Qiang2018,Yin:2018:MCS:3219819.3220094}, whereas others
consider  a document to be a collection of short posts that are aggregated by some criteria like the same author~\cite{Weng:2010:TFT:1718487.1718520}, temporal or geographical proximity~\cite{Bauer2012}, or content similarity  that indicates some relevance (\ie\ hashtag or keyword)~\cite{Mehrotra2013}. 
Fig~\ref{fig:lda-topics} shows the top ten words of some \lda\ topics resulting from tweets that we collected during the 2016 U.S. presidential debates (produced by \textsc{t}witter\-\lda~\cite{twitterLDA}).
Some approaches use the co-occurrence of  words in the posts to capture meaningful phrases and use these bi-terms in the generative process ~\cite{yan2013biterm,chen-etal-2015-user,10.1007/978-981-10-6893-5_2}.
The  determination of the predefined number of topics for large collections of tweets that are contributed by numerous people can be quite difficult.

More recently, word-embeddings learned from the Twitter public stream and other corpora (\ie\ Wikipedia articles) have been used to improve \lda\ based~\cite{Li:2017:ETM:3133943.3091108,Fang2019} and \nmf\ based~\cite{8993771} topics.
In some cases,  word-embeddings are used to enhance the posts with semantically relevant words
~\cite{10.1007/978-3-319-57529-2_29,Bicalho:2017:GFE:3062405.3062584}.
Another utility of word-embeddings is in assessing the coherence of topics by determining the semantic similarity of their terms~\cite{Li:2016:TMS:2911451.291149}.

Alternatively, in micropost topic identification, some approaches consider topics as a set of similar posts, such as those based on the fluctuation in the frequency of the terms of interest (\ie\ words and hashtags)~\cite{Alvanaki:2012:SWE:2247596.2247636,Cataldi:2010:ETD:1814245.1814249}.
The evolution of such trending topics can be traced using the co-occurrences of terms. 
Marcus \etal~\cite{marcus2011twitinfo} created an end-user system that presents frequently used words and representative tweets that occur during peak posting activity which they consider as indicators of relevant events.
Petrović \etal~\cite{Petrovic:2010:SFS:1857999.1858020} use term frequency and inverse document frequency (\tfidf) to determine similar posts and select the first post (temporally) to represent topics. 
An alternative similarity measure is utilized by Genc \etal~\cite{Genc:2011:DCC:2021773.2021833} who compute the semantic distances between posts based on the distances of their linked entities in Wikipedia's link graph.
In these approaches, the interpretation of what a topic represents is also considered a separate task.

Sharifi \etal~\cite{sharifi2014summarization} produce human-readable topics in the form of a summary phrase that is constructed from common consecutive words within a set of posts.
\bounti~\cite{bounti} represents topics as a list of Wikipedia page titles (that are designed to be human-readable) which are most similar (cosine similarity of \tfidf\ values) to a set of posts.
While these topics are human-comprehensible, they are less suitable for automated processing.

The approaches mentioned thus far have been domain-independent, however, in some cases domain-specific topics may be of interest.
In the health domain, Prieto \etal\ explore posts related to specific sicknesses to track outbreaks and epidemics of diseases~\cite{Prieto2014} by matching tweets with illness-related terms that are manually curated.
Similarly, Parker \etal\ utilize sickness-related terms which are automatically extracted from Wikipedia~\cite{parker2013framework}. 
Eissa \etal~\cite{Eissa:2018:TRU:3184558.3191562} extract topic related words from resources such as \dbp\ and WordNet to identify topics related to user profiles.
As opposed to machine learning-based approaches that generate topics, these approaches map a collection of posts to  pre-defined topics.

Entity linking approaches~\cite{gruetze2016} have been proposed to identify meaningful fragments in short posts and link them to external resources such as Wikipedia pages or \dbp\ resources~\cite{8695381EntitiLinking1,sakor-etal-2019-oldEntityLinking2,ferragina2012tagme,Gattani2013entity}.
These approaches identify topics related to single posts.
Such approaches may not adequately  capture topics of general interest since they miss  contextual information present in crowd-sourced content.

Semantic Web technologies~\cite{kumar2019knowledge} are frequently utilized to interpret documents with unique concepts that are machine-interpretable and interlinked with other data~\cite{Liao:2019:UAT:3359984.3324473,10.1007/978-3-030-21395-4_7,Matentzoglu2018}.
For example, events and  news documents have been semantically annotated with ontologically represented elements (time, location, and  persons) to obtain machine-interpretable data~\cite{rospocher2019boosting, gottschalk2019eventkg,ABEBE2019}.
Parliamentary texts are semantically annotated with concepts from \dbp\ and Wikipedia using look-up rules and entity linking approaches~\cite{Aggelen2015,madoc49597}.
Biomedical references have been normalized with an unsupervised method that utilizes the OntoBiotope ontology~\cite{ontobiotop} that models microorganism habitats and word-embeddings~\cite{Karadeniz2019}.
In this manner, they can map text  like \textit{children less than 2 years of age} to the concept \textit{pediatric patient} (\textit{OBT:002307}) that bears no syntactic similarity.

Our work utilizes semantic Web technologies to identify topics from domain-independent collections of microposts and to express them.
Like many of the other approaches, we aggregate numerous posts.
The ontology specification language \owl~\cite{W3COWL} is used to specify \topico\ to represent topics.
The elements of topics are identified via entity linking using \lod\ resources.
The collective information gathered from sets of posts is utilized in conjunction with the information within \lod\ resources to improve the topic elements.
The elements of topics are related based on having co-occurred in several posts.
In other words, numerous posters have related these elements by posting them together.
This can result in topics that may seem peculiar, such as the \textit{FBI} and the U.S. presidential candidate  \textit{Hillary Clinton}, which  became a hot subject  on Twitter as a result of public reaction.
A co-occurrence graph is processed to determine the individual topics. 
The topic elements are the \textsc{URI}s of web resources that correspond to fragments of posts. 
Various fragments may be associated with the same resource since our approach aims to capture the meaning (\ie\ ``FBI'', ``feds''  and ``Federal Bureau of Investigation'' to \url{http://dbpedia.org/resource/Federal_Bureau_of_Investigation}).
Semantically represented topics offer vast opportunities for processing since  short unstructured posts are mapped to ontologically represented topics consisting of elements within  a rich network of similarly represented information.

\section{Background}
\label{sec:background}
\label{sec:background-relevant-ontologies}
\label{sec:linkeddata}
\label{sec:background-entity-linking}

This section describes the basic concepts and tools related to semantic Web and ontologies, entity linking, and Linked Open Data that are used in this work.

\subsection{Ontologies and Linked Open Data}
Ontology is an explicit specification that formally represents a domain of knowledge~\cite{vanHarmelenSemanticWebBook,thesemanticweb}.
It defines inter-related concepts in a domain.
The concepts are defined as a hierarchy of \textit{classes} that are related through {\em properties}.
Ontologies often refer to definitions of concepts and properties in other ontologies, which is important for reusability and interoperability.
Resource Description Framework Schema (\rdfs) definitions are used to define the structure of the data.
Web ontology language (\owl)~\cite{W3COWL} is used to define semantic relationships between concepts and the data.
Ontology definitions and the data expressed with ontologies are published on the Web and referred to with their unified resource identifiers (\uri).
To easily refer to the ontologies and data resources, the beginning of \uri s are represented with namespace prefixes.
For example {\em dbr:} is often used to refer to {\em http://dbpedia.org/resource/}.
A specific entity is referred to with its namespace prefix and the rest of the \uri\ following it such as  \instance{dbr}{Federal\_Bureau\_of\_Investigation} for \url{http://dbpedia.org/resource/Federal_Bureau_of_Investigation} (which is the definition of FBI in \dbp).

We use \owl\ language to define the \topico\ ontology to express microblog topics.
Other ontologies that \topico\ refers to are \dbp\ to express encyclopedic concepts, (\foaf)~\cite{foafspecweb} to express agents (with emphasis on people), \wwwc\ basic Geo vocabulary~\cite{wgsweb,wgsweb2} and Geonames~\cite{geonames} to express geolocations, {\em Schema.org}~\cite{Schema.Org,schemaorgvoc} to express persons and locations, and \wwwc\ time~\cite{W3CTime} to express intervals of topics.
The namespace prefixes that are referred to in this paper are given in~\citeSupporting{S1_Table}. 

\subsection{Entity linking}
\begin{figure}[!h]
\hspace{-2cm}\includegraphics[scale=1.1]{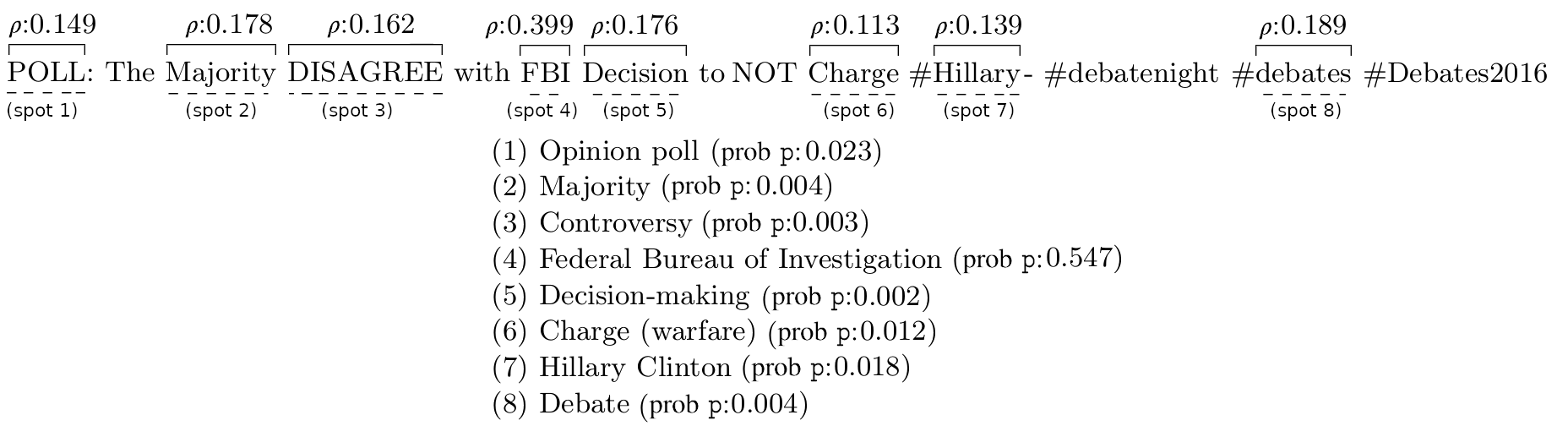}
\caption{{\bf Entity linking results from \tagme\ for a short text.}
\label{fig:spot-link-ex}}
\end{figure}

{\em Entity linking} is used to identify fragments within text documents (surface forms or spots) and link them to external resources that represent real-world entities (\ie\ dictionaries and/or encyclopedias such as Wikipedia)~\cite{gruetze2016}.
Entity linking for microposts is challenging due to the use of unstructured and \messy\ language as well as the limited context of short texts. 
We use \tagme~\cite{ferragina2012tagme} for this purpose as it offers a fast and well-documented application programming interface (\api)~\cite{TagmeApiDoc}.
\tagme\ links text to Wikipedia articles to represent entities.
This is suitable for our purposes since the articles are cross-domain and up-to-date.
Given a short text, \tagme\ returns a set of linked entities corresponding to its spots.
For each result, \tagme\ provides a goodness value ($\rho$) and a probability ($\tagmeP$) which are used to select those that are desirable.
The chosen entities are treated as candidate topic elements.
Fig~\ref{fig:spot-link-ex} illustrates the response of \tagme\ for a short text. 
Here, the spot {\em FBI} is linked to \url{https://en.wikipedia.org/wiki/Federal_Bureau_of_Investigation} with $\rho=0.399$ and $\tagmeP=0.547$.
We use $\rho$ and $\tagmeP$ to determine the viability of a topic element.

\subsection{Linked Data}

Linked Data~\cite{LinkedDataOrg,bizer2011linked} specifies the best practices for creating linked knowledge resources.
Linked Open Data (\lod) refers to the data published using Linked Data principles under an open license.
It is an up-to-date collection of interrelated web resources that spans all domains of human interests, such as music, sports, news, and life sciences~\cite{lodAuer2014,Schmachtenberg2014}. 
\lod\ contains \num{1,255} datasets with \num{16,174} links among them (as of May 2020)~\cite{LodCloud}. 
With its rich set of resources, \lod\ is suitable for representing the elements of topics, such as  \res{http://dbpedia.org/resource/Federal\_Bureau\_of\_Investigation} to represent ``FBI''.
Among the most widely used data resources in \lod\ are \dbp~\cite{Bizer2009154}  with more than 5.5 million articles derived from Wikipedia (as of September  2018~\cite{dbpediadatasetsize}) and 
Wikidata~\cite{Vrandecic:2014:WFC:2661061.2629489,WikidataOrg} with more than \num{87} million items (as of June 2020~\cite{wikidatadatasetsize}).
\dbp\ is a good resource for identifying entities such as known persons, places, and events that often occur in microposts.
For example, in the short text: \textit{``POLL: The Majority DISAGREE with FBI Decision to NOT Charge \#Hillary - \#debatenight \#debates \#Debates2016''} the entity linking task identifies the spots \textit{Hillary} and \textit{FBI} that are linked to \instance{dbr}{Hillary\_Clinton} and \instance{dbr}{Federal\_Bureau\_of\_Investigation} respectively.
Both Wikidata and \dbp\ support semantic queries by providing \sparql\ endpoints~\cite{WikidataSparql,DbpediaSparql}.
\sparql~\cite{Seaborne:13:SQL} (the recursive acronym for \sparql\ Protocol and \rdf\ Query Language) is a query language recommended by W3C for extracting and manipulating information stored in the Resource Description Framework (\rdf) format. 
It utilizes graph-matching techniques to match a query pattern against data.
 \sparql\ supports networked queries over web resources which are identified with URIs~\cite{10.1145/1567274.1567278}.
This work uses \lod\ resources in \sparql\ queries to demonstrate the utility of the proposed approach and to identify if topic elements  are persons or locations.

\section{Approach to identifying semantic topics}
\label{sec:model}
\label{sec:approach}

This work focuses on two main aspects related to extracting topics from collections of microposts:  their identification and their representation. 
More specifically, the determination of whether \lod\ is suitable for capturing information from microposts and if semantically represented topics offer the expected benefits. 
The key tasks associated with our approach are (1) identifying the elements of topics from collections of microposts, 
(2) determining which elements belong to which topics, and (3) semantically representing the topics.
This section presents a topic identification approach and describes its prototype implementation which is used for evaluation and validation purposes.
First, we describe the ontology developed to represent topics since it models the domain of interests and, thus, clarifies the context of our approach.
Then, we present a method for identifying topics from micropost collections, which will be represented using this ontology.
While describing this method, aspects relevant to the prototype implementation are introduced in context. 
Finally, various implementation details are provided at the end of this section.

\subsection{Topico ontology for representing topics}
\label{sec:model-ontology}

\begin{figure}[tbh!]
\hspace{-0.6cm}\includegraphics[scale=0.37]{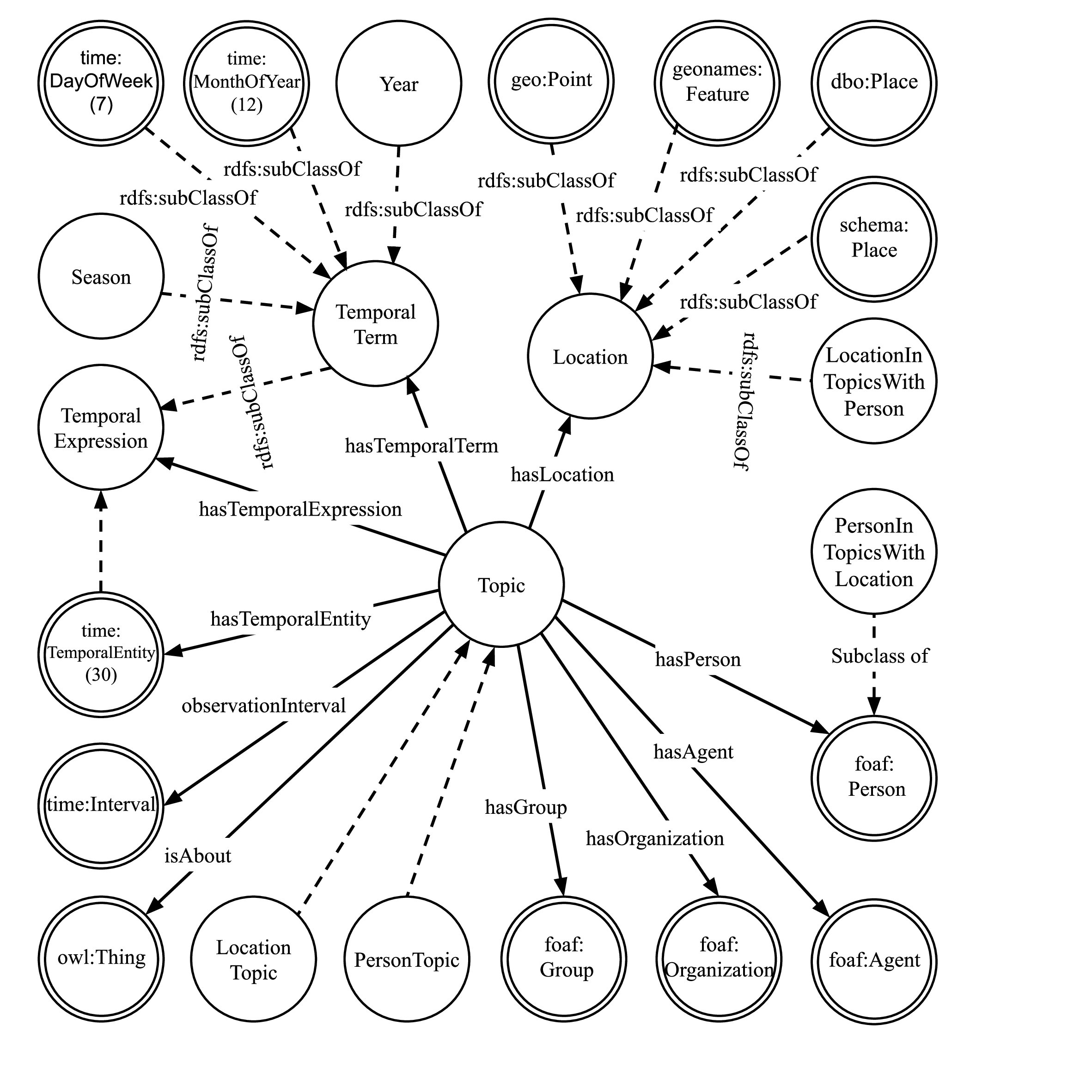}\vspace{-1cm}
\caption{{\bf The fundamental classes and object properties of \topico.}
The classes defined in other ontologies are shown with double circles and labeled with the prefixes of the namespaces such as \foaf.
\prop{rdfs}{subClassOf} relationships are represented with dashed lines.
\label{fig:appendix-ontology-class-diagram}}
\end{figure}

In the context of this work, a topic is considered to be a set of elements that are related when numerous people post about them in the same context (post).
Here, we focus on the elementary aspects of the topics that are most common in social media.
For this purpose, we define an ontology called \topico\ that is specified with the Web Ontology Language (\owl)~\cite{W3COWL} using Protègè~\cite{ProtegeWeb} according to the Ontology 101 development process~\cite{noy2001ontology}. 
The main classes and object relations of the ontology (\topico) are shown in Fig~\ref{fig:appendix-ontology-class-diagram}.
Further object properties are shown in \citeSupporting{S1_Appendix}.
This section describes some of the design decisions and characteristics relevant to \topico.
In doing so, the references to the guidelines recommended for specifying ontologies are shown in italic font.
The prefix {\em topico} is used to refer to the \topico\ namespace.

The first consideration is to  \textit{determine the domain and scope of the ontology}.
Representing topics that emerge from collections of microblog posts is at the core of our domain.
Thus, the ontology must reflect the concepts (classes) and properties (relations) common to microblogs.
It aims to serve as a basic ontology to represent general topics that could  be extended for domain-specific cases if desired.
The simplicity is deliberate to create a baseline for an initial study and to avoid premature detailed design.

To \textit{enumerate the important terms in the ontology}, we  inspected a large volume of tweets.
We observed the presence of well-known people, locations, temporal expressions across all domains since people seem to be interested in the ``who, where, and when'' aspects of topics. What the topic is about varies greatly, as one would expect.
As a result we decided to focus on the agents (persons or organizations), locations, temporal references, related issues, and meta-information of topics.
Based on this examination, \textit{a definition of classes and a class hierarchy} was developed.
The main class of \topico\ is \class{topico}{Topic} which is the domain of the object properties \prop{topico}{hasAgent}, \prop{topico}{hasLocation}, and \prop{topico}{hasTemporalExpression} which relate a topic to people/organizations, locations, and temporal expressions. 
To include all other kinds of topic elements we introduce the \prop{topico}{isAbout} property (\ie\ \textit{Topic1} \prop{topico}{isAbout} \instance{dbr}{Abortion}).
We defined several temporal terms as instances of the \class{topico}{TemporalExpression} class.
Also, since the subjects of conversation change rapidly in microblogs, the required property \prop{topico}{observationInterval}  is defined which corresponds to the time interval corresponding a collection (timestamps of the earliest and latest posts).
This information enables tracking how topics emerge and change over time, which is specifically interesting for event-based topics like political debates and news.
A topic may be related to zero or more of elements of each type.
Topics with no elements would indicate that no topics of collective interest were identified.
Such information may be of interest to those tracking the topics in microblogs.
In our prototype, however, an approach that yields topics with at least two elements was implemented since we were interested in the elements of the topics.

With respect to the \textit{consider reusing existing ontologies} principle,  we utilize the classes and properties of existing ontologies whenever possible such as \timeo~\cite{W3CTime}, \foaf, Schema.org, and Geonames.
\foaf\ is used for agents and persons. 
The classes \class{schema}{Place}, \class{dbo}{Place}, \class{geonames}{Feature}, and \class{geo}{Point} are defined as subclasses of \class{topico}{Location}.
Temporal expression of \timeo~\cite{W3CTime} are grouped under  \class{topico}{TemporalExpression}.
The temporal expressions of interest which were not found are specified in \topico.

A topic related to the first 2016 U.S. presidential debate (27 September 2016) is shown in Fig~\ref{fig:exampleTopic}.
This topic is related to the 2016 U.S. presidential candidate Donald Trump, the journalist Lester Holt (\prop{topico}{hasPerson}), racial profiling, and terry stopping (\prop{topico}{isAbout}) in the United States (\prop{topico}{hasLocation}) in 2016 (\prop{topico}{hasTemporalTerm}). The subject of racial profiling and terry stopping (stop and frisking mostly of African American men)  frequently emerged during the election. Further information about \topico\ may be found at~\cite{ontologyDetails} and the ontology itself is published at \url{http://soslab.cmpe.boun.edu.tr/ontologies/topico.owl}.

\begin{figure}[tbh!]
\begin{lstlisting}[language=XML,frame=none]
<owl:NamedIndividual rdf:about="http://soslab.cmpe.boun.edu.tr/sbounti/topics_algorithmV0.5.owl#2016_first_presidential_debate_50_52_topic_23">
    <rdf:type rdf:resource="http://soslab.cmpe.boun.edu.tr/ontologies/topico.owl#Topic"/>
    <foaf:maker rdf:resource="http://soslab.cmpe.boun.edu.tr/sbounti/topics_algorithmV0.5.owl#algorithmV0.5"/>
    <rdfs:label xml:lang="en">2016 First Presidential Debate 50-52 minutes topic: 23 </rdfs:label>
    <topicCreatedAt rdf:datatype="http://www.w3.org/2001/XMLSchema#dateTimeStamp">2017-06-26T13:05:37Z </topicCreatedAt>
    <observationInterval rdf:resource="http://soslab.cmpe.boun.edu.tr/sbounti/topics_algorithmV0.5.owl#interval_2016-09-27T01-50-00Z-2016-09-27T01-51-59Z"/>
    <hasPerson rdf:resource="http://dbpedia.org/resource/Lester_Holt"/>
    <hasPerson rdf:resource="http://dbpedia.org/resource/Donald_Trump"/>
    <isAbout rdf:resource="http://dbpedia.org/resource/Racial_profiling"/>
    <isAbout rdf:resource="http://dbpedia.org/resource/Terry_stop"/>
    <hasTemporalTerm rdf:resource="http://soslab.cmpe.boun.edu.tr/sbounti/topics_algorithmV0.5.owl#year_2016"/>
</owl:NamedIndividual>
\end{lstlisting}
\vspace{-0.2cm}\caption{A semantic topic extracted from 50-52nd minutes of the first presidential debate of U.S. 2016 elections. This topic is related to Lester Holt (Journalist) and Donald Trump (presidential candidate) regarding racial profiling and terry stopping (stop and frisk) in the U.S. in 2016. Automatic enumeration gave the topic number 23 to this topic.
\label{fig:exampleTopic}}
\end{figure}

\subsection{Identifying topics}
\label{sec:model-topic-identification}

\begin{figure}[tbh!]
\begin{center}
\includegraphics[scale=1]{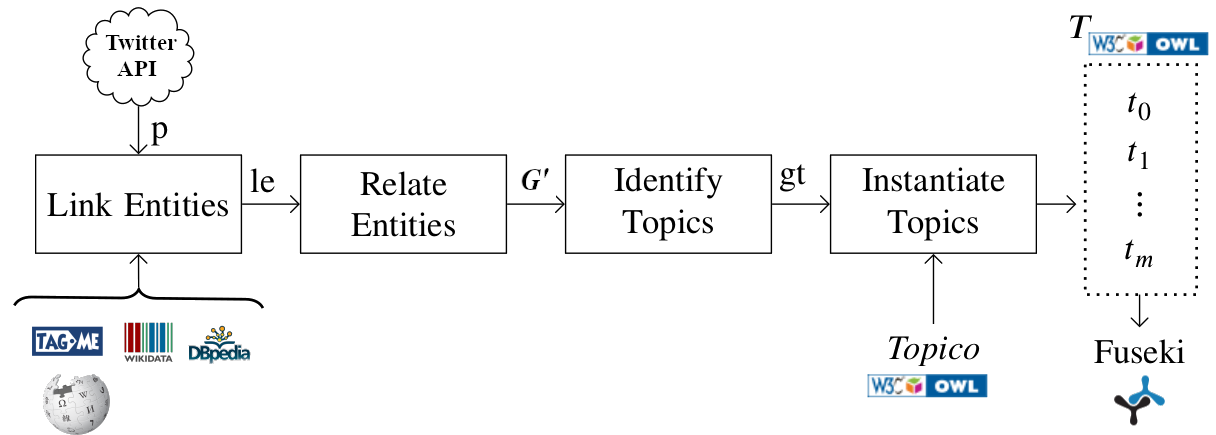}
\caption{{\bf An overview of identifying semantic topics from collections of tweets.}
An overview of identifying semantic topics from collections of tweets where \textit{p} is the set of tweets; 
\textit{le} is the set of linked entities (candidate elements); 
$\mathit{G'}$ is the co-occurrence graph of candidate topic elements; 
\textit{gt} is the set of sub-graphs of $\mathit{G'}$ whose elements belong to the same topic; and
\textit{T} is the set of semantic topics represented in \owl. 
\textit{TagMe, Wikidata, DBpedia,} and \textit{Wikipedia} are external resources used during entity linking.
\topico\ is the ontology we specified to express semantic topics. 
All topics are hosted on a Fuseki \sparql\ endpoint.
\label{fig:modelOverview}}
\end{center}
\end{figure}

The task of topic identification consists of identifying significant elements within posts and determining which of them  belong to the same topic. 
An overview for identifying topics is shown in Fig~\ref{fig:modelOverview}, which takes a set of microposts and results in 
a set of topics represented with \topico\ (\stopics).
Semantic topics are stored in \rdf\ repositories to facilitate  processing.
Algorithm~\ref{alg:GeneralAlgorithm} summarizes the process of generating semantically represented topics given a collection of microposts. 
It  has three phases: the determination of candidate topic elements, topic identification, and topic representation.

\begin{algorithm}[!h]
\caption{Topic extraction from microposts}
\label{alg:GeneralAlgorithm}
\begin{algorithmic}[1]
\State Input: $P$ \Comment{micropost set} 
\State Output: $T$ : OWLDocument \Comment{semantic topics}
\State $\textit{unlinkedSpots} \leftarrow []$ \Comment{unlinked spots}
\State $\textit{elements} \leftarrow []$ \Comment{candidate elements}
\State $\textit{types} \leftarrow []$ \Comment{types of elements}
\State $le \leftarrow []$ \Comment{linked entities}
\State $G$, $G'$ : \texttt{graph} \Comment{initial and pruned graphs}
\State $\textit{gt}$ : $\{\}$ \Comment{set of sub-graphs} 
\State $\textit{sTopic}$ : \texttt{OWL} \Comment{\sbounti\ topic} 
\Algphase{Phase 1 -- Identify Candidate Elements}
\For{\textbf{each} $p$ \textbf{in} $P$ } \label{algl:a}
\State $\textit{elements}[p] \leftarrow$ entities($p$) $\cup$ temporalExpressions($p$) 
\State $\textit{unlinkedSpots}[p] \leftarrow$ unlinkedSpots($p$) 
\EndFor \label{algl:b}
\For{\textbf{each} $p$ \textbf{in} P} \label{algl:c}
\State $\textit{le}[p] \leftarrow$ reLink($\textit{elements}[p],\textit{elements}$) 
\LineComment{Add the newly linked entities}
\State $\textit{le}[p] \leftarrow \textit{le}[p]\ \cup\ $ linkSpots($\textit{unlinkedSpots}[p]\textit{,elements,unlinkedSpots}$)
\EndFor \label{algl:d}
\Algphase{Phase 2 -- Identify topics}
\State $G \leftarrow $  relate($le$) \label{algl:e} \Comment{construct co-occurrence graph} 
\State $G' \leftarrow $  prune($G,\tau_{e}$) \Comment{prune the graph using $\tau_e$} \label{algl:f}
\State $gt \leftarrow $  subgraphsOfRelatedNodes($G'$,$\tau_{sc}$ ) \Comment{sub-graphs of related nodes of G'} \label{algl:i}
\State $\textit{observationInterval} \hspace{-0.1cm} \leftarrow$  \hspace{-0.2cm} getObservationInterval($P$)\Comment{timestamps of the earliest and latest posts} \label{algl:m}
\For{\textbf{each} $v$ \textbf{in} $G'$} \label{algl:g}
\State $types[v] \leftarrow $  getType($v,P,\tau_{loc}$)\Comment{determine the type of element ($v$)}
\EndFor \label{algl:h}
\Algphase{Phase 3 -- Represent topics with \topico}
\For{\textbf{each} $topic$ \textbf{in} $gt$} \label{algl:j}
\State $\textit{sTopic} \leftarrow $  sem-topic($topic,types,\textit{observationInterval}$)\Comment{represent as \sbounti\ topic}
\State \textit{T}.add($\textit{sTopic}$) \Comment{Add to topics for collection P}
\EndFor \label{algl:k}
\\ \Return \textit{T}
\end{algorithmic}
\end{algorithm}

\label{sec:model-candidate-element-extraction}
\label{sec:model-agent-identification}
\label{sec:model-location-identification}
\label{sec:model-processing-collective}

The first phase determines the candidate topic elements that are extracted from each post (Lines \ref{algl:a}-\ref{algl:b}).
$P$ is a set of posts.
The function \textit{entities}$(p)$ returns the entities within a post $p$. 
We denote a post and its corresponding entities as $\langle p, l \rangle$ where $l$ are linked entities.
Determining the candidate elements entails the use of an entity linker that links elements of microposts to external resources and a rule-based temporal term linker. 
We defined temporal term linking rules~\cite{TopicDownload} to detect frequently occurring terms like the days of the week, months, years, seasons, and relative temporal expressions (\ie\ {\em tomorrow}, {\em now}, and {\em tonight}) to handle the various ways in which they are expressed in social media.
We denote linked entities as \lentity{spots}{\uri}, where all the spots that are linked to an entity are shown as a list of lowercase terms and the entities are shown as \uri s.
For example, \lentity{north dakota, n. dakota}{\instance{dbr}{North\_Dakota}} indicates the two spots \textit{north dakota} and \textit{n. dakota} that are linked to \instance{dbr}{North\_Dakota} where some posts refer to the state North Dakota with its full name and others have abbreviated the word north as ``n.''.
The entity linking process may yield alternatively linked spots or unlinked spots.
For example, the spot ``Clinton'' may be linked to any of \instance{dbr}{Hillary\_Clinton}, \instance{dbr}{Bill\_Clinton}, \instance{dbr}{Clinton\_Foundation} or not at all (unlinked spot).
Such  candidate topic elements may be improved by examining the use of patterns within the collective information for agreement among various posters.
Thus, the linked entities retrieved from all the posts are used to improve the candidate elements attempting to link previously unlinked spots or altering the linking of a previously linked spot  (Lines \ref{algl:c}-\ref{algl:d}).
In our prototype,  entities are retrieved using \tagme\ which links spots to Wikipedia pages. 
We map these entities to \dbp\ resources which are suitable for the semantic utilization goals of our approach (see the \nameref{sec:background} section).
At the end of this phase, any remaining unlinked spots are eliminated, yielding the final set of candidate topic elements.

%





\label{sec:model-graph}

The second phase decides which elements belong to  which topics. 
We consider the limited size of microposts to be significant when relating elements since the user chose to refer to them in the same post.
The more often a co-occurrence is encountered the more significant that relation is considered since the aim is to capture what is of collective interest.
In this work, the term co-occurring elements/entities is defined to be the co-occurrence of the spots within a post to which these elements are linked.

To identify topics, a co-occurrence graph the candidate elements is constructed.
Let:\\[3pt]
$\mathit{LEP}$ = \{$\langle p, l \rangle |\  p \in P \land\ l = entities(p) \}$ be the set of entities extracted from the posts.\\
$G = (V,E)$ be the co-occurrence graph of the entities, where \\
$V = \bigcup\limits_{p \in P}^{} entities(p)$ and\\
$E = \{ (v_i,v_j) | v_i,v_j \in V \land v_i \neq v_j \land \{v_i,v_j\} \subseteq l \land \langle p,l \rangle \in \mathit{LEP} \}$.\\
And, let $w \colon E \to \mathbb{R}_{[0,1]}$ be a function that returns the weight of an edge defined as:

\begin{equation}
w(e) =  \frac{|\{\ p\ |\ e = (v_i,v_j) \in E \land   v_i \neq v_j \land \{v_i,v_j\}  \subseteq l\  \land\ l = entities(p) \} |}{|P|}
\end{equation}

Fig~\ref{fig:co-occurrence-construction-ex} shows an example co-occurrence graph constructed from four micropost texts. 
The linked entities obtained from these posts are \instance{dbr}{Donald\_Trump}, \instance{dbr}{Lester\_Holt}, 
\instance{dbr}{Social\_Profiling}, \instance{dbr}{Terry\_Stop}, \instance{dbr}{Constitutionality}. 
Within four posts, the co-occurrence between some of these entities ranged between $0.25$ to $0.75$.
Thus, we have extracted a significantly rich set of information from the posts in terms of relating them to web resources which themselves are related to other resrouces via data and object properties.

\begin{figure}[tbhp]
\begin{center}
{\small
\begin{tabular}{|p{2.5in}|p{2in}|}
\hline
\multicolumn{1}{|c|}{Post text} & \multicolumn{1}{c|}{Entities} \\ \hline
\autour{Lester} reminds \autour{Trump} that \autour{Stop-and-Frisk} was ruled \autour{unconstitutional}. \#debatenight &  \makecell[l]{\instance{dbr}{Lester\_Holt}, \instance{dbr}{Donald\_Trump}\\ \instance{dbr}{Terry\_stop}, \instance{dbr}{Constitutionality}} \\[3pt] \hline
\autour{Lester Holt} got the receipts on this \autour{stop and frisk} s..t. \#debatenight & 
\instance{dbr}{Lester\_Holt}, \instance{dbr}{Terry\_stop} \\ \hline
Add \autour{stop and frisk} to the mix of \autour{racial profiling} ...that's a great idea \#Debates. & \instance{dbr}{Terry\_stop}, \instance{dbr}{Racial\_profiling} \\ \hline
\autour{Donald Trump} tells \autour{Lester Holt} he is wrong on \#\autour{stopandfrisk} in regards to it being a form of \autour{racial profiling} & \makecell[l]{\instance{dbr}{Donald\_Trump}, \instance{dbr}{Lester\_Holt}\\ \instance{dbr}{Terry\_stop}, \instance{dbr}{Racial\_profiling}} \\ \hline
\end{tabular}}

\vspace{0.2cm}
Co-occurrence Graph

\framebox{\begin {tikzpicture}[-,-latex ,node distance=1cm and 1.5cm ,semithick ,on grid,state/.style ={}]
\node[state,] (E0) {\instance{dbr}{{\small \instance{dbr}{Donald\_Trump}}}};
\node[state] (E1) [above=of E0,xshift=0cm] {{\small \instance{dbr}{Lester\_Holt}}};
\node[state] (E4) [left=of E1,xshift=-1.7cm] {{\small\instance{dbr}{Social\_Profiling}}};
\node[state] (E3) [above=of E4] {{\small\instance{dbr}{Terry\_Stop}}};

\node[state] (E2) [above=of E1,xshift=1.7cm] {{\small\instance{dbr}{Constitutionality}}};

\path [-] (E0) edge  node[above,sloped] {\scriptsize $0.5$} (E1);
\path [-] (E0) [bend right=30] edge node[right] {\scriptsize $0.25$ } (E2);
\path [-] (E0) [bend left=10] edge node[left,xshift=-0.4cm] {\scriptsize $0.25$} (E4);
\path [-] (E1) [bend left=8]  edge node[above,sloped,xshift=-0.2cm] {\scriptsize $0.25$ } (E2);
\path [-] (E1) [bend right=8] edge node[above,sloped,xshift=0.2cm] {\scriptsize $0.75$ } (E3);
\path [-] (E1) edge node[sloped,above] {\scriptsize $0.25$} (E4);
\path [-] (E2) edge node[above] {\scriptsize $0.25$ } (E3);
\path [-] (E3) edge  node[left] {\scriptsize $0.5$} (E4);
  
\end{tikzpicture}}

\end{center}
\caption{\label{fig:co-occurrence-construction-ex} A sample co-occurrence graph ($G$) of topic elements.
Nodes are candidate topic elements and edge labels represent weights. 
The table at the top show the linked entities.
Here, the entities are \dbp\ resources, shown using the namespace \textit{dbr} is \url{http://dbpedia.org/resource} (\ie\ \instance{dbr}{Terry\_stop} is \url{http://dbpedia.org/resource/Terry\_stop}).
The spots within the microposts are encircled with a box. 
The entities are shown in the second column.  
Note that alternative forms of terry stopping have been linked to the same entity (\instance{dbr}{Terry\_stop}). 
The co-occurrence graph captures the elements of topics within posts. 
Its nodes represent the entities and the edges represent the degree to which their corresponding spots co-occurred in the posts. }
\end{figure}

To represent collective topics (those of interest to many people) the weak elements are eliminated prior to identifying the topics (Line~\ref{algl:f}).
The  weak edges ($w(e)<\tau_{e}$) are removed.
All vertices that become disconnected due to edge removal are also removed. 
The following equations describe how $G=(V,E)$ is pruned to obtain the final co-occurrence graph  $G'=(V',E')$: 
\begin{equation}
E'=\{e|e\in E\wedge w(e)\geq \tau_{e}\} \label{eq:EdgeTransformation}
\end{equation}
\begin{equation}
V'=\{v|\exists x [ (x,v) \in E' \vee (v,x) \in E']\}
\label{eq:VertexTransformation}
\end{equation}
The co-occurrence graph $G'$  represents all related topic elements (Line~\ref{algl:f}). 
\citeSupporting{S2_Fig} shows a co-occurrence  graph obtained at this step.


\label{sec:model-identifying-topics}

$G'$ is  processed to yield sets of related topic elements, each of which will represent a topic (Line~\ref{algl:i}). 
The criteria for determining the topics (sub-graphs of $G'$) are (1) an element may belong to several topics since it may be related to  many topics, (2) topics with more elements are preferable as they are likely to convey richer information, (3)
topics with few elements are relevant if their relationships are strong (\ie\ topics of intense public interest such as the death of a public figure).

\begin{figure}[!h]
\begin{center}
\includegraphics[scale=0.2]{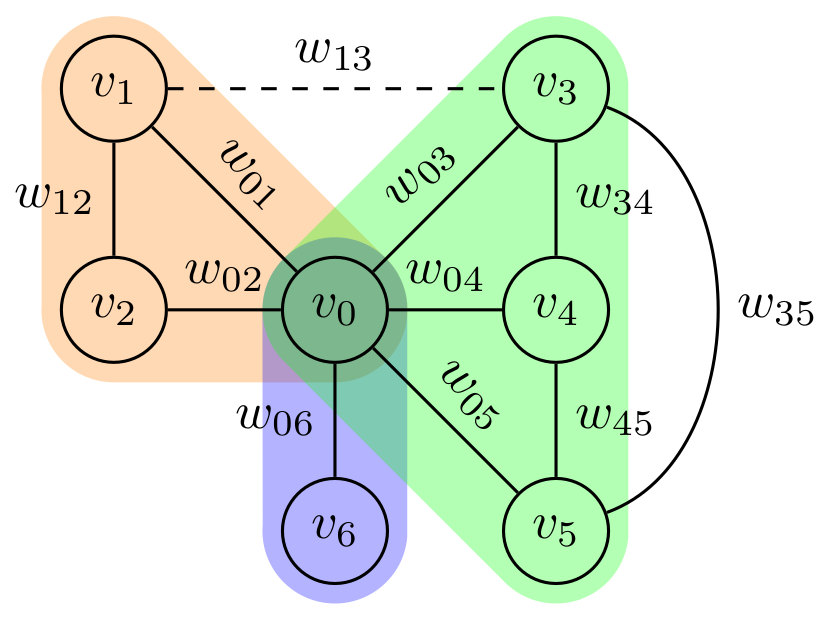}
\caption{{\bf An example entity co-occurrence graph and its corresponding topics.}
A co-occurrence graph, where  $v_i$ are candidate topic elements and $w_{ij} = w(e_{ij})$.
The edge  $e_{ij}$ is eliminated for being weak ($w_{13}< \tau_{e}$).  
Three maximal cliques (topics) emerge: $\{{v}_0,{v}_1,{v}_2$\},  $\{{v}_0,{v}_6\}$,  and  $\{{v}_0,{v}_3,{v}_4,{v}_5\}$. 
If $\textit{freq}({v}_0)<\tau_{kc}$ or $\textit{freq}({v}_6)<\tau_{kc}$ then $\{{v}_0,{v}_6\}$ will also be eliminated.
\label{fig:model-topic-ex}} 
\end{center}
\end{figure}

The maximal cliques algorithm~\cite{DBLP:journals/corr/abs-1006-5440} is used to determine the sub-graphs, where for a graph $G=(V,E)$, $\mathbb{C} \subseteq V$ is a clique $\iff \forall_{v,v' \in \mathbb{C }} (v,v') \in E, v \neq v'$ . 
Maximal cliques are sub-graphs that are not subsets of any larger clique.
They ensure that all elements in a sub-graph are related to each other.
An examination of the maximal cliques obtained from co-occurrence graphs revealed that many of them had a few (two or three) vertices.
This is not surprising since it is unlikely that many elements become related through many posts.
Another observation is that some elements (vertices) occur with a far lower frequency than the others. 
Since topics with few very weak elements are not likely to be of great interest, they are 
eliminated with the use of the $\tau_{sc}$ threshold: 
$ \frac{\textit{freq}(v)}{|P|} < \tau_{sc} $ where  $v \in V$ and $ \textit{freq}(v) = | \{ \langle p,l \rangle | v \in l \land \langle p,l\rangle \in \mathit{LEP} \} |$.
Fig~\ref{fig:model-topic-ex} shows an example of how topics are identified from graphs. 

After the elements of the topics are determined, additional information necessary to represent the final topics is obtained.
The observation interval is computed using the posts with the earliest and the latest timestamps in $P$ (Line~\ref{algl:m}). 
Since our \stopics\ represent \textit{persons}, \textit{locations}, and \textit{temporal expressions} the entity types of elements are resolved (Lines~\ref{algl:g}-\ref{algl:h}).
The type of \textit{temporal expression}s is determined while they are being extracted.
The types of other elements are identified with semantic queries. 
For example, if the  value of the \instance{rdf}{type} property includes \class{foaf}{Person} or \class{dbo}{Person} its type is considered to be a person.
To determine if an entity is a location, first, the value of \prop{rdf}{type} is checked for a location indicator (\class{schema}{Place}, \class{dbo}{PopulatedPlace}, \class{dbo}{Place}, \class{dbo}{Location}, \class{dbo}{Settlement}, \class{geo}{SpatialThing} and \class{geonames}{Feature}).
Locations are quite challenging as they may be ambiguous and be used in many different manners. 
Then, the contexts of the spots corresponding to entities are inspected for \textit{location indicator}s (succeeding the prepositions {\em in}, {\em on}, or {\em at}) within the post-collection.
Again, we employ a threshold ($\tau_{loc}$) to eliminate weak elements of location type.
Finally, an entity $v$ is considered a location if $\frac{ | \mathit{locationPrepositions}(v,\mathit{LEP}) |}{ |P|} > \tau_{loc}$.
For example, the entity {\em FBI} in {\em ``FBI reports to both the Attorney General and the Director of National Intelligence.''} is considered  as an agent, whereas in {\em ``I'm at FBI''} it is considered as a location.



At this stage, we have the sub-graphs of $G'$ (topics) along with the types of topic elements.
The final phase is to represent these topics with the \topico\ ontology.
Each $t \in T$ is mapped to an instance of \class{topico}{Topic} (Lines~\ref{algl:j}-\ref{algl:k}).
The topics are related with their elements in accordance in accordance with their types.
For example, elements  of type \textit{person} are associated with a topic with the \prop{topico}{hasPerson} property.
The property \prop{topico}{isAbout} is used for all elements of type other than the types of \textit{person}, \textit{location}, and \textit{temporal expression}.
The observation intervals  are associated with the \prop{topico}{observationInterval} property. 
The instantiated topics are referred to as \stopics\  and are ready for semantic processing.

\label{sec:approach-discussion}
Note that other graph algorithms could be used to obtain topics. 
Also, alternative pre- and post-processing steps could be utilized.  
For example, it may be desirable to eliminate or merge some topics to yield better results.
For illustration purposes, let's consider the consequence of using the maximal-cliques algorithm on pruned graphs.
The maximal-cliques algorithm requires all of its elements to be related. 
The pruning of weak bonds introduces the potential of severing the relations necessary to be identified as a topic.
In such cases, very similar topics  may emerge, such as those that differ in only a single element. 
This conflicts with our desire to favor a variety of topics with higher numbers of elements.
Not pruning the graph to prevent such cases would unreasonably increase the cost of computation since the original graphs are very large and consist of many weak relations.
A post-processing step could be introduced to merge similar topics with the
 use of two thresholds: $\tau_{c}$ for topic similarity and $\tau_{e_{min}}$ for an absolute minimum edge relevancy weight.
Let $T_0$ be the set of cliques ($T_0 \subset P(V')$). 
The set of merged cliques, $T$, is obtained by:
 \begin{equation}
    T =
    \begin{cases}
      (T_0\! \setminus\{t_i,t_j\}) \ \ccup\ \{t_k\}  & \text{if}\ \text{jaccard}(t_i,t_j)> \tau_{c} \ \lland\  \forall v_x,v_y\  [w((v_x,v_y)) > \tau_{e_{min}} ]\\
      T_0 & \text{otherwise}
    \end{cases}
  \end{equation}
\noindent where $t_k= t_i \ccup t_j, t_i,t_j\!\!\in\!\textsc{t}_0, v_x\!\!\in\!t_i, v_y\!\!\in\!t_j$.
Higher values for $\tau_{c}$ or $\tau_{e_{min}}$ lead to more topics that are similar to one another.

\subsection{Prototype infrastructure}
\label{sec:implementation-candidate-element-identification}
\label{sec:post-processing-cliques}
\label{sec:implementation}
\label{sec:implementation-semantic-topic-instantiation}

The services used to acquire external information are: the \tagme\ \api\ for entity linking suggestions, the \dbp\ and \wikidata\ for fetching semantic resources to be used as topic elements, and the Phirehose Library~\cite{phirehoseWeb} for continuously fetching posts from the Twitter streaming \api\ filter endpoint~\cite{FilterEndpointWeb}.
\tagme\ and Twitter have granted us access tokens to make \api\ requests.
\dbp\ makes resources available under the Creative Commons Attribution-ShareAlike 3.0 License and the GNU Free Documentation License. 
\wikidata\ resources are under CC0 license which is equivalent to public domain access (both of which provide a public \sparql\ endpoint).
Our implementation has complied by all the terms and conditions related to the use of all services.
The prototype was deployed on a virtual machine based on VMWare infrastructure running on Intel Xeon hardware with 2 \textsc{gb}s of \textsc{ram} and Linux operating system (Ubuntu).
The implementation of the maximal-cliques algorithm~\cite{DBLP:journals/corr/abs-1006-5440} is run within the R~\cite{CiteR} statistical computation environment.
In all of the processes mentioned above, we  use a local temporary cache to reduce unnecessary  \api\ calls to reduce network traffic.


Finally, all topics are represented as instances of \class{topico}{Topic}, serialized into \owl, and stored in a Fuseki~\cite{FusekiWeb} server (a \sparql\ server with a built-in \owl\ reasoner) for further processing.

\section{Experiments and results}
\label{sec:evaluation}

The main focus of this work is to examine the feasibility of using \lod\ resources to identify useful processable topics.
Accordingly, our evaluation focuses on the examination of the characteristics and the utility of the generated \sbounti\ topics. 
We considered it important to generate topics from real data, which we gathered from Twitter.
The quality and utility of the resulting topics are examined by: 
\begin{itemize}
\item{inspecting the characteristics of their elements (\nameref{sec:resulting-semantic-topics} section)},
\item{comparing  the effort required to perform various tasks  in comparison to topics generated by words-list-based (\kwb) approaches (\nameref{sec:semantic-processing} section)}, and
\item{manually assessing their relevancy (\nameref{sec:user-evaluation} section)}.
\end{itemize}

Furthermore, to gain insight into the similarity of topics with topics generated by other methods we compared them with human-readable (\nameref{sec:comparison-with-bounti} section) and \kwb\ topics (\nameref{sec:comparing-with-lda} section).

\subsection{Datasets}
\label{sec:datasets}

For evaluation purposes  \stopics\ were generated from  $11$  datasets consisting of $1,076,657$ tweets collected during significant events~\cite{TopicDownload}.
The Twitter Streaming API was used to fetch the tweets via queries which are summarized in Table~\ref{tab:datasetsSubject}.
The first four sets were fetched during the 2016 U.S. election debates, which are significantly greater than the others.
We expected that there would be a sufficient quantity of interesting tweets during the debates, which was indeed yielded plenty of divers tweets ($\sim$48 tweets/second) resulting in very large datasets.
The remainder of the datasets (except  \public) were collected during other notable events. 
These are focused due to a particular person such as Carrie Fisher or a concept such as \textit{concert}.
The debates related sets were collected for the duration of the televised debates and the remainder were collected until they reached at least \num{5000} posts. 
The \public\ dataset was collected to inspect the viability of topics emerging from tweets arriving at the same time but without any query (\textit{public stream}).
Note that the Twitter \api\ imposes rate limits on the number of tweets it returns during heavy use.
Although they do not disclose their selection criteria, the tweets are considered to be a representative set. 

\begin{table}[!ht]
\small
\centering
\caption{{\bf The queries to fetch the datasets from Twitter and information about the collections.}}
\begin{tabular}{|c|p{2.4cm}|p{2.9cm}|r|c|}
\hline 
\multicolumn{1}{|c|}{\bf Dataset} & \multicolumn{1}{c|}{\bf Explanation} & \multicolumn{1}{c|}{\bf Twitter Query} & \multicolumn{1}{c|}{\bf Start time (UTC)} & \multicolumn{1}{c|}{\bf$\Delta$t (m)}\\ \thickhline
\pdone & 2016 First presidential debate & election2016, 2016election, @HillaryClinton, @realDonaldTrump, \#trump, \#donaldtrump, \#trumppence2016, hillary, hillaryclinton, hillarykaine, @timkaine, @mike\_pence, \#debates2016, \#debatenight & 2016-09-27T01:00:00Z & 90 \\ \hline
\pdtwo & 2016 Second presidential debate & same as \pdone & 2016-10-10T01:00:00Z& 90 \\ \hline 
\pdthree & 2016 Third presidential debate & same as \pdone & 2016-10-20T01:00:00Z& 90  \\ \hline 
\vp & 2016 Vice presidential debate & keywords in \pdone, \#vpdebate2016, \#vpdebate\ & 2016-10-05T01:00:00Z& 90\\ \hline 
\brangelina & The divorce of Angelina Jolie and Brad Pitt & \#Brangelina & 2016-09-20T23:38:38Z& 21 \tabularnewline \hline 
\carriefisher & Carrie Fisher's death & Carrie Fisher & 2016-12-28T13:59:50Z& 15 \\ \hline 
\concert &  Concerts & concert & 2016-12-02T19:00:00Z & 60 \\ \hline 
\northdakota & North Dakota demonstrations& north dakota & 2016-12-03T06:59:48Z & 14\\ \hline 
\tonibraxton  & Toni Braxton (trending) & Toni Braxton &2017-01-08T07:08:56Z& 765 \\ \hline
\inauguration & Inauguration of President Trump & \#inauguration, Trump, @realDonaldTrump & 2017-01-21T20:41:44Z& 6\\
\hline 
\public & A sample of public English tweets& \textit{(no keyword)} & 2016-12-02T20:29:53Z& 8 \\ \hline 
\end{tabular}
\label{tab:datasetsSubject}
\end{table}

Table~\ref{tab:datasets} shows the number of posts and the ratios of distinct posters.
The number of posts during the debates (\pdone,\pdtwo,\pdthree, and \vp) are fairly similar. 
For all datasets, the percentage of unique contributors is generally greater than  $70\%$, which is desirable since our approach aims to capture topics from a collective perspective.

\begin{table}[!ht]
\centering
\caption{{\bf The datasets used to create \stopics.}}
\begin{tabular}{|c|r|r|r|}
\hline 
\multicolumn{1}{|c}{\bf Dataset}& \multicolumn{1}{|c|}{\bf Posts} & \multicolumn{2}{c|}{\bf Distinct-Poster} \\ \hline
\multicolumn{1}{|c|}{(label)} & \multicolumn{1}{c|}{(\#)}& \multicolumn{1}{c|}{(\#)} & \multicolumn{1}{c|}{\bf (\%)} \\ \thickhline
\pdone& \num{259200} & \num{206827} & 79\\ \hline
\pdtwo &\num{259203} & \num{187049} & 72 \\ \hline 
\pdthree &\num{258227} & \num{181436} & 70 \\ \hline 
\vp & \num{256174} & \num{135565} & 52\\ \hline 
\brangelina  & \num{5900} & \num{4777} & 79 \tabularnewline \hline 
\carriefisher & \num{7932} & \num{6753} & 85 \\ \hline 
\concert & \num{5326} & \num{4743} & 89\\ \hline 
\northdakota &\num{7466} & \num{6231} & 83 \\ \hline 
\tonibraxton & \num{5948} & \num{4506} &75\\ \hline
\inauguration & \num{5809} & \num{5425} & 93 \\
\hline 
\public & \num{5472} & \num{5365} & 98 \\ \hline 
\end{tabular}
\label{tab:datasets}
\end{table}

\sbounti\ topics are generated from collections of tweets.
The debate datasets were segmented into sets of tweets posted within a time interval to capture the temporal nature of topics.
Throughout the remainder of this paper, a collection of  posts will be denoted as
$[ds_{id}][t_s,t_e)$ where  $ds_{id}$ is the name of the dataset, and $t_s$ and $t_e$ are the starting and ending times of a time interval. 
For example, \pdone\ $[10,12)$ refers to the tweets in \pdone\ that were posted between the \nth{10} to the \nth{12} minutes of the 90-minute long debate. 
The earliest tweet is considered to be posted at the \nth{0} minute, thus  $t_s=0$ for the first collection of a dataset.

\subsection{Experiment setup}
\label{sec:prototypesetup}

The first consideration is to determine the size of the collections. 
Streams of posts can be very temporally relevant, as is the case during events of high interest (\ie\ natural disaster, the demise of a popular person, political debates).
Furthermore, the subjects of conversation can vary quite rapidly. 
Short observation intervals are good at capturing temporally focused posts. 
Processing time is also significant in determining the size of the collections.
When the rate of posts is high, the \api\ returns approximately \num{5800} tweets per two minutes.
Under the best of circumstances  (when all  required data  is retrieved from a local cache), the processing time required for a collection of this size is approximately four minutes. 
Whenever  \api\ calls are required the processing time increases.
We experimented with generating  \stopics\ with collections of different sizes and decided on limiting the size of collections to about \num{5000}--\num{8000} posts. 
This range resulted in meaningful topics with reasonable processing time.
During heavy traffic, it corresponds to approximately 2--3 minutes of tweets, which is reasonable when topics tend to vary a lot.

As described earlier, our approach favors topics with a higher number of elements of significant strength. 
Table~\ref{tab:prototypeSetup} shows the values of the thresholds we used to generate the topics where all values are normalized by the collection size.

\begin{table}[!ht]
\centering
\caption{{\bf The values of thresholds for generating topics.}}
\begin{tabular}{|l|S[table-format=1.4,round-mode=off]|l|}
\hline
& \textbf{Value} & \textbf{Description} \tabularnewline \thickhline
$\tau_\tagmeP$ & 0.15 & entity link confidence\tabularnewline \hline
$\tau_\rho$ & 0.35 & spot confidence \tabularnewline \hline
$\tau_{e}$ & 0.001 & weak edge pruning weight \tabularnewline \hline
$\tau_{sc}$ & 0.01 & small clique removal \tabularnewline \hline
$\tau_{c}$ & 0.8 &  clique merge similarity  \tabularnewline \hline
$\tau_{e_{min}}$ & 0.0005 & minimum edge weight for clique merge  \tabularnewline \hline
$\tau_{loc}$ & 0.01 & weight of location entities with preposition \tabularnewline \hline

\end{tabular}
\label{tab:prototypeSetup}
\end{table}

The thresholds $\tau_\rho$ and $\tau_p$ are confidence values used to link entities as defined by the \tagme\ \api, are set to the recommended default values. 
Higher values yield fewer candidate topic elements, thus fewer topics.

Crowd-sourcing platforms typically exhibit long tails (a few items having relatively high frequencies and numerous items having low frequencies), which is also observed for  entities we identified.
For example, highly interconnected and dominant six entities in a co-occurrence graph extracted from \pdone\ are:  \textit{Debate, Donald\_Trump, Hillary\_Clinton, year:2016, Tonight}, and \textit{Now} with weights of  $0.12$, $0.11$, $0.11$, $0.10$, $0.07$, and $0.03$ respectively (maximum edge weight is $0.12$).
Similar distributions are observed in other collections.

All the thresholds are set heuristically based on experimentation.
The threshold for eliminating weak edges ($\tau_e$)  is set to $0.001$ based on the desire to capture entities with some  agreement among posters.
The threshold $\tau_{sc}$ that is used to eliminate small cliques consisting of  weak elements is set to $0.01$.
The threshold for  clique similarity $\tau_c$ is set to $0.8$.
Similar  cliques were merged as a post-processing step (as explained in the \nameref{sec:approach} section) where  $\tau_{e_{min}}$ is set to $0.0005$  (~$\tau_e/2$).
Finally, $\tau_{loc}=0.01$ to decide whether an entity is collectively used as a location.
The cliques  that remain after applying these thresholds are considered as collective topics.

To examine the impact of pruning applied before identifying topics, we traced the topic elements  to the  posts from which they were extracted.
The percentage of posts that end up in the topics vary according to the dataset, with an average of 58\% for vertices and 43\% for edges (see~\citeSupporting{S2_Table}). 
Since the remaining vertices and edges are relatively strong, the resulting topics are considered to retain the essential information extracted from large sets of tweets.

\subsection{Semantic topic characteristics}
\label{sec:resulting-semantic-topics}

Table~\ref{tab:TopicElementTypes} summarizes the type of elements of the generated \sbounti\ topics.
Most of them have persons, which is not surprising since tweeting about people is quite common.
Topics with persons emerged regardless of whether the query used to gather the dataset included persons.
Temporal expressions occurred more frequently in topics that were generated from  datasets that correspond to events where  time is more relevant (\ie\ \concert). 

\begin{table}[!ht]
\centering
\caption{{\bf The frequencies of the types of topic elements}}
\begin{tabular}{|l|r|r@{ }|r|r@{ }|r|r@{ }|r|r@{ }|r|}
\hline 
\multicolumn{1}{|c|}{} &\multicolumn{1}{c|}{\bf Topic} & \multicolumn{2}{c|}{\bf Person} &\multicolumn{2}{c|}{\bf Location} & \multicolumn{2}{c|}{\bf Temp.} & \multicolumn{2}{c|}{\bf isAbout}\\ \hline
\multicolumn{1}{|c|}{\bf Set} &\multicolumn{1}{c|}{\bf \#} & \multicolumn{1}{c|}{\bf \#} & \multicolumn{1}{c|}{\bf \%} & \multicolumn{1}{c|}{\bf \# }& \multicolumn{1}{c|}{\bf \%} & \multicolumn{1}{c|}{\bf \#} & \multicolumn{1}{c|}{\bf \%} & \multicolumn{1}{c|}{\bf \#} & \multicolumn{1}{c|}{\bf \%} \\ \thickhline 
\pdone & \num{1221} & \num{1121} & 91 & 8 & 0.6 & 808 & 66 & \num{1129} & 92\\ \hline
\pdtwo & \num{1120} & \num{1068} & 95 & 32 & 2 & 559 & 49 & 1,010 & 90\\ \hline
\pdthree & \num{1214} & \num{1130} & 93 & 11 & 0.9 & 265 & 21 & \num{1118} & 92 \\ \hline
\vp & \num{1511} & \num{1377} & 91 & 50 & 3 & 395 & 26 & \num{1380} & 91\\ \hline
\brangelina & 9 & 6 & 66 & 0 & 0 & 7 & 77 & 7 & 77 \\ \hline
\carriefisher & 35 & 34 & 97 & 0 & 0 & 18 & 51 & 27 & 77 \\ \hline
\concert & 31 & 7 & 22 & 2 & 6 & 19 & 61 & 29 & 93 \\ \hline
\northdakota & 43 & 5 & 11 & 40 & 93 & 11 & 25 & 43 & 100 \\ \hline
\tonibraxton & 46 & 46 &100 & 0 & 0 & 1 & 2 & 43 & 93\\ \hline
\inauguration& 18 & 18 & 100 & 8 & 44 & 9 & 50 & 17 & 94\\ \hline
\end{tabular}
\label{tab:TopicElementTypes}
\end{table}

The viability of our method requires the ability to link posts to linked data, thus the linked entities must be examined.
From our datasets, some of the spots and its corresponding topic element that we observe are:
\lentity{donald, trump, donald trump, donald j. trump, donald j.trump}{\instance{dbr}{Donald\_Trump}}, 
\lentity{stop and frisk, stopandfrisk, stop-and-frisk}{\instance{dbr}{Terry\_stop}}, 
and \lentity{racial divide, racial profiling, racial profile, racial segment, racial violence}{\instance{dbr}{Racial\_profiling}}.
As seen, the topic captures the intended meaning regardless of how it was expressed by the contributors. 


Tweets can be very useful in tracking the impact of certain messaging since people tend to post what is on their mind very freely.
Considering political campaigns, much effort is expended on deciding their talking points and how to deliver them. 
The ability to track the impact of such choices is important since it is not easily observable from the televised event or its transcripts.
To give an example, during the \nth{86} minute of the 2016 U.S. vice presidential debate, the candidates were talking about abortion and its regulation.
The topics of \vp\ $[86,88)$ relate to the vice-presidential candidates Tim Kaine and Mike Pence and the subjects of  law, faith, and religion.
This shows that the audience resonated with the  debate at that time.
Counterexamples can be seen in some of the topics generated from the \inauguration\ dataset, which were gathered with query terms the inauguration of Donald Trump as the U.S. President.
It so happened that the {\em Women's March} event that was to take  place the subsequent day was trending and was related to the  inauguration through those who posted tweets during this time.
As a result, the topics included people such as \textit{Madonna} and \textit{Michael Moore} who were very active in the Women's March.
Besides, the locations London, France and Spain appeared in topics from tweets expressing support for the Women's March. 
As a result, the topics that were captured rather accurately reflected what was on the mind of the public during the inauguration.

Finally, we examined whether public streams (\public) would yield topics, which we expected they would not since there would not be sufficient alignment among posts. 
Indeed no topics were generated, however, some entities were identified.
We speculate that in public datasets collected during major events, such as earthquakes and terrorist attacks, the strength of ties could be strong enough to yield topics, although this must be verified.

\subsection{The utility of semantic topics in comparison to \kwb\ topics}
\label{sec:semantic-processing}

The main purpose of this work is to  produce topics that lend themselves to semantic processing.
To demonstrate the utility of \sbounti\ topics, we provide a comparative analysis with topics represented as lists of terms (\kwb ) in terms of the effort required to  perform various topic related tasks.
The effort is described with the use of the helper functions shown in Table~\ref{tab:evaluation-subtasks}.
This section describes how various tasks are achieved with \sbounti\ topics and what would need to be done if \kwb\ topics were used.
For readability purposes, the queries are described in natural language.
The \sparql\ queries may be found in the supporting material. 

\begin{table}[!ht]
\centering
\caption{{\bf Helper functions for performing topic related tasks.}}
\begin{tabular}{|l|c|}
\hline
\multicolumn{1}{|c|}{\textbf{Description}} & \textbf{Abbreviation} \tabularnewline \thickhline
Entity identification & \ei \tabularnewline\hline 
Type resolution & \tr \tabularnewline \hline
External resource utilization & \ex \tabularnewline \hline
Time of contribution & \ti \tabularnewline \hline
Rule definition & \rd \tabularnewline \hline
Query optimization & \qo \tabularnewline \hline
Semantic analysis & \sa \tabularnewline \hline
\end{tabular}
\label{tab:evaluation-subtasks}
\end{table}

\noindent \newline \textbf{Task 1: Who occurs how many times in the topics related to Hillary Clinton:} 

This is a simple task that can be achieved by querying the topics for persons who co-occur with Hillary Clinton (and counting them) as shown in Listing~\ref{lst:PeopleWithHillaryClinton}.

\begin{lstfloat}[!h]
\caption{{\bf Query: Who occurs how many times in the topics related to Hillary Clinton?} \label{lst:PeopleWithHillaryClinton}}
\begin{lstlisting}[language=SPARQL]
SELECT ?person (COUNT(?topic) AS ?C) 
WHERE {
   ?topic topico:hasPerson dbr:Hillary_Clinton;
          topico:hasPerson ?person.
   FILTER (?person NOT IN (dbr:Hillary_Clinton ) ) } 
GROUP BY ?person 
ORDER BY DESC(?C)
\end{lstlisting}
\end{lstfloat}

\noindent The first three results (out of 41) are:
\vspace{0.2cm}

\begin{tabular}{ll} 
\multicolumn{1}{c}{\textbf{Person}}& \multicolumn{1}{c}{\textbf{Count}}\\
\instance{dbr}{Donald\_Trump} & "2205"\textasciicircum\textasciicircum xsd:integer\\
\instance{dbr}{Bill\_Clinton} & "338" \textasciicircum\textasciicircum xsd:integer\\
\instance{dbr}{Tim\_Kaine}& "314"\textasciicircum\textasciicircum xsd:integer
\end{tabular}

\vspace{0.2cm}

Other persons include Barack Obama,  Ruth Bader Ginsburg, Al Gore, Ronald Reagan, Anderson Cooper, Abraham Lincoln, and George Washington. 
Since \sbounti\ topics represent persons (with \prop{topico}{hasPerson}) retrieving those who co-occur with Hillary Clinton is trivial.

To achieve the same results with \kwb\ representations, it is necessary to determine the words that represent the person Hillary Clinton, other people, and count them.
Thus, a type-resolution (\tr) task to identify persons is needed which requires entity identification (\ei) using an external resource (\ex) with information about people.

\noindent \newline \textbf{Task 2 - When do the topics related to women's issues occur?} 

This type of query is useful to know  when certain topics emerge and to track whether they trend, persist, or diminish within  streaming content.
The time of observation of subjects is significant since they tend  to rapidly change   on social media. 

The query in Listing~\ref{lst:WomenIssues} identified \num{166} topics in \num{66} time intervals related to topics about  \textit{abortion}, \textit{rape}, and \textit{women's health}.
Retrieving the time intervals  is straightforward, since the \prop{topico}{observationInterval} property  captures this information.
An inspection of the linked entities revealed that the posters used different terms related to these concepts, such as 
\lentity{rape, raped, rapist, rapists, raping, sexual violence, serial rapist}{\instance{dbr}{Rape}}.

\begin{lstfloat}[!h]
\caption{{\bf Query: When do the topics related to women's issues occur?} \label{lst:WomenIssues}}
\begin{lstlisting}[language=SPARQL]
SELECT DISTINCT ?startTime ?endTime WHERE {
   ?topic topico:observationInterval ?interval.
   ?interval time:hasBeginning ?begin.
   ?interval time:hasEnd ?end.
   ?begin time:inXSDDateTime ?startTime.
   ?end time:inXSDDateTime ?endTime.
   {?topic topico:isAbout dbr:Rape .}
   UNION {?topic topico:isAbout dbr:Abortion .}
   UNION {?topic topico:isAbout dbr:Women\'s_health.}}
\end{lstlisting}
\end{lstfloat}

To achieve the same results with \kwb\ representations, the terms related to the women's issues and the time intervals (\ti) corresponding to the time they appeared must be determined.

\noindent \newline \textbf{Task 3 - When do the top 50 issues related to topics including Hillary Clinton and/or Donald Trump appear?} 

This type of query is relevant for tracking how a particular messaging resonates with the public, such as  for political and marketing campaigns that involve immense preparation.

\begin{sidewaysfigure}[!h]
\begin{center}
\includegraphics[scale=0.6]{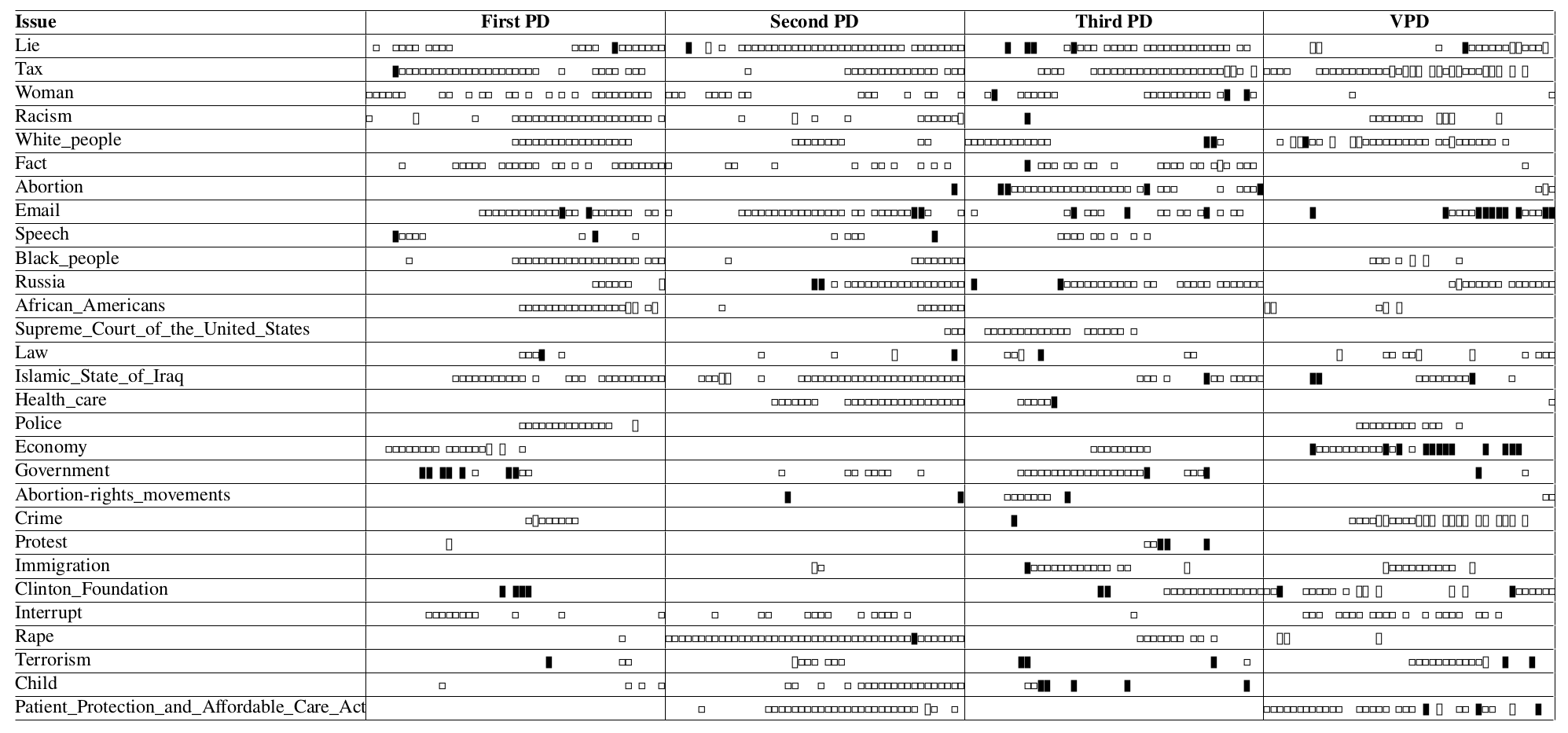}
\caption{{\bf The time intervals of topic elements that co-occur with the 2016 U.S. presidential candidates Donald Trump and Hillary Clinton during the two-minute intervals of the four debates.}
The symbols ($ \vrectangle$), ($\vrectangleblack$) and ($\mdsmwhtsquare$) respectively mark those that included Donald Trump, Hillary Clinton, or both.
\label{fig:semantic-query-results} }
\end{center}
\end{sidewaysfigure}

The query in Listing~\ref{lst:tableQuery} retrieves the top 50 issues (topic elements) associated with the topics including Donald Trump and/or Hillary Clinton and when they were observed.
To do this, first, the top 50 issues (\prop{topico}{isAbout}) in the topics related to  \instance{dbr}{Donald\_Trump} or \instance{dbr}{Hillary\_\-Clinton} are retrieved.
Then, when these issues occurred is determined.
This query returned  \num{3061} results, among which:

\begin{description}[itemsep=0cm,labelwidth=6em,align=parleft,labelsep=0.5cm,font=\normalfont,labelindent=\parindent]
\item[time:]{"\texttt{2016-10-10T01:38:00Z}"\textasciicircum\textasciicircum \class{xsd}{dateTime}} 
\item[issueOfInterest:]{\instance{dbr}{Patient\_Protection\_and\_Affordable\_Care\_Act}}
\item[person:]{\instance{dbr}{Don\-ald\_Trump}}
\end{description}

\noindent indicate that the issue of \textit{patient protection and affordable care act} occurred in topics with Donald Trump on 09 October 2016 at 21:38 EST (during the \nth{2} presidential debate).
Fig~\ref{fig:semantic-query-results} summarizes some of the issues that co-occurred  with \textit{Hillary Clinton} and/or \textit{Donald Trump} for each two-minute interval during the 90-minute long debates (\pdone,\pdtwo,\pdthree,\vp).

\begin{lstfloat}[!h]
\caption{{\bf Query: When do the top 50 issues related to topics including Hillary Clinton and/or Donald Trump appear?} \label{lst:tableQuery}}
\begin{lstlisting}[language=SPARQL]
SELECT ?time ?issueOfInterest ?person {
SERVICE <http://193.140.196.97:3030/topic/sparql>{
   SELECT ?issueOfInterest (COUNT(?topic) AS ?C)
   WHERE {
      ?issueOfInterest topico:inTopic ?topic.
      {dbr:Hillary_Clinton topico:isAPersonOf ?topic}
       UNION
      {dbr:Donald_Trump topico:isAPersonOf ?topic}}
      GROUP BY ?issueOfInterest
      ORDER BY DESC(?C) LIMIT 50}
SERVICE <http://193.140.196.97:3030/topic/sparql>{
   SELECT ?time ?about ?person 
   WHERE {
      ?topic topico:hasPerson ?person.
      ?topic topico:isAbout ?about.
      ?topic topico:observationInterval ?interval.
      ?interval time:hasBeginning ?intervalStart.
      ?intervalStart time:inXSDDateTime ?time.
      FILTER(?person IN 
        (dbr:Hillary_Clinton, dbr:Donald_Trump))}
   GROUP BY ?time ?about ?person}
FILTER (?about=?issueOfInterest)}
\end{lstlisting}
\end{lstfloat}

To gain some insight regarding how the resulting issues corresponded with the actual debates we inspected them along with their transcripts~\cite{FirstDebate2016Transcript,SecondDebate2016Transcript,ThirdDebate2016Transcript,VPDebate2016Transcript}.
For example, racism was an issue that was mostly discussed by the candidates during the second half of the first presidential debate and the first half of the third debate.
The topics we identified also revealed that racism was mostly posted during the same time (see the rows labeled {\em White\_people} and {\em Black\_people} in Fig~\ref{fig:semantic-query-results}).
Furthermore, an inspection of the tweets posted during the same time also included tweets related to racism that were posted in pro-Republican and pro-Democratic contexts.
For the topics related to {\em Tax} and  Donald Trump only (\vp$[48,50)$ and  \pdthree$[80,82)$)  the corresponding tweets  were indeed  related   to only Donald Trump.
On the other hand, while the candidates were talking about ISIS, Iraq, and the position of the United States in the Middle East 
(\pdthree$[68,70)$) the  identified topics were related to illegal immigration and income tax.
In this case, we observe a lack of resonance between what was transpiring during the debate and the topics of interest to the posters (who preferred to post about other matters).

This query demonstrates the use of the inverse relationships \prop{topico}{isAPersonOf} and \prop{topico}{inTopic}.
These relationships are inferred trough reasoning according to the definitions in \topico\ ontology (see descriptions of object relationships in Description Logic in \citeSupporting{S1_Appendix}).

To achieve the same result with \kwb\ topics, the terms indicating {\em Hillary Clinton}, {\em Donald Trump}, and the terms corresponding to the top 50 issues must be identified (\ei) which requires reference to  external resources (\ex).
Finally,  when the issues emerged must be identified (\ti).

\label{sec:eval:task4}
\noindent \newline \textbf{Task 4 - Which politicians occur in the topics?}

This task requires determining the occupation of persons, which may be of interest to known people. 
While \sbounti\ topics include the persons, the  \dbp\ entities that represent them often do not include the \prop{dbo}{occupation} or worse are related to incorrect entities.
However, Wikidata entities utilize \prop{wdt}{P106} (occupation) property with persons quite systematically.
Since \dbp\ refers to equivalent Wikidata entities with the \prop{owl}{sameAs} property, the occupation of a person can be retrieved from Wikidata. 
The query in Listing~\ref{lst:threeQueries} fetches the politicians  in topics extracted from the debates by: (1) fetching all persons in the \stopics\ from our endpoint, (2) retrieving the Wikidata identifiers of these persons from the \dbp\ endpoint, and (3) identifying the persons whose occupation (\prop{wdt}{P106}) is  a {\em Politician} (\instance{wd}{$Q82955$}) from the Wikidata endpoint.
Among the results are: \instance{dbr}{Abraham\_Lincoln}, \instance{dbr}{Bill\_Clinton}, \instance{dbr}{Colin\_Powell}, \instance{dbr}{Bernie\_Sanders}, and \instance{dbr}{Saddam\_Hussein}.
Query optimization (\qo) is performed to reduce the search space by prioritizing the sub-queries according to their expected response sizes. 
This example shows the benefits of using \lod\ in our topics, where the links within the entities lead to a multitude of options.

\begin{lstfloat}[!h]
\caption{{\bf Query: Which politicians occur in the topics?} This query performs three queries to the \sbounti\, \dbp, and Wikidata endpoints. {\em wdt:P106} refers to the occupation property. {\em wd:Q82955} refers to the {\em politician} concept. \label{lst:threeQueries}}
\begin{lstlisting}[language=SPARQL]
%{\#Query 1:}%
SELECT DISTINCT ?person WHERE {
   ?topic topico:hasPerson ?person}

%{\#Query 2:}%
SELECT ?DbPediaPerson ?wikidataPerson WHERE {
   ?DbPediaPerson owl:sameAs ?wikidataPerson .
   FILTER (?DbPediaPerson IN
   (<http://dbpedia.org/resource/Donald_Trump>,
    <http://dbpedia.org/resource/Lester_Holt>,
    ... )).
FILTER regex(str(?wikidataPerson),"^.*wikidata\\.org.*$")}

%{\#Query 3:}%
SELECT DISTINCT ?person WHERE {
   ?person wdt:P106 wd:Q82955 .
   FILTER (?person IN (wd:Q22686, wd:Q3236790,
    ... )) }
\end{lstlisting}
\end{lstfloat}

Performing this task with \kwb\ topics requires the identification of persons who are politicians (\ei, \ex, \tr).

\label{sec:eval:task5}
\noindent \newline \textbf{Task 5 - Which rock music band is performing where?}

Social media is often used to get information about events.
The query shown in Listing~\ref{lst:RockMusicConcert} fetches the bands and locations of rock music concerts.
This requires fetching the names of the bands performing rock concerts and the locations of these concerts.
The type of concert is determined with the use of \dbp\ resources (\ex).
This query returns results like \instance{dbr}{Guns\_N'\_Roses} and \instance{dbr}{Mexico\_City}.
Country music concerts are found by appropriately revising the query (\instance{dbc}{Country\_music\_genres}), which returns results like  \instance{dbr}{Luke\_Bryan} and \instance{dbr}{Nashville,\_Ten\-nessee}.
The locations of the concerts are retrieved from the tweets, whereas the genres of bands are determined from \dbp.
This query demonstrates the use of \class{topico}{LocationTopic} which is a subclass of \class{topico}{Topic}. 
The individuals of this type are determined trough reasoning according to the definitions in \topico\ ontology.

\begin{lstfloat}[!h]
\caption{{\bf Query: Which rock music band is performing where?} \label{lst:RockMusicConcert}}
\begin{lstlisting}[language=SPARQL]
SELECT ?musicGroup ?location {
SERVICE <http://193.140.196.97:3030/topic/sparql>{
   SELECT ?topic ?musicGroup ?location WHERE {
      ?topic topico:isAbout dbr:Concert .
      ?topic a topico:LocationTopic.
      ?location topico:isLocationOf ?topic.
      {?topic topico:isAbout ?musicGroup .}
      UNION 
      {?topic topico:hasPerson ?musicGroup .}}}
SERVICE <http://dbpedia.org/sparql>{
   SELECT ?musicGroup2 WHERE {
      ?musicGroup2 a schema:MusicGroup .
      ?musicGroup2 dbo:genre ?musicGenre .
      ?musicGenre dct:subject dbc:Rock_music_genres }}
FILTER (?musicGroup = ?musicGroup2)}
\end{lstlisting}
\end{lstfloat}

To achieve this with \kwb\ topics, the terms referring to the bands and locations (\ei\ and \tr) are needed. 
External resources (\ex) are needed to identify bands, locations, and the genre of the bands.
Also, as explained in the \nameref{sec:model-location-identification} section, the context of the location terms must be examined to determine if they were indeed used as locations.

\label{sec:eval:task6}
\noindent \newline \textbf{Task 6 - Which of the issues related to Barack Obama during the 2012 and 2016 U.S. election debates  are the same?}

In politics, it is useful to know which issues persist over time.
The query shown in Listing~\ref{lst:20122016Obama} queries the topics identified during the 2012 and the 2016 U.S. election debates.
For this purpose, topics were generated from the tweets gathered during the 2012 U.S. Presidential debates~\cite{dataset2012}.  
The resulting elements are: 
\instance{dbr}{Debate}, 
\instance{dbr}{President\_of\_the\_United\_States}, 
\instance{dbr}{Debt}, \instance{dbr}{Question}, 
\instance{dbr}{Tax}, 
\instance{dbr}{Tax\_cut}, 
\instance{dbr}{Golf}, 
\instance{dbr}{Economy}, 
\instance{dbr}{Black\_people}, 
\instance{dbr}{Racism}, 
\instance{dbr}{Violence}, 
\instance{dbr}{Birth\_certificate}, 
\instance{dbr}{Lie}, 
\instance{dbr}{Muslim}, 
\instance{dbr}{Barack\_Oba\-ma\_presi\-dential\_campaign,\_2008}, 
\instance{dbr}{Rus\-sia}, 
\instance{dbr}{Iraq}, 
\instance{dbr}{Immi\-gration}, 
\instance{dbr}{Blame}, and 
\instance{dbr}{Central\_Intelligence\_Agency}.
There are many issues one would expect to see in a presidential debate such as taxes, violence, and the economy.
Among other issues that appear in both years are racism, immigration, Muslim, Iraq, and Russia.
One might be surprised to see golf (\instance{dbr}{Golf}) in this list; alas, the amount of golf played by candidates seems to be a matter of public interest. 
An inspection of the tweets confirms that the amount of golf that Barack Obama played became a topic of discussion.

\begin{lstfloat}[!h]
\caption{{\bf Query: Which of the issues related to Barack Obama are the same in the 2016 and 2016 U.S. election debates?} 
This is a federated query that queries two endpoints, one for the debates in 2012 and one for the debates in 2016.
\label{lst:20122016Obama}}
\begin{lstlisting}[language=SPARQL]
SELECT ?about1 {
SERVICE <http://193.140.196.97:3031/topic/sparql>{
   SELECT DISTINCT ?about1 WHERE {
      ?topic1 topico:isAbout ?about1 .
      ?topic1 topico:hasPerson dbr:Barack_Obama}}
SERVICE <http://193.140.196.97:3032/topic/sparql>{
   SELECT DISTINCT ?about2 WHERE {
      ?topic1 topico:isAbout ?about2 .
      ?topic1 topico:hasPerson dbr:Barack_Obama}}
FILTER( ?about1=?about2)}
\end{lstlisting}
\end{lstfloat}

To obtain a similar result with \kwb\ representations, words common to topics of 2012 and 2016 must be retrieved.
The results would be terms rather than concepts. 
For conceptual results, entity identification (\ei\ and \ex) could be used.

\label{sec:eval:task7}
\noindent \newline \textbf{Task 7 - Which religions and ethnicity were mentioned during the 2012 and 2016 debates?}

\topico\ explicitly represents only persons, location, and temporal elements. 
To detect other types, external resources must be used.
The query shown in Listing~\ref{wikidataQuery1} utilizes knowledge about religions and ethnicities in Wikidata to 
retrieve related topics in the 2012 and the 2016 U.S. elections' debate topics with 
Query~1 that retrieves all religions from the Wikidata endpoint and
Query~2 that retrieves the topics that include any of the items fetched in Query~1. 
A program that optimizes this query by feeding the output of Query~1 to Query~2 is used for this task  (\qo).
The same process is repeated for ethnic groups.
The tweets themselves refer to specific religions or ethnicities (\ie\ Christian and Mexican).
This query enables retrieving information about religions and ethnicities independent of any specific instance.

\begin{lstfloat}[!h]
\caption{{\bf Query: Which religions were mentioned during the 2012 and 2016 debates?}
This query is issued using two queries. Query~1: Get the religions from Wikidata, where the property
$P279$* means all subclasses and $Q9174$ is the identifier for the religion class.
Query~2: Get the topics that include religions.
\label{wikidataQuery1}}
\begin{lstlisting}[language=SPARQL]
%{\#Query 1:}%
PREFIX wdt: <http://www.wikidata.org/prop/direct/>
PREFIX wd:  <http://www.wikidata.org/entity/>
SELECT DISTINCT ?item ?article
WHERE {
    ?item wdt:P279* wd:Q9174 .
    ?article schema:about ?item .
FILTER (
    SUBSTR(str(?article),9,17)="en.wikipedia.org/") .}

%{\#Query 2:}%
SELECT ?about (COUNT(?about) as ?C)
WHERE {
   ?topic topico:isAbout ?about .
   FILTER (?about IN (
     dbr:Buddhism, dbr:Jainism,
     ...
     dbr:Tapa_Gaccha, dbr:Zen))}
GROUP BY ?about
ORDER BY DESC(?C)
\end{lstlisting}
\end{lstfloat}

The query for 2012 returned only \instance{dbr}{Catholicism}, whereas for  2016 it returned   \instance{dbr}{Islam\_\-in\_\-the\_\-United\_\-States}, \instance{dbr}{Islam}, and \instance{dbr}{Sunni\_\-Islam}.
A manual inspection of tweets confirms the difference in tweeting about religion.
In 2012, Catholicism was a subject of concern related to abortion and in 2016 Islam became an issue in the context of the Iraq War  and the 9/11 terrorist attacks.

The issues regarding ethnicity 
in 2012 were 
\instance{dbr}{African\_Americans}, 
\instance{dbr}{Russians}, 
\instance{dbr}{Egyptians}, 
\instance{dbr}{Jews}, 
\instance{dbr}{Mexican\_Americans}, 
\instance{dbr}{Arabs}, and 
\instance{dbr}{Is\-raelis}.
Ethic references were also present during 2016, however with differing emphasis:
\instance{dbr}{Rus\-sians}, 
\instance{dbr}{Hispanic}, 
\instance{dbr:Asian\_Americans}, 
\instance{dbr}{Chinese\_Americans}, 
\instance{dbr}{Hispanic\_and\_Latino\_Americans}, 
\instance{dbr}{Mex\-ican\_Americans}, 
and \instance{dbr}{Mexicans}.

Furthermore, we observed that the topic elements that co-occur with \instance{dbr}{African\_Americans} also varied.
With the support and opposition to the \textit{black lives matter} movement, the elements \instance{dbr}{Police} and \instance{dbr}{Racism} were observed in 2016.

To accomplish this task with \kwb\ representations, the identification of religions and ethnic groups are needed (\tr) that requires entity identification (\ei) using an external resource (\ex).

\label{sec:task:new-relation}
\noindent \newline \textbf{Task 8 - Which people are related to the same topics?}

Semantic representation enables inference from present information, such as 
introducing the \prop{vcard}{hasRelated} property that specifies relationships among people and organizations (see vCard ontology~\cite{vcardprefix}).
This property can be used to relate people that occur in the same topic using the Semantic Web Rule Language (\swrl)~\cite{swrl-w3c} by defining the rule (\rd):

\begin{small}
\begin{verbatim}
Topic(?topic)^hasPerson(?topic,?person1)^hasPerson(?topic,?person2) 
             -> vcard:hasRelated(?person1,?person2)
\end{verbatim}
\end{small}

A reasoner can be used to relate all people who are in the same topic with the \prop{vcard}{hasRelated} relation, such as  \instance{dbr}{Donald\_Trump}, \instance{dbr}{Hillary\_Clinton} and \instance{dbr}{Lester\_Holt}.
Such rules support the introduction of subjective inquiries of interest.
Also, software agents aware of the vCard ontology may reason about this information.

In the \kwb\ case, persons in topics must be identified (\tr), which requires entity identification (\ei) using external resources (\ex).
There must be some way of expressing this relation so it can be referenced.

\noindent \newline \textbf{Task 9 - What are  the categories of topics?}

Topic  enrichment~\cite{TACL485,Newman:2007:SME:1255175.1255248,Tuarob:2013:ATR:2467696.2467706} through  external resources (\ex) may greatly enhance the utility of topics.
A useful enrichment for \stopics\ would be to relate them to their \dbp\ subject categories through the \prop{dct}{subject} property.
For example, the category of the \instance{dbr}{Job} is \instance{dbc}{Employment}, which indirectly relates
all  \stopics\  having  \instance{dbr}{Job} as a topic element to \instance{dbc}{Employment}. 
The following \textsc{swrl} rule (\rd) enriches \stopics\ with \prop{topico}{isAbout} relations to the categories of their elements:

\begin{small}
\begin{verbatim}
Topic(?topic)^isAbout(?topic,?element)^dct:subject(?element,?category) 
             -> isAbout(?topic,?category) 
\end{verbatim}
\end{small}

Thus, all topics related to the \dbp\ category \instance{dbc}{Employment} can be fetched with:

\begin{lstlisting}[language=SPARQL]
SELECT ?topic WHERE { 
       ?topic topico:isAbout dbc:Employment}
\end{lstlisting}

\vspace{0.5cm}

In this query, when the \dbp\ category \instance{dbc}{Employment} is replaced with \instance{dbc}{Law\_enforcement\_operations\_in\_the\_United\_States} the results include topics with the element \instance{dbr}{Stop-and-frisk\_in\_New\_York\_City}.
Therefore, with this simple rule definition, it becomes possible to relate the topics with their categories and query the topics according to these categories.

A similar enrichment for \kwb\ topics requires external resources (\ex) and functionality such as semantic analysis (\sa) of topics that could require considerable programming.

\label{sec:documentclassification}




Table~\ref{tab:task-summary} summarizes the effort to perform  \task{1} through \task{9} for \sbounti\ and \kwb\ approaches in terms of the subtasks that must be performed. 
The subtasks are described in terms of the helper functions, where those that must be performed numerous times are indicated with a subsequent parenthesized number (\ie~\tr(2), for  two type resolutions).
For the sake of brevity, we assume the existence of primitive functions (\ie~string, set, and list operations) and query support, which are not indicated in the comparison.

\begin{table}[!ht]
\centering
\caption{{\bf Subtasks to perform \task{1}-\task{9} for \kwb\ vs \sbounti\ topics .}}
\begin{tabular}{|S[table-format=2]|l|l|}
\hline
\textbf{Task}& \multicolumn{1}{c|}{\textbf{\kwb}} & \multicolumn{1}{|c|}{\textbf{\sbounti}} \\ \thickhline
1& \tr, \ex, \ei & \multicolumn{1}{c|}{--} \\ \hline
2& \ti & \multicolumn{1}{c|}{--} \\ \hline
3& \ei, \ti, \ex & \multicolumn{1}{c|}{--} \\ \hline
4& \tr, \ex, \ei & \ex\ (2), \qo \\ \hline
5& \ei (2), \tr\ (2), \ex\ & \ex \\ \hline
6& \multicolumn{1}{c|}{--} & \multicolumn{1}{c|}{--} \\ \hline
7& \tr, \ex, \ei & \ex, \qo \\ \hline
8& \tr, \ex, \ei & \rd \\ \hline
9& \sa, \ex & \rd, \ex \\ \hline
\end{tabular}
\label{tab:task-summary}
\end{table}

Semantically represented topics offer many opportunities when they are utilized in conjunction with resources and ontologies within \lod. 
The utility of \stopics\ is most apparent when it yields results that are not directly accessible in the source content. 
The use of semantic rules enables enriching topics with general or highly domain-specific information. 
The latter being quite lucrative for domain-specific applications.

\subsection{Topic relevancy assessment}
\label{sec:user-evaluation}

A comparative evaluation of the relevancy of \sbounti\ topics is difficult since the proposed approach has no precedence and produces topics that are significantly different from other approaches.
The effort required for manual evaluation is complex, highly time-consuming, and error-prone since it involves the simultaneous examination of large sets of tweets (approximately \num{5800}  per collection) and many semantic resources for every topic.
The level of effort and diligence required to evaluate topics through surveys or services such as Amazon Mechanical Turk \cite{mechanical-turk} that rely on human intelligence was deemed prohibitive.
However, to gain insight regarding the relevancy of the topics, a  meticulous evaluation was performed by the authors of this work with the
assistance of a web application we developed for this purpose (see \citeSupporting{S1_Fig}).
This tool presents a set of topics to be annotated as \textit{very satisfied}, \textit{satisfied}, \textit{minimally satisfied}, \textit{not satisfied}, or \textit{error} (when \textsc{uri}s are no longer accessible) along with optional comments  to document  noteworthy  observations.
During annotation, an evaluator may view  the tweets from which the topics were generated as well as a word cloud that presents the words in proportion to their frequency.
Also, the linked entities and temporal expressions extracted from the tweets can be inspected.

For evaluation purposes, $10$ topics from randomly selected $36$ intervals ($9$ from each debate) were annotated (\citeSupporting{S3_Table}).
Two annotators evaluated $24$ intervals, $12$ of which were identical to compute the inter-annotator agreement rate.
The topics to be evaluated were selected based on a higher number of topic elements since they result from higher levels of alignment among posters. 
As such, they were deemed more significant to evaluate.
Of the topics shown to annotators, there were  $3$ of size $8$,  $13$ of size $7$, $66$ of size $6$, $162$ of size $5$, $147$ of size $4$, $87$ of size $3$, and $2$ topics of size $2$.


The topics are presented per interval since they are all identified from the same collection.
The evaluator is expected to inspect each topic to determine if it is related to  tweet collection from which it was generated (by also inspecting the tweets).
Each element of each topic is inspected by visiting their  \dbp\ resources to determine their relevancy to the collection in the \textit{context} of the other elements of the topic.
An element that is related to the tweet set, but not in the context of the other elements  is considered  irrelevant. 
Each topic is labeled as: {\em very satisfied} only if all of the topic elements are valid; 
{\em satisfied} only if one of the topic elements is incorrect; {\em minimally satisfied} if more than one element is incorrect while retaining significantly valuable information; and {\em not satisfied} if several topic elements are incorrect (\ie\ the relative temporal expression may be true but does not convey sufficiently useful information).
Note that the evaluation was performed in a  strict manner, where a penalty is given for any kind of dissatisfaction -- regardless of  the source of the error.
For example, if a web resource on \dbp\ has incorrect information (which happens), the annotation of that topic is penalized.
This was done to avoid subjective and relative evaluation as well as to assess the viability of the resources being used.
Furthermore, since \sbounti\ topics are produced for machine interpretation the accuracy of topic elements is quite significant.
It is also easier to identify mistaken elements in contrast to assessing a whole document as an error.

The results are examined in two ways: for topics marked either {\em Very satisfied} or {\em Satisfied} (assuming general satisfaction) and for
topics annotated exclusively as {\em Very satisfied}.
The evaluation resulted in the precision and $F_1$ scores of  $74.8\%$,  $92.4\%$ when considering only those marked as \textit{Very satisfied}, and $81.0\%$,  $93.3\%$  when \textit{Very satisfied} or \textit{Satisfied}.
The $F_1$ scores (computed as defined by Hripcsak and Rothschild~\cite{hripcsak2005agreement}) indicate a high degree of agreement among annotators.


\subsection{Comparison with human-readable topics}
\label{sec:comparison-with-bounti}

In an earlier work (\bounti~\cite{bounti}), we identified human-readable topics  from collections of microblogs (Wikipedia page titles).
\bounti\ models collections of posts as bags of words and compares their \tfidf\ vector with the content of Wikipedia pages to identify a ranked list of topics.
The titles of the pages represent topics that are easily human-interpretable. 
\bounti\ topics are satisfactory for human consumption, especially since they are descriptive titles produced by the prolific contributors of Wikipedia.
Misleading topics can result when several subjects are posted about with similar intensities, such as the topic {\em Barack Obama citizenship conspiracy theories} derived from the words {\em Barack} and {\em citizen} whereas context of  \textit{citizen} was in \tweettext{Hillary is easily my least favorite citizen in this entire country} - clearly not related to Barack Obama.
Such cases occur as a consequence of using bag-of-words to model the documents. 
\sbounti\ overcomes this issue by considering both the wider context of the collection and the local context of posts while identifying topics. 
The context of individual tweets is used to determine potential topic elements, while the context of collections to capture the collective interest and patterns of use.  

We inspected and compared \bounti\ and \sbounti\ topics by deriving them from the same datasets (see \citeSupporting{S1_Fig}).
To give example, for \pdone\ [26-28), some of \bounti\ topics are: \textit{Donald Trump}, \textit{Hillary Clinton}, \textit{Bill Clinton}, \textit{Barack Obama's Citizenship}, and \textit{Laura Bush}.
Since  \sbounti\ topics include many elements, we will suffice by mentioning some of the topic elements: persons \instance{dbr}{Hillary\_\-Clinton},  \instance{dbr}{Donald\_\-Trump}, \instance{dbr}{Lester\_\-Holt} (the moderator of the debate) and other elements \instance{dbr}{Debate}, \instance{dbr}{ISIS}, \instance{dbr}{Fact},  \instance{dbr}{Interrupt}, \instance{dbr}{Watching}, and \instance{dbr}{Website}.
These elements are identified because people were talking about ISIS, a high level of interruptions during the debate, and Hillary Clinton's fact-checking website.

The evaluation of \bounti\ topics yielded  $79.3$\% and $89.0$\%   for precision and $F_{1}$ scores for those marked \textit{Very Satisfied} only and $88.9$\% and $94.0$\%  if annotated as \textit{Very Satisfied} or \textit{Satisfied}.
The scores for \bounti\ are higher, which is largely influenced by two factors. 
Firstly, \bounti\ topics are single titles that tend to be high level, thus they tend to be relevant even when not very specific.
For example, the \textit{Debate} topic would be considered relevant to a collection of posts about a specific debate at a specific time, even though it is very general.
In \sbounti, topics are more granular with more elements, thus the evaluator scrutinized each element to determine relevancy in a manner that penalizes mistakes. 
Since \sbounti\ topics are intended for machine processing, a harsher judgment is called for.

In summary, we have found some similarities between \sbounti\ and \bounti\ topics. 
In some cases, the corresponding \dbp\ resources of \bounti\ topics (Wikipedia pages) were elements of \sbounti\ topics that indicate a similarity between the results of \bounti\ and \sbounti.
Both approaches produce relevant topics, while \sbounti\ produces a greater variety and more granular topics in comparison to \bounti\ topics.
In general, \bounti\ captures higher-level human-readable (encyclopedic) topics, while \sbounti\ picks up on lower-level elements that provide conceptual information that lend themselves to a greater variety of machine-interpretation such as \textit{Barack Obama is a person and was a president}.

\subsection{Comparison with \kwb\ topics}
\label{sec:comparing-with-lda}

Latent Dirichlet allocation (\lda) is one of the most popular topic models, which makes a comparison with \sbounti\ interesting.
To perform a comparison, \lda\ topics are generated with \textsc{t}witter\-\lda~\cite{twitterLDA} with the two-minute intervals of the datasets \pdone, \pdtwo, \pdthree, and \vp\ that are used for generating \sbounti\ topics (using the default values of \lda\ $\alpha=0.5$, $\beta=0.01$, number of iteration$=100$).
Topics were generated for alternative values of \lda\ parameters for the expected number of topics: $N={2{-}10 ,20, 30, 40, 50}$. 

The topic representations of \sbounti\ and \lda\ are very different where
\lda\ topics capture terms expressed by contributors as words-list-based (\kwb) topics, \sbounti\ topics map original content to instances in \lod\ which are expressed with \lod\ resources and the \owl\ language~\cite{W3COWL}.
With \sbounti\ a set of alternative words that are contributed may be mapped to the same semantic entity, capturing the intended meaning rather than how it was articulated.

To get a rough idea about the similarity of topics, we utilized the label (\prop{rdfs}{label}) of the \sbounti\ topic elements.
The union of the lowercase form of words in the labels of all elements is compared using Jaccard similarity with topmost 10 terms of \lda\ topics (according to their distributions).
We observed that there are cases that \lda\ and \sbounti\ topic elements are the same but not matching due to some syntactic difference.
For example, if an \lda\ topic element is ``emails'', and \sbounti\ topic element is {\em dbr:Email}, the strings ``emails'' and ``email'' do not match which results in lower similarity scores. 
To address similar issues, we assumed that the cases that one term is a substring of the other are matching.
Each \sbounti\ topic is element-wise compared with \lda\ topics that are generated for the same input set.

The maximum similarity of an \sbounti\ topic in an interval is considered its similarity.
The average of such similarities in an interval is the similarity measure obtained from that interval.
And, the average similarity for all intervals is the average for a dataset, which ranges between 60-70\% with a maximum of 77\%.
Since the comparison of the topics is performed on elements of different levels, the results give a very rough idea. 
The semantic similarity is expected to be higher. 
We would have been concerned if the comparisons resulted in very low values since that would indicate a significantly different relation among topic elements.
As a result, we observe considerable coverage between the topics identified by these approaches, which is interesting for future work towards alternative methods for identifying topic elements.
For both methods, and in the case of using any other comparison methods to compare \sbounti\ topics with words-list-based topics, there is still the issue of word versus entity comparison.
Automatically assessing the relevancy of topics without a gold standard is a challenging issue that requires domain knowledge and understanding of ``topics'' in the domain.
We address these operations for future work.

\subsection{Evaluation summary}
\label{sec:overall-evaluation}

To assess the proposed approach, \sbounti\ topics were generated from sets of tweets and examined by inspecting their characteristics, using them in processing tasks, and comparing them with topics generated from \bounti. 
Our main inquiry was to assess the viability of generating topics from collections of microposts with the use of resources on \lod.
We found that considerable links between tweets and \lod\ resources were identified and that identifying topics from the constructed entity co-occurrence graph yielded relevant topics.
With semantic queries and reasoning, we saw that it was possible to reveal information that is not directly accessible in the source (tweets), which could be very useful for those (\ie\ campaign managers, marketers, journalists) who are following information from social media.

The proposed approach is a straightforward one aimed at gaining a basic understanding of the feasibility of mapping sets of tweets to semantically-related entities. 
If possible, this would facilitate a vast number of applications that harvest the richly connected web of data.
Our observations lead us to believe that this is possible.
Furthermore, this approach would improve by enhancing the techniques used to identify and relate topic elements, refining the topic representation, and with the increasing quality of data on \lod, which have been improving in terms of quantity and quality during the span of this work, a most encouraging prospect.
Potential improvements are elaborated in the following section.

\section{Discussion and future work}
\label{sec:discussion}


In this section, we discuss some of our observations regarding the approach we proposed and present some future directions. 
The  main objective of this work was to examine the feasibility of linking informal, noisy, and distributed micropost content to semantic resources in \lod\ to produce relevant machine-interpretable topics.
We specifically focused on subjects of significant interest from a collective perspective.
Topics of general interest lead to vast numbers of microposts.
We generated semantic topics from a variety of tweets collections and represented them with an ontology that we developed for this purpose.
The semantic topics were subjected to various tasks to examine their utility.
The results show that relevant topics were identified for a diverse set of subjects.
In the \nameref{sec:evaluation} section, we presented the semantic topics generated from collections of tweets with emphasis on a complete set collected during the four major debates of U.S.elections (a total of \num{1036800} tweets).
The utility of the resulting topics (respectively \num{1221}, \num{1120}, \num{1214}, \num{1511} number of topics) was demonstrated through various tasks that facilitated the understanding of the issues relevant to the debate watchers, such as the persons, the locations, the temporal and other aspects of interest. 
Furthermore, issues at higher conceptual levels such as violence, ethnicities, and religions were revealed.

\begin{figure}[!h]
\hspace{-3cm}\includegraphics[scale=0.3]{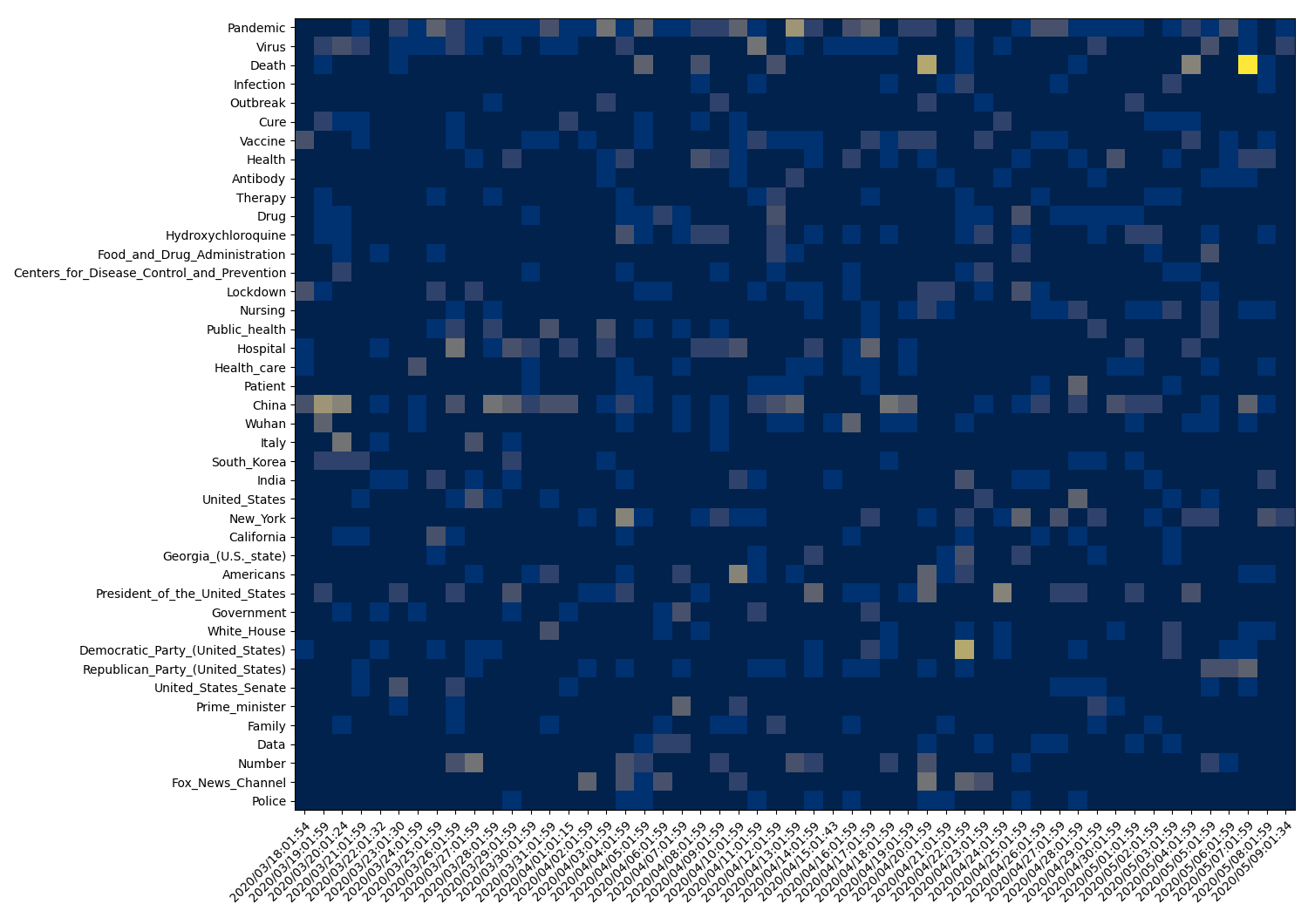}
\caption{{\bf Topic elements for \prop{topico}{isAbout} in topics generated from \covidnineteen\ dataset.}
The more yellow the color is the more topics exist including that topic element.
\label{fig:covid_topics}
} 
\end{figure}

In our experiments, we observed that our approach produces relevant topics for diverse contexts.
The topics of an entirely different context can be observed in a subject that is of great  interest during the final preparations of this article, namely the coronavirus 2 (SARS-CoV-2) pandemic (a.k.a. \covidnineteen) that is  widely reflected on social media. 
A preliminary exploration of topics generated from collections of tweets related to \covidnineteen\  also yielded relevant topics.
In this case collections of tweets posted during the same time for 53 consecutive days were inspected to get a general sense of the issues of relevance. 
There were \num{140} people, \num{32} locations, \num{46} temporal expressions, \num{1097} issues that distinctly occurred in the topics. 
Among the occupations of people are politicians (several heads of states), journalists, singers, and athletes. 
The locations were dominated by China, Wuhan, Italy. 
While many locations were across the globe (\ie\ Germany, West Bengal, and London) others were regions within the U.S. (\ie\ Texas, Michigan, and Louisiana).
This is reasonable since the time intervals of the tweets correspond to midday in the United States and \covidnineteen\ cases were spiking in various parts of the country.
There were also numerous temporal references, the most frequent ones being \textit{now}, \textit{today}, \textit{tonight}, the months of January through May. 
The occurrence and the frequency of the specific temporal terms are significantly different from those encountered in the debate related sets, which did not have such a diverse set of temporal expressions (mostly the year 2016 \textit{now},  \textit{tonight}).
Such differences capture the nature of contributions where the temporal aspect of the pandemic is indeed of much more significance due to the interest in how fast the rate of cases change and speculation about when things would improve.
The resulting topics were processed to see when various issues emerged.
Fig~\ref{fig:covid_topics} shows when the \textit{about} topic elements  (\prop{topico}{isAbout}) were observed daily.
Upon observing the references to drugs, we checked if other drugs were also referenced simply by querying \dbp\ if the element type is \prop{dbo}{Drug}. 
This identified the other drugs referenced in tweets as: BCG vaccine, Cocaine, Doxycycline, Favipiravir, Generic drug, Pharmaceutical drug, Polio vaccine, Ibuprofen,  Paracetamol, Chloroquine, Azithromycin, Antiviral drug, Hydroxychloroquine.
The most frequently referenced one was Hydroxychloroquine -- a drug mentioned by the president of the United States several times.
Obviously, tweets from such small intervals are insufficient to inspect such a vast issue.
A more comprehensive examination with collections that covers all time zones would be required.
Nevertheless, even with this small set, it is evident that this approach produces relevant topics.

The results we obtained are encouraging leaving us with many future directions to pursue, which we elaborate in the remainder of this section.

\subsection{Topic element detection}
\label{sec:discussion-improving-elements-of-topics}
\label{sec:discussion-linkeddata}

Since semantic topics consist of topic elements, correctly identifying them is important. 
Here, the main challenges are the inability to link to anything at all and incorrect linking.
Obviously, entity linking fails when suitable entities are not represented on \lod, such as when new subjects emerge.
In recent years the significance given to the creation and accessibility to open data resources has led to a rapid increase in the data represented on \lod~\cite{LinkedDataOrg} (see \nameref{sec:background} section for information about \lod). 
In this work, we eliminate unlinked spots, which could be particularly problematic for spots with high frequencies since that indicates common interest.
To alleviate this matter, such spots could be linked to an instance of \class{owl}{Thing} indicating that there is some \textit{thing} of significance whose type is unknown.  

For entities that exist but not successfully linked,  better approaches are required.
Named entity recognition and linking are active research areas that are improving across all domains and languages.
Another approach to determining the correct entity type ranks all relevant types using taxonomies and ontologies such as \textsc{yago}, and \textsc{Freebase}~\cite{Garigliotti2019,Tonon2016170}.
Also, additional pre-processing steps can be taken prior to entity linking, such as  tweet normalization~\cite{sonmez_emnlp_2014} and hashtag segmentation~\cite{celebi2017}.

As discussed in the \nameref{sec:evaluation} section, identifying locations is challenging since many entities can be  considered a location in some context.
We imposed some rules to determine if such elements qualified as a location in the context it was used.
Our evaluation revealed that although all elements we deemed to be locations were correct.
Unfortunately, we missed identifying some of them since they did not match our rules -- mostly due to how tweets were articulated.
Location prediction on Twitter is known to be challenging and is of significant interest since there are many areas of application~\cite{LocationPredictionTwitterSurvey,KUMAR2019365}.
It is of interest for many purposes, such as disaster tracking and mitigation and with the emergence of the \covidnineteen\ pandemic crisis, this work has been intensified. 
 Many studies appear as preprints, which are not yet vetted, however with the immense motivation improved location detection is expected.
Our rules for detecting locations must be revisited. 
Also, indirect location indicators such as those found in profiles and geotagged content~\cite{Chi2016GeolocationPI}, and in co-occurrence patterns~\cite{OZDIKIS20191280} could offer hints that improve the detection of locations. 
For ethical reasons, we do not (and do not intend to) use any profile information, but could consider utilizing other indirect signals.

A more troublesome issue stems from ambiguous terms, which is most prevalent in person names.
For example, the spot \textit{Clinton} was inaccurately linked to the \nth{42} U.S. president Bill Clinton instead of the 2016 U.S. presidential candidate Hillary Clinton in tweets regarding lying about Obamacare. 
In this case, both persons are politicians, one was a U.S. president and the other a U.S. presidential candidate, and they are spouses. 
Although this example is particularly challenging, the ambiguity of person names is generally challenging.
A similar issue arises since the titles of songs, movies, albums, and books are terms of ordinary conversation, such as time (\textit{Time} magazine),  cure (the music band \textit{The Cure}), and WHO (the television character \textit{Dr. Who}).
We encountered such cases in our experiments,  albeit not frequently, since the entity linker typically assigns low confidence rates for such links. 
Furthermore, our approach eliminates the links to entities that occur infrequently (see the \nameref{sec:model-topic-identification} section) in a collaborative filtering manner.

Recently, word-embedding techniques that capture semantic similarity among terms are being applied to named entity recognition (NER) and disambiguation~\cite{tacla00104,YamadaS0T16}, and entity linking~\cite{FrancisLandauD16,bertEL2019}.
These techniques represent terms as vectors in a high dimensional vector space and obtain them via machine learning given a corpus.
The semantics of terms are captured from the context of the terms. 
The vectors of semantically similar terms are close in the vector space.
Since emerging entities are expected to be included in the knowledge bases, the topic and/or topic elements could be periodically revisited for opportunities for improvement.
These advances in named entity detection and linking are very promising and are expected to positively improve the detection of topic elements.

In our experiments, we focused on English tweets and named entity recognition so that we could interpret the results.
Several tools work for other languages (including TagMe that we used in our prototype).
Furthermore, the natural language processing community strongly emphasizes work on low resource languages, which is resulting in additional knowledge resources and tools. 
The goal is to work with multiple languages, which link to the same conceptual entity. Thereby, being able to glean information regarding content that is globally produced. This is important for many tasks of global interest, such as pandemic diseases, disasters, news, entertainment, and learning material.

\subsection{Semantic topic identification}
\label{sec:discussion-improving-topics}

In this work, we chose maximal cliques to identify the topics so as to assure that all the elements are related by virtue of having been posted  together.
The co-occurrence graphs from which topics are extracted have relatively few nodes with high degree centrality (\ie\ Hillary Clinton and  Donald Trump in the debate sets) with the remaining node being relatively weak (see~\citeSupporting{S2_Fig}).
Thus, several topics extracted from such graphs tend to share the dominant nodes, which reflect the narratives related to the dominant nodes.
On the other hand, the nodes that are connected to the dominant nodes tend to fall into different topics since they are usually not  connected to each other.
This fairly accurately reflects micropost content (\ie\ many different topics involve Donald Trump).
However, this results in some topics seeming very similar or repetitive.
It is worth investigating more relaxed graph algorithms to increase the elements of topics while preserving the context.
For example, $k$-cliques (the maximal sub-graphs where the largest geodesic distance between any two vertices is $k$) constrained by  type rules could  yield richer topics. 
However, caution must be exercised, since the volume of microposts and their limited context is likely to yield many potential yet unrelated candidates for $k>1$, which is  computationally challenging and  costly. 
Note that the goal is not to simply increase the number of topics or their elements since we are aiming to reduce large sets of tweets to higher-level topics. 
Rather, the aim to increase the quality of topics by associating related elements.

The size of the post collections we used was limited by the rate limits of the Twitter streaming \api\ and  our computational resources.
For the debates, this corresponded to the tweets posted withing 2-minute intervals. 
During heavy posting conditions the subjects change frequently and short windows are suitable since the topics change and the number of tweets to detect collective interest is sufficient.
During slower posting conditions the subjects don’t change as fast, thus sets collected over longer durations are appropriate. 
Dynamically varying collection durations based on how frequently the subjects of topics change over time would be valuable. 
\topico\ is capable of representing such intervals, however, they must be determined through time series analysis.

\subsection{Semantic topic representation}

\topico\ specifies an elementary set of topic element types, namely person, location, temporal expressions, and other entities (those related by \prop{topico}{isAbout}.
It encompasses basic classes, object properties, and data properties to represent commonly occurring elements.
Inferred relations and classes support convenient processing.
This ontology could be extended to cover additional types (such as events, art, currency, character, products, drug,  and natural objects) as well as refine existing types (such as facility, address, astral body, organization, and market)~\cite{sekine-etal-2002-extended}.
While covering a wider type of cross-domain entities is of interest, we expect that customized for a specific domain will be quite interesting.
For this purpose,  ontologies relevant to the domain of interest and associated data resources are required. 
There are many useful ontologies and resources, especially in the life-sciences domain. 
Naturally, domain-specific tasks would also be defined.
The tasks shown in the \nameref{sec:evaluation}  illustrate the kinds of topic-related tasks that could be of interest to campaign managers, journalists, and political enthusiasts.

\subsection{Semantic topics utilization}

The purpose of focusing on semantic topics is for their semantic processing potentials.
We demonstrated how semantic topics can be utilized through various semantic tasks in  the \nameref{sec:evaluation} section.
The vision is to deliver this power to an end-user who is following the rapidly flowing distributed microposts.
Towards this end, higher-level tasks should be defined such as similarity, sentiment analysis, and recommendations.
Tracking topics, such as when they emerge, if they persist, if and when they spike, and if they exhibit some pattern is useful information.
Reports that provide statistical information will enable those who are interested in the topics to take action.
Other interesting tasks are tracking the evolution of and predicting topics.
One of the future directions is an explorer for  \sbounti\ topics which requires the generation of human-interpretable topics.
Using such an explorer users could search and browse topics, view ranked topics, graphs, and charts that provide relational and temporal information (trends), view social network analysis.
They must be able to view the results of the processing of semantic topics, which may be predefined domain-specific.
Such a system should recommend topics and topic related observations such as trends, newly emerging issues. 
Furthermore, multimedia presentations that depict the lifecycles of topics persist over a given time can be generated with dynamic summarization techniques~\cite{ragunath2015ontology,KalenderEWCKGK18}. 

Eventually, a tool that is customizable with domain-specific knowledge resources for detecting and processing topics.
The specific nature of the subject and desired processing will vary depending on the context.
A domain-specific topic detection system customized for diseases with knowledge bases like the International Classification of Diseases (ICD)~\cite{icd-who-www} for diseases  and SNOMED-CT (Systematized Nomenclature of Medicine -- Clinical Terms)~\cite{SNOMEDCTURL} would be useful in tracking the for pandemics related topics of public interest.
Such topic explorers would enable users to glean domain-specific insights that are very difficult to obtain by direct experience with a vast number of microposts.

One of the most interesting potentials of semantic resources is revealed via {\em federated queries} that search across distributed resources. 
Unfortunately, the performance of federated queries can be quite inefficient.
The order in which queries are executed must be carefully designed to achieve reasonable response times.
Generally, electing to execute the more restrictive queries prior to others which restricts the search space is considered a good approach. 
Finally, generating streams of semantic topics could be facilitated with stream reasoning~\cite{Pham2019Enhancing} and queried with a stream query language such as \textsc{c-sparql}~\cite{Barbieri2009} and \textsc{c-sprite}~\cite{Bonte:2019:CEH:3328905.3329502}. 

\section{Conclusions}
\label{sec:conclusion}

This work investigates the viability of extracting semantic topics from collections of microposts via processing their corresponding linked entities that are \lod\ resources.
To this end, an ontology (\topico) to represent topics is designed, an approach to extracting topics from sets of microposts is proposed, a prototype of this approach is implemented, and topics are generated from large sets of posts from Twitter.

The main inquiry of this work was to examine whether an approach based on linking microposts to \lod\ resources could be utilized in generating semantically represented and machine-interpretable topics.
The proposed approach extracts a significantly rich set of information from the posts in terms of relating them to web-resources which themselves are related to other resources via data and object properties.
We demonstrate the benefits of using \lod\ and ontologies while identifying as well as utilizing the topics.
During the identification phase, we were able to identify candidate elements and resolve their types.
The ontologically represented topics consisting of entities enabled processing opportunities that  revealed information about collections of microposts that are not readily observable even if each post were to be manually inspected.
Also, we notice an increase in the quality of the generated topics over time thanks to the efforts related to the continued expansion and correction of  \lod\ resources. 

Our main goal of producing machine-interpretable topics was for their utilization in further processing.
We demonstrated such utilization with several examples of various levels of complexity, where information that is not readily available in the original posts is revealed.
A user evaluation (with 81.0\% precision and $F_1$ of 93.3\%) and regularly performed manual inspections  show that the identified topics are relevant.
In summary, we are encouraged by the results we obtained  and list many research opportunities  to improve the topic identification approach and process topics in general and in domain-specific manners.

\section*{Acknowledgments}

\vspace{-0.2cm}We thank Dr. T. B. Dinesh and Dr. Jayant Venkatanatha for valuable contributions during the preparation of this work.
We are grateful for the feedback received from the members of SosLab (Department of Computer Engineering, Boğaziçi University) during the development of this work.

\bibliographystyle{elsarticle-harv}

\bibliography{sbountiPersonalVersion4ArxivVersion}

\begin{thebibliography}{126}
\expandafter\ifx\csname natexlab\endcsname\relax\def\natexlab#1{#1}\fi
\expandafter\ifx\csname url\endcsname\relax
  \def\url#1{\texttt{#1}}\fi
\expandafter\ifx\csname urlprefix\endcsname\relax\def\urlprefix{URL }\fi

\bibitem[{Abebe et~al.(2019)Abebe, Tekli, Getahun, Chbeir, and
  Tekli}]{ABEBE2019}
Abebe, M.~A., Tekli, J., Getahun, F., Chbeir, R., Tekli, G., 2019. Generic
  metadata representation framework for social-based event detection,
  description, and linkage. Knowledge-Based Systems.
\newline\urlprefix\url{http://www.sciencedirect.com/science/article/pii/S0950705119302928}

\bibitem[{Alvanaki et~al.(2012)Alvanaki, Michel, Ramamritham, and
  Weikum}]{Alvanaki:2012:SWE:2247596.2247636}
Alvanaki, F., Michel, S., Ramamritham, K., Weikum, G., 2012. {See What's
  enBlogue: Real-time} emergent topic identification in social media. In:
  Proceedings of the 15th International Conference on Extending Database
  Technology. EDBT '12. ACM, New York, NY, USA, pp. 336--347, iSBN
  978-1-4503-0790-1.

\bibitem[{{Amazon Inc.}(2020)}]{mechanical-turk}
{Amazon Inc.}, 2020. {Amazon Mechanical Turk}. Accessed: 2020-05-15.
\newline\urlprefix\url{https://www.mturk.com}

\bibitem[{Antoniou et~al.(2012)Antoniou, Groth, Harmelen, and
  Hoekstra}]{vanHarmelenSemanticWebBook}
Antoniou, G., Groth, P., Harmelen, F.~v., Hoekstra, R., 2012. A Semantic Web
  Primer. The MIT Press.

\bibitem[{{Apache Jena}(2020)}]{FusekiWeb}
{Apache Jena}, 2020. Fuseki. Accessed: 2020-05-15.
\newline\urlprefix\url{https://jena.apache.org/documentation/fuseki2/index.html}

\bibitem[{Auer(2014)}]{lodAuer2014}
Auer, S., 2014. Introduction to {LOD2}. In: Auer, S., Bryl, V., Tramp, S.
  (Eds.), Linked Open Data -- Creating Knowledge Out of Interlinked Data:
  Results of the LOD2 Project. Springer International Publishing, pp. 1--17,
  iSBN 978-3-319-09846-3.

\bibitem[{Bailey(2018)}]{phirehoseWeb}
Bailey, F., 2018. Phirehose. Accessed: 2020-05-15.
\newline\urlprefix\url{https://github.com/fennb/phirehose}

\bibitem[{Barbieri et~al.(2009)Barbieri, Braga, Ceri, Della~Valle, and
  Grossniklaus}]{Barbieri2009}
Barbieri, D.~F., Braga, D., Ceri, S., Della~Valle, E., Grossniklaus, M., 2009.
  {C-SPARQL}: {SPARQL} for continuous querying. In: Proceedings of the 18th
  International Conference on World Wide Web. WWW '09. ACM, New York, NY, USA,
  pp. 1061--1062, iSBN 978-1-60558-487-4.

\bibitem[{Bauer et~al.(2012)Bauer, Noulas, S\'{e}aghdha, Clark, and
  Mascolo}]{Bauer2012}
Bauer, S., Noulas, A., S\'{e}aghdha, D.~O., Clark, S., Mascolo, C., Sept 2012.
  {Talking Places}: Modelling and analysing linguistic content in {Foursquare}.
  In: 2012 International Conference on Privacy, Security, Risk and Trust and
  2012 International Confernece on Social Computing. pp. 348--357.

\bibitem[{Bicalho et~al.(2017)Bicalho, Pita, Pedrosa, Lacerda, and
  Pappa}]{Bicalho:2017:GFE:3062405.3062584}
Bicalho, P., Pita, M., Pedrosa, G., Lacerda, A., Pappa, G.~L., Jul 2017. A
  general framework to expand short text for topic modeling. Inf. Sci. 393~(C),
  66--81.
\newline\urlprefix\url{https://doi.org/10.1016/j.ins.2017.02.007}

\bibitem[{Bizer et~al.(2011)Bizer, Heath, and Berners-Lee}]{bizer2011linked}
Bizer, C., Heath, T., Berners-Lee, T., 2011. Linked data: The story so far. In:
  Semantic services, interoperability and web applications: emerging concepts.
  IGI Global, pp. 205--227.

\bibitem[{Bizer et~al.(2009)Bizer, Lehmann, Kobilarov, Auer, Becker, Cyganiak,
  and Hellmann}]{Bizer2009154}
Bizer, C., Lehmann, J., Kobilarov, G., Auer, S., Becker, C., Cyganiak, R.,
  Hellmann, S., 2009. {DBpedia} - a crystallization point for the web of data.
  Web Semantics: Science, Services and Agents on the World Wide Web 7~(3),
  154--165, the Web of Data.

\bibitem[{Blei et~al.(2003)Blei, Ng, and Jordan}]{blei2003latent}
Blei, D.~M., Ng, A.~Y., Jordan, M.~I., 2003. Latent {Dirichlet} allocation.
  Journal of machine Learning research 3~(Jan), 993--1022.

\bibitem[{Bonte et~al.(2019)Bonte, Tommasini, De~Turck, Ongenae, and
  Valle}]{Bonte:2019:CEH:3328905.3329502}
Bonte, P., Tommasini, R., De~Turck, F., Ongenae, F., Valle, E.~D., 2019.
  C-sprite: Efficient hierarchical reasoning for rapid rdf stream processing.
  In: Proceedings of the 13th ACM International Conference on Distributed and
  Event-based Systems. DEBS '19. ACM, New York, NY, USA, pp. 103--114.
\newline\urlprefix\url{http://doi.acm.org/10.1145/3328905.3329502}

\bibitem[{Brickley and Miller(2014)}]{foafspecweb}
Brickley, D., Miller, L., 2014. {FOAF}. Accessed: 2020-05-15.
\newline\urlprefix\url{http://xmlns.com/foaf/spec/}

\bibitem[{Cataldi et~al.(2010)Cataldi, Di~Caro, and
  Schifanella}]{Cataldi:2010:ETD:1814245.1814249}
Cataldi, M., Di~Caro, L., Schifanella, C., 2010. Emerging topic detection on
  {Twitter} based on temporal and social terms evaluation. In: Proceedings of
  the Tenth International Workshop on Multimedia Data Mining. MDMKDD '10. ACM,
  New York, NY, USA, pp. 4:1--4:10, iSBN 978-1-4503-0220-3.

\bibitem[{Celebi and Uskudarli(2012)}]{celebi2012content}
Celebi, H.~B., Uskudarli, S., 2012. Content based microblogger recommendation.
  In: 2012 International Conference on Privacy, Security, Risk and Trust and
  2012 International Confernece on Social Computing. IEEE, pp. 605--610.

\bibitem[{Chen et~al.(2015)Chen, Wang, Zhang, Yan, and
  Li}]{chen-etal-2015-user}
Chen, W., Wang, J., Zhang, Y., Yan, H., Li, X., Jul 2015. User based
  aggregation for biterm topic model. In: Proceedings of the 53rd Annual
  Meeting of the Association for Computational Linguistics and the 7th
  International Joint Conference on Natural Language Processing (Volume 2:
  Short Papers). ACL, Beijing, China, pp. 489--494.
\newline\urlprefix\url{https://www.aclweb.org/anthology/P15-2080}

\bibitem[{Chen et~al.(2019)Chen, Zhang, Liu, Ye, and Lin}]{CHEN20191}
Chen, Y., Zhang, H., Liu, R., Ye, Z., Lin, J., 2019. Experimental explorations
  on short text topic mining between lda and nmf based schemes. Knowledge-Based
  Systems 163, 1--13.
\newline\urlprefix\url{http://www.sciencedirect.com/science/article/pii/S0950705118304076}

\bibitem[{{Cheng} et~al.(2014){Cheng}, {Yan}, {Lan}, and {Guo}}]{6778764}
{Cheng}, X., {Yan}, X., {Lan}, Y., {Guo}, J., Dec 2014. Btm: Topic modeling
  over short texts. IEEE Transactions on Knowledge and Data Engineering
  26~(12), 2928--2941.

\bibitem[{Chi et~al.(2016)Chi, Lim, Alam, and Butler}]{Chi2016GeolocationPI}
Chi, L., Lim, K.~H., Alam, N., Butler, C.~J., 2016. Geolocation prediction in
  {Twitter} using location indicative words and textual features. In:
  {NUT@COLING}, 2nd Workshop on Noisy User-generated Text. pp. 227--234.

\bibitem[{Chiu and Nichols(2016)}]{tacla00104}
Chiu, J.~P., Nichols, E., 2016. Named entity recognition with bidirectional
  lstm-cnns. Transactions of the Association for Computational Linguistics 4,
  357--370.
\newline\urlprefix\url{https://doi.org/10.1162/tacl_a_00104}

\bibitem[{Chodhary et~al.(2016)Chodhary, Jagdale, and
  Deshmukh}]{chodhary2016semantic}
Chodhary, S., Jagdale, R.~S., Deshmukh, S.~N., 2016. Semantic analysis of
  tweets using lsa and svd. International Journal of Emerging Trends and
  Technology in Computer Science 5~(4).

\bibitem[{Cox et~al.(2020)Cox, Little, Hobbs, and Pan}]{W3CTime}
Cox, S., Little, C., Hobbs, J.~R., Pan, F., Mar 2020. Time ontology in {OWL}.
  {W3C} candidate recommendation, W3C.
\newline\urlprefix\url{https://www.w3.org/TR/2020/CR-owl-time-20200326/}

\bibitem[{{DBpedia}(2018)}]{dbpediadatasetsize}
{DBpedia}, 2018. {The Release Circle - A Glimpse behind the Scenes}. Accessed:
  2020-05-15.
\newline\urlprefix\url{https://blog.dbpedia.org/2018/10/17/the-release-circle-a-glimpse-behind-the-scenes/}

\bibitem[{{DBpedia}(2020)}]{DbpediaSparql}
{DBpedia}, 2020. Virtuoso {SPARQL} query editor. Accessed: 2020-05-15.
\newline\urlprefix\url{http://dbpedia.org/sparql}

\bibitem[{Degirmencioglu and Uskudarli(2010)}]{degirmencioglu2010exploring}
Degirmencioglu, E.~A., Uskudarli, S., 4 2010. Exploring area-specific
  microblogging social networks. In: WebSci10: Extending the Frontiers of
  Society On-Line. Web Science, Raleigh, North Carolina, USA, on-Line.

\bibitem[{Eisenstein(2013)}]{eisenstein2013bad}
Eisenstein, J., June 2013. What to do about bad language on the internet. In:
  Proceedings of the 2013 Conference of the North American Chapter of the
  Association for Computational Linguistics: Human Language Technologies. ACL,
  Atlanta, Georgia, pp. 359--369.
\newline\urlprefix\url{http://www.aclweb.org/anthology/N13-1037}

\bibitem[{Eissa et~al.(2018)Eissa, El-Sharkawi, and
  Mokhtar}]{Eissa:2018:TRU:3184558.3191562}
Eissa, A. H.~B., El-Sharkawi, M.~E., Mokhtar, H. M.~O., 2018. Towards
  recommendation using interest-based communities in attributed social
  networks. In: Companion Proceedings of the The Web Conference 2018. WWW '18.
  International World Wide Web Conferences Steering Committee, Republic and
  Canton of Geneva, Switzerland, pp. 1235--1242.

\bibitem[{Eppstein et~al.(2010)Eppstein, L{\"{o}}ffler, and
  Strash}]{DBLP:journals/corr/abs-1006-5440}
Eppstein, D., L{\"{o}}ffler, M., Strash, D., 2010. Listing all maximal cliques
  in sparse graphs in near-optimal time. CoRR abs/1006.5440.
\newline\urlprefix\url{http://arxiv.org/abs/1006.5440}

\bibitem[{Fang(2019)}]{Fang2019}
Fang, A., 2019. Analysing political events on {Twitter}: topic modelling and
  user community classification. doctoral dissertation, university of glasgow.
\newline\urlprefix\url{https://theses.gla.ac.uk/41135/}

\bibitem[{Ferragina and Scaiella(2012)}]{ferragina2012tagme}
Ferragina, P., Scaiella, U., 2012. Fast and accurate annotation of short texts
  with {Wikipedia} pages. IEEE Software 29~(1), 70--75.

\bibitem[{Francis{-}Landau et~al.(2016)Francis{-}Landau, Durrett, and
  Klein}]{FrancisLandauD16}
Francis{-}Landau, M., Durrett, G., Klein, D., 2016. Capturing semantic
  similarity for entity linking with convolutional neural networks. CoRR
  abs/1604.00734.
\newline\urlprefix\url{http://arxiv.org/abs/1604.00734}

\bibitem[{Garigliotti et~al.(2019)Garigliotti, Hasibi, and
  Balog}]{Garigliotti2019}
Garigliotti, D., Hasibi, F., Balog, K., Aug 2019. Identifying and exploiting
  target entity type information for ad hoc entity retrieval. Information
  Retrieval Journal 22~(3), 285--323.
\newline\urlprefix\url{https://doi.org/10.1007/s10791-018-9346-x}

\bibitem[{Gattani et~al.(2013)Gattani, Lamba, Garera, Tiwari, Chai, Das,
  Subramaniam, Rajaraman, Harinarayan, and Doan}]{Gattani2013entity}
Gattani, A., Lamba, D.~S., Garera, N., Tiwari, M., Chai, X., Das, S.,
  Subramaniam, S., Rajaraman, A., Harinarayan, V., Doan, A., Aug. 2013. Entity
  extraction, linking, classification, and tagging for social media: A
  {Wikipedia}-based approach. Proc. VLDB Endow. 6~(11), 1126--1137.

\bibitem[{Genc et~al.(2011)Genc, Sakamoto, and
  Nickerson}]{Genc:2011:DCC:2021773.2021833}
Genc, Y., Sakamoto, Y., Nickerson, J.~V., 2011. Discovering context:
  Classifying tweets through a semantic transform based on {Wikipedia}. In:
  Proceedings of the 6th international conference on Foundations of augmented
  cognition: directing the future of adaptive systems. FAC'11. Springer-Verlag,
  pp. 484--492, iSBN 978-3-642-21851-4.
\newline\urlprefix\url{http://dl.acm.org/citation.cfm?id=2021773.2021833}

\bibitem[{{GeoNames}(2010)}]{geonames}
{GeoNames}, 2010. Geonames ontology - geo semantic web. Accessed: 2020-05-15.
\newline\urlprefix\url{http://www.geonames.org/ontology/documentation.html}

\bibitem[{Gottschalk and Demidova(2019)}]{gottschalk2019eventkg}
Gottschalk, S., Demidova, E., 2019. Eventkg--the hub of event knowledge on the
  web--and biographical timeline generation. Semantic Web~(Pre-press), 1--32.

\bibitem[{Gruetze et~al.(2016)Gruetze, Kasneci, Zuo, and Naumann}]{gruetze2016}
Gruetze, T., Kasneci, G., Zuo, Z., Naumann, F., 2016. {CohEEL}: Coherent and
  efficient named entity linking through random walks. Web Semantics: Science,
  Services and Agents on the World Wide Web 37~(0).
\newline\urlprefix\url{http://www.websemanticsjournal.org/index.php/ps/article/view/463}

\bibitem[{Guha et~al.(2016)Guha, Brickley, and Macbeth}]{Schema.Org}
Guha, R.~V., Brickley, D., Macbeth, S., Jan. 2016. {Schema.Org}: Evolution of
  structured data on the {Web}. Commun. ACM 59~(2), 44--51.

\bibitem[{{Han} et~al.(2019){Han}, {Viriyothai}, {Lim}, {Lameter}, and
  {Mussell}}]{8695381EntitiLinking1}
{Han}, H., {Viriyothai}, P., {Lim}, S., {Lameter}, D., {Mussell}, B., March
  2019. Yet another framework for tweet entity linking (yaftel). In: 2019 IEEE
  Conference on Multimedia Information Processing and Retrieval (MIPR). pp.
  258--263.

\bibitem[{Horrocks et~al.(2004)Horrocks, Patel-Schneider, Boley, Tabet, Grosof,
  and Dean}]{swrl-w3c}
Horrocks, I., Patel-Schneider, P.~F., Boley, H., Tabet, S., Grosof, B., Dean,
  M., May 2004. {SWRL}: A semantic web rule language combining {OWL} and
  {RuleML}. Tech. rep., W3C.
\newline\urlprefix\url{https://www.w3.org/Submission/2004/SUBM-SWRL-20040521/}

\bibitem[{Hripcsak and Rothschild(2005)}]{hripcsak2005agreement}
Hripcsak, G., Rothschild, A.~S., 2005. Agreement, the f-measure, and
  reliability in information retrieval. Journal of the American Medical
  Informatics Association 12~(3), 296--298.

\bibitem[{{Internet Live Stats}(2020)}]{Twitterstatistics}
{Internet Live Stats}, 2020. Twitter statistics. Accessed: 2020-05-15.
\newline\urlprefix\url{http://www.internetlivestats.com/twitter-statistics/}

\bibitem[{Kalender et~al.(2018)Kalender, Eren, Wu, Cirakman, Kutluk, Gultekin,
  and Korkmaz}]{KalenderEWCKGK18}
Kalender, M., Eren, M.~T., Wu, Z., Cirakman, O., Kutluk, S., Gultekin, G.,
  Korkmaz, E.~E., 2018. Videolization: knowledge graph based automated video
  generation from web content. Multimedia Tools Appl. 77~(1), 567--595.
\newline\urlprefix\url{https://doi.org/10.1007/s11042-016-4275-4}

\bibitem[{Karadeniz and Özgür(2019)}]{Karadeniz2019}
Karadeniz, I., Özgür, A., 2019. Linking entities through an ontology using
  word embeddings and syntactic re-ranking. BMC Bioinformatics 20~(1), 156.
\newline\urlprefix\url{https://doi.org/10.1186/s12859-019-2678-8}

\bibitem[{Kasiviswanathan et~al.(2011)Kasiviswanathan, Melville, Banerjee, and
  Sindhwani}]{kasiviswanathan2011emerging}
Kasiviswanathan, S.~P., Melville, P., Banerjee, A., Sindhwani, V., 2011.
  Emerging topic detection using dictionary learning. In: Proceedings of the
  20th ACM International Conference on Information and Knowledge Management.
  CIKM '11. ACM, New York, NY, USA, pp. 745--754, iSBN 978-1-4503-0717-8.

\bibitem[{Kumar and Singh(2019)}]{KUMAR2019365}
Kumar, A., Singh, J.~P., 2019. Location reference identification from tweets
  during emergencies: A deep learning approach. International Journal of
  Disaster Risk Reduction 33, 365--375.
\newline\urlprefix\url{http://www.sciencedirect.com/science/article/pii/S2212420918307799}

\bibitem[{Kumar(2019)}]{kumar2019knowledge}
Kumar, V.~K., 2019. Knowledge representation technologies using semantic web.
  In: Web Services: Concepts, Methodologies, Tools, and Applications. IGI
  Global, pp. 1068--1076.

\bibitem[{Li et~al.(2017)Li, Duan, Wang, Zhang, Sun, and
  Ma}]{Li:2017:ETM:3133943.3091108}
Li, C., Duan, Y., Wang, H., Zhang, Z., Sun, A., Ma, Z., Aug. 2017. Enhancing
  topic modeling for short texts with auxiliary word embeddings. ACM Trans.
  Inf. Syst. 36~(2), 11:1--11:30.
\newline\urlprefix\url{http://doi.acm.org/10.1145/3091108}

\bibitem[{Li et~al.(2016)Li, Wang, Zhang, Sun, and
  Ma}]{Li:2016:TMS:2911451.291149}
Li, C., Wang, H., Zhang, Z., Sun, A., Ma, Z., 2016. Topic modeling for short
  texts with auxiliary word embeddings. In: Proceedings of the 39th
  International ACM SIGIR Conference on Research and Development in Information
  Retrieval. SIGIR '16. ACM, New York, NY, USA, pp. 165--174.
\newline\urlprefix\url{http://doi.acm.org/10.1145/2911451.2911499}

\bibitem[{Liao and Zhao(2019)}]{Liao:2019:UAT:3359984.3324473}
Liao, X., Zhao, Z., Aug 2019. Unsupervised approaches for textual semantic
  annotation, a survey. ACM Comput. Surv. 52~(4), 66:1--66:45.
\newline\urlprefix\url{http://doi.acm.org/10.1145/3324473}

\bibitem[{Lin et~al.(2014)Lin, Tian, Mei, and
  Cheng}]{Lin:2014:DTM:2566486.2567980}
Lin, T., Tian, W., Mei, Q., Cheng, H., 2014. The dual-sparse topic model:
  Mining focused topics and focused terms in short text. In: Proceedings of the
  23rd International Conference on World Wide Web. WWW '14. ACM, New York, NY,
  USA, pp. 539--550.
\newline\urlprefix\url{http://doi.acm.org/10.1145/2566486.2567980}

\bibitem[{{Linked Data community}(2020)}]{LinkedDataOrg}
{Linked Data community}, 2020. Linked data | linked data - connect distributed
  data across the web. [cited 25 may 2020]. in: [internet] - . [about 1
  screen].
\newline\urlprefix\url{http://linkeddata.org/}

\bibitem[{Marcus et~al.(2011)Marcus, Bernstein, Badar, Karger, Madden, and
  Miller}]{marcus2011twitinfo}
Marcus, A., Bernstein, M.~S., Badar, O., Karger, D.~R., Madden, S., Miller,
  R.~C., 2011. Twitinfo: Aggregating and visualizing microblogs for event
  exploration. In: Proceedings of the SIGCHI Conference on Human Factors in
  Computing Systems. CHI '11. ACM, New York, NY, USA, pp. 227--236, iSBN
  978-1-4503-0228-9.

\bibitem[{Martinez-Rodriguez et~al.(2019)Martinez-Rodriguez, Lopez-Arevalo,
  Rios-Alvarado, Hernandez, and Aldana-Bobadilla}]{10.1007/978-3-030-21395-4_7}
Martinez-Rodriguez, J.~L., Lopez-Arevalo, I., Rios-Alvarado, A.~B., Hernandez,
  J., Aldana-Bobadilla, E., 2019. Extraction of rdf statements from text. In:
  Villazón-Terrazas, B., Hidalgo-Delgado, Y. (Eds.), Knowledge Graphs and
  Semantic Web. Springer International Publishing, Cham, pp. 87--101.

\bibitem[{Matentzoglu et~al.(2018)Matentzoglu, Malone, Mungall, and
  Stevens}]{Matentzoglu2018}
Matentzoglu, N., Malone, J., Mungall, C., Stevens, R., 2018. Miro: guidelines
  for minimum information for the reporting of an ontology. Journal of
  Biomedical Semantics 9~(1), 6.
\newline\urlprefix\url{https://doi.org/10.1186/s13326-017-0172-7}

\bibitem[{Mathioudakis and
  Koudas(2010)}]{TwitterMonitor:Mathioudakis:2010:TTD:1807167.1807306}
Mathioudakis, M., Koudas, N., 2010. {TwitterMonitor}: Trend detection over the
  {Twitter} stream. In: Proceedings of the 2010 ACM SIGMOD International
  Conference on Management of Data. SIGMOD '10. ACM, New York, NY, USA, pp.
  1155--1158, iSBN 978-1-4503-0032-2.

\bibitem[{{McCrae} et~al.(2020){McCrae}, Abele, Buitelaar, Cyganiak, Jentzsch,
  Andryushechkin, Debattista, and Nasir}]{LodCloud}
{McCrae}, J.~P., Abele, A., Buitelaar, P., Cyganiak, R., Jentzsch, A.,
  Andryushechkin, V., Debattista, J., Nasir, J., 2020. The linked open data
  cloud diagram. Accessed: 2020-05-15.
\newline\urlprefix\url{http://lod-cloud.net}

\bibitem[{McKinney and Iannella(2014)}]{vcardprefix}
McKinney, J., Iannella, R., May 2014. {vCard} ontology - for describing people
  and organizations. {W3C} note, W3C.
\newline\urlprefix\url{http://www.w3.org/TR/2014/NOTE-vcard-rdf-20140522/}

\bibitem[{Mehrotra et~al.(2013)Mehrotra, Sanner, Buntine, and
  Xie}]{Mehrotra2013}
Mehrotra, R., Sanner, S., Buntine, W., Xie, L., 2013. Improving {LDA} topic
  models for microblogs via tweet pooling and automatic labeling. In:
  Proceedings of the 36th International ACM SIGIR Conference on Research and
  Development in Information Retrieval. SIGIR '13. ACM, New York, NY, USA, pp.
  889--892, iSBN 978-1-4503-2034-4.

\bibitem[{Nanni et~al.(2019)Nanni, Menini, Tonelli, and Ponzetto}]{madoc49597}
Nanni, F., Menini, S., Tonelli, S., Ponzetto, S.~P., 2019. Semantifying the uk
  hansard (1918-2018). In: Proceedings of the 19th ACM/IEEE Joint Conference on
  Digital Libraries : JCDL {\textquoteright}19, June 2019, Urbana-Champaign,
  Illinois. ACM, New York, NY, pp. 1--2.
\newline\urlprefix\url{https://madoc.bib.uni-mannheim.de/49597/}

\bibitem[{Newman et~al.(2007)Newman, Hagedorn, Chemudugunta, and
  Smyth}]{Newman:2007:SME:1255175.1255248}
Newman, D., Hagedorn, K., Chemudugunta, C., Smyth, P., 2007. Subject metadata
  enrichment using statistical topic models. In: Proceedings of the 7th
  {ACM/IEEE-CS} Joint Conference on Digital Libraries. JCDL '07. ACM, New York,
  NY, USA, pp. 366--375.

\bibitem[{Noy and McGuinness(2001)}]{noy2001ontology}
Noy, N.~F., McGuinness, D.~L., 2001. Ontology development 101: A guide to
  creating your first ontology.
  Https://protege.stanford.edu/publications/\mbox{ontology\_develop}\-ment/ontology101.pdf,
  Accessed: 2017-12.

\bibitem[{Nédellec(2018)}]{ontobiotop}
Nédellec, C., 2018. Ontobiotope. [cited 15 june 2020]. database: Inra
  [internet].
\newline\urlprefix\url{10.15454/1.4382640528105164E12}

\bibitem[{Ozdikis et~al.(2019)Ozdikis, Ramampiaro, and
  Nørvåg}]{OZDIKIS20191280}
Ozdikis, O., Ramampiaro, H., Nørvåg, K., 2019. Locality-adapted kernel
  densities of term co-occurrences for location prediction of tweets.
  Information Processing \& Management 56~(4), 1280--1299.
\newline\urlprefix\url{http://www.sciencedirect.com/science/article/pii/S0306457318309063}

\bibitem[{Ozer et~al.(2016)Ozer, Kim, and
  Davulcu}]{0604371a2d4c430ea137e8d4086734b6}
Ozer, M., Kim, N., Davulcu, H., 11 2016. Community detection in political
  twitter networks using nonnegative matrix factorization methods. In:
  Proceedings of the 2016 IEEE/ACM International Conference on Advances in
  Social Networks Analysis and Mining, ASONAM. Institute of Electrical and
  Electronics Engineers Inc., pp. 81--88.

\bibitem[{Parker et~al.(2013)Parker, Wei, Yates, Frieder, and
  Goharian}]{parker2013framework}
Parker, J., Wei, Y., Yates, A., Frieder, O., Goharian, N., 2013. A framework
  for detecting public health trends with {Twitter}. In: Proceedings of the
  2013 IEEE/ACM International Conference on Advances in Social Networks
  Analysis and Mining. ASONAM '13. ACM, New York, NY, USA, pp. 556--563, iSBN
  978-1-4503-2240-9.

\bibitem[{Perrier(2015)}]{AlexisPerrier2015}
Perrier, A., 2015. Segmentation of twitter timelines via topic modeling.
\newline\urlprefix\url{https://alexisperrier.com/nlp/2015/09/16/segmentation_twitter_timelines_lda_vs_lsa.html}

\bibitem[{Petrović et~al.(2010)Petrović, Osborne, and
  Lavrenko}]{Petrovic:2010:SFS:1857999.1858020}
Petrović, S., Osborne, M., Lavrenko, V., 2010. Streaming first story detection
  with application to {Twitter}. In: Human Language Technologies: The 2010
  Annual Conference of the North American Chapter of the Association for
  Computational Linguistics. ACL, pp. 181--189.

\bibitem[{Pham et~al.(2019)Pham, Ali, and Mileo}]{Pham2019Enhancing}
Pham, T.-L., Ali, M.~I., Mileo, A., 2019. Enhancing the scalability of
  expressive stream reasoning via input-driven parallelization. Semantic Web
  Journal 10~(3), 457--474.
\newline\urlprefix\url{http://www.semantic-web-journal.net/content/enhancing-scalability-expressive-stream-reasoning-input-driven-parallelization}

\bibitem[{{Politico Staff}(2016{\natexlab{a}})}]{VPDebate2016Transcript}
{Politico Staff}, 2016{\natexlab{a}}. Full transcript: 2016 vice presidential
  debate. Accessed: 2020-05-15.
\newline\urlprefix\url{https://www.politico.com/story/2016/10/full-transcript-2016-vice-presidential-debate-229185}

\bibitem[{{Politico Staff}(2016{\natexlab{b}})}]{FirstDebate2016Transcript}
{Politico Staff}, 2016{\natexlab{b}}. Full transcript: First 2016 presidential
  debate. Accessed: 2020-05-15.
\newline\urlprefix\url{https://www.politico.com/story/2016/09/full-transcript-first-2016-presidential-debate-228761}

\bibitem[{{Politico Staff}(2016{\natexlab{c}})}]{SecondDebate2016Transcript}
{Politico Staff}, 2016{\natexlab{c}}. Full transcript: Second 2016 presidential
  debate. Accessed: 2020-05-15.
\newline\urlprefix\url{https://www.politico.com/story/2016/10/2016-presidential-debate-transcript-229519}

\bibitem[{{Politico Staff}(2016{\natexlab{d}})}]{ThirdDebate2016Transcript}
{Politico Staff}, 2016{\natexlab{d}}. Full transcript: Third 2016 presidential
  debate. Accessed: 2020-05-15.
\newline\urlprefix\url{https://www.politico.com/story/2016/10/full-transcript-third-2016-presidential-debate-230063}

\bibitem[{Prieto et~al.(2014)Prieto, Matos, Àlvarez, Cacheda, and
  Oliveira}]{Prieto2014}
Prieto, V.~M., Matos, S., Àlvarez, M., Cacheda, F., Oliveira, J.~L., 01 2014.
  {Twitter}: A good place to detect health conditions. {PLoS} ONE 9~(1), 1--11.

\bibitem[{Pérez et~al.(2009)Pérez, Arenas, and
  Gutierrez}]{10.1145/1567274.1567278}
Pérez, J., Arenas, M., Gutierrez, C., Sep. 2009. Semantics and complexity of
  sparql. ACM Trans. Database Syst. 34~(3).
\newline\urlprefix\url{https://doi.org/10.1145/1567274.1567278}

\bibitem[{Qiang et~al.(2017)Qiang, Chen, Wang, and
  Wu}]{10.1007/978-3-319-57529-2_29}
Qiang, J., Chen, P., Wang, T., Wu, X., 2017. Topic modeling over short texts by
  incorporating word embeddings. In: Kim, J., Shim, K., Cao, L., Lee, J.-G.,
  Lin, X., Moon, Y.-S. (Eds.), Advances in Knowledge Discovery and Data Mining.
  Springer International Publishing, Cham, pp. 363--374.

\bibitem[{Qiang et~al.(2018)Qiang, Li, Yuan, and Wu}]{Qiang2018}
Qiang, J., Li, Y., Yuan, Y., Wu, X., Jul 2018. Short text clustering based on
  pitman-yor process mixture model. Applied Intelligence 48~(7), 1802--1812.
\newline\urlprefix\url{https://doi.org/10.1007/s10489-017-1055-4}

\bibitem[{Qiang et~al.(2019)Qiang, Qian, Li, Yuan, and Wu}]{abs-1904-07695}
Qiang, J., Qian, Z., Li, Y., Yuan, Y., Wu, X., 2019. Short text topic modeling
  techniques, applications, and performance: {A} survey. CoRR abs/1904.07695.
\newline\urlprefix\url{http://arxiv.org/abs/1904.07695}

\bibitem[{Qiu(2017)}]{twitterLDA}
Qiu, M., 2017. Latent dirichlet allocation (lda) model for microblogs
  ({Twitter}, weibo etc.).
\newline\urlprefix\url{https://github.com/minghui/Twitter-LDA}

\bibitem[{{R Core Team}(2013)}]{CiteR}
{R Core Team}, 2013. R: A Language and Environment for Statistical Computing. R
  Foundation for Statistical Computing, Vienna, Austria.
\newline\urlprefix\url{http://www.R-project.org/}

\bibitem[{Ragunath and Sivaranjani(2015)}]{ragunath2015ontology}
Ragunath, R., Sivaranjani, N., 2015. Ontology based text document summarization
  system using concept terms. ARPN J. Eng. Appl. Sci 10, 2638--2642.

\bibitem[{Rospocher et~al.(2019)Rospocher, Corcoglioniti, and
  Dragoni}]{rospocher2019boosting}
Rospocher, M., Corcoglioniti, F., Dragoni, M., 2019. Boosting document
  retrieval with knowledge extraction and linked data. Semantic Web 10~(4),
  753--778.

\bibitem[{Sakor et~al.(2019)Sakor, Onando~Mulang\', Singh, Shekarpour,
  Esther~Vidal, Lehmann, and Auer}]{sakor-etal-2019-oldEntityLinking2}
Sakor, A., Onando~Mulang\', I., Singh, K., Shekarpour, S., Esther~Vidal, M.,
  Lehmann, J., Auer, S., Jun 2019. Old is gold: Linguistic driven approach for
  entity and relation linking of short text. In: Proceedings of the 2019
  Conference of the North {A}merican Chapter of the Association for
  Computational Linguistics: Human Language Technologies, Volume 1 (Long and
  Short Papers). ACL, Minneapolis, Minnesota, pp. 2336--2346.
\newline\urlprefix\url{https://www.aclweb.org/anthology/N19-1243}

\bibitem[{Sayyadi and Raschid(2013)}]{Sayyadi:2013:GAA:2542214.2542215}
Sayyadi, H., Raschid, L., Dec 2013. A graph analytical approach for topic
  detection. ACM Trans. Internet Technol. 13~(2), 4:1--4:23.

\bibitem[{{Schema.org}(2020)}]{schemaorgvoc}
{Schema.org}, 2020. Home - schema.org. Accessed: 2020-05-15.
\newline\urlprefix\url{http://schema.org/}

\bibitem[{Schmachtenberg et~al.(2014)Schmachtenberg, Bizer, and
  Paulheim}]{Schmachtenberg2014}
Schmachtenberg, M., Bizer, C., Paulheim, H., 2014. Adoption of the Linked Data
  Best Practices in Different Topical Domains. Springer International
  Publishing, Cham, Ch. The Semantic Web - ISWC 2014, pp. 245--260.

\bibitem[{Seaborne and Harris(2013)}]{Seaborne:13:SQL}
Seaborne, A., Harris, S., Mar 2013. {SPARQL} 1.1 query language. {W3C}
  recommendation, W3C.
\newline\urlprefix\url{http://www.w3.org/TR/2013/REC-sparql11-query-20130321/}

\bibitem[{Sekine et~al.(2002)Sekine, Sudo, and
  Nobata}]{sekine-etal-2002-extended}
Sekine, S., Sudo, K., Nobata, C., May 2002. Extended named entity hierarchy.
  In: Proceedings of the Third International Conference on Language Resources
  and Evaluation ({LREC}{'}02). European Language Resources Association (ELRA),
  Las Palmas, Canary Islands - Spain, pp. 1818--1824.
\newline\urlprefix\url{http://www.lrec-conf.org/proceedings/lrec2002/pdf/120.pdf}

\bibitem[{{Shadbolt} et~al.(2006){Shadbolt}, {Berners-Lee}, and
  {Hall}}]{1637364}
{Shadbolt}, N., {Berners-Lee}, T., {Hall}, W., Jan 2006. The semantic web
  revisited. IEEE Intelligent Systems 21~(3), 96--101.

\bibitem[{Sharifi et~al.(2014)Sharifi, Inouye, and
  Kalita}]{sharifi2014summarization}
Sharifi, B.~P., Inouye, D.~I., Kalita, J.~K., March 2014. Summarization of
  twitter microblogs. The Computer Journal 57~(3), 378--402.

\bibitem[{{SNOMED International}(2020)}]{SNOMEDCTURL}
{SNOMED International}, 2020. {SNOMED CT}. Accessed: 2020-05-15.
\newline\urlprefix\url{http://www.snomed.org/snomed-ct/}

\bibitem[{{SoSLab}(2020)}]{TopicExplorer}
{SoSLab}, 2020. Query and browse semantic topics. Accessed: 2020-05-15.
\newline\urlprefix\url{http://soslab.cmpe.boun.edu.tr/sbounti/}

\bibitem[{{Stanford Center for Biomedical Informatics
  Research}(2018)}]{ProtegeWeb}
{Stanford Center for Biomedical Informatics Research}, 2018. Protègè.
  Accessed: 2020-05-15.
\newline\urlprefix\url{https://protege.stanford.edu/}

\bibitem[{Sönmez and Özgür(2014)}]{sonmez_emnlp_2014}
Sönmez, {\c{C}}., Özgür, A., 2014. A graph-based approach for contextual
  text normalization. In: Proceedings of the 2014 Conference on Empirical
  Methods in Natural Language Processing (EMNLP). ACL, pp. 313--324.
\newline\urlprefix\url{http://aclweb.org/anthology/D14-1037}

\bibitem[{TagMe(2020)}]{TagmeApiDoc}
TagMe, 2020. Tagme api documentation. Accessed: 2020-05-15.
\newline\urlprefix\url{https://sobigdata.d4science.org/web/tagme/tagme-help}

\bibitem[{Tonon et~al.(2016)Tonon, Catasta, Prokofyev, Demartini, Aberer, and
  Cudrè-Mauroux}]{Tonon2016170}
Tonon, A., Catasta, M., Prokofyev, R., Demartini, G., Aberer, K.,
  Cudrè-Mauroux, P., 2016. Contextualized ranking of entity types based on
  knowledge graphs. Web Semantics: Science, Services and Agents on the World
  Wide Web 37-38, 170--183, 10.1016/j.websem.2015.12.005.
\newline\urlprefix\url{http://www.sciencedirect.com/science/article/pii/S1570826815001468}

\bibitem[{Tuarob et~al.(2013)Tuarob, Pouchard, and
  Giles}]{Tuarob:2013:ATR:2467696.2467706}
Tuarob, S., Pouchard, L.~C., Giles, C.~L., 2013. Automatic tag recommendation
  for metadata annotation using probabilistic topic modeling. In: Proceedings
  of the 13th ACM/IEEE-CS Joint Conference on Digital Libraries. JCDL '13. ACM,
  New York, NY, USA, pp. 239--248.

\bibitem[{Twitter(2020{\natexlab{a}})}]{FilterEndpointWeb}
Twitter, 2020{\natexlab{a}}. Filter realtime tweets. Accessed: 2020-05-15.
\newline\urlprefix\url{https://developer.twitter.com/en/docs/tweets/filter-realtime/api-reference/post-statuses-filter}

\bibitem[{Twitter(2020{\natexlab{b}})}]{TwitterCom}
Twitter, 2020{\natexlab{b}}. Twitter. Accessed: 2020-05-15.
\newline\urlprefix\url{https://twitter.com}

\bibitem[{van Aggelen et~al.(2017)van Aggelen, Hollink, Kemman, Kleppe, and
  Beunders}]{Aggelen2015}
van Aggelen, A., Hollink, L., Kemman, M., Kleppe, M., Beunders, H., 2017. The
  debates of the {European Parliament} as {Linked Open Data}. Semantic Web
  Journal 8~(2), 271--281.
\newline\urlprefix\url{http://www.semantic-web-journal.net/content/debates-european-parliament-linked-open-data-0}

\bibitem[{Vrandečić and Krötzsch(2014)}]{Vrandecic:2014:WFC:2661061.2629489}
Vrandečić, D., Krötzsch, M., Sep. 2014. Wikidata: A free collaborative
  knowledgebase. Commun. ACM 57~(10), 78--85.

\bibitem[{{W3C}(2009)}]{wgsweb}
{W3C}, 2009. {WGS84} geo positioning: an {RDF} vocabulary. Accessed:
  2020-05-15.
\newline\urlprefix\url{http://www.w3.org/2003/01/geo/wgs84_pos}

\bibitem[{{W3C}(2015)}]{thesemanticweb}
{W3C}, 2015. Semantic web. Accessed: 2020-05-15.
\newline\urlprefix\url{https://www.w3.org/standards/semanticweb/}

\bibitem[{{W3C}(2020)}]{W3COWL}
{W3C}, 2020. {Web Ontology Language (OWL)}. Accessed: 2020-05-15.
\newline\urlprefix\url{https://www.w3.org/OWL/}

\bibitem[{{W3C Semantic Web Interest Group}(2004)}]{wgsweb2}
{W3C Semantic Web Interest Group}, 2004. Basic geo ({WGS84} lat/long)
  vocabulary. Accessed: 2020-05-15.
\newline\urlprefix\url{https://www.w3.org/2003/01/geo/}

\bibitem[{Wang et~al.(2015)Wang, Bansal, Gimpel, Ziebart, and Yu}]{TACL485}
Wang, J., Bansal, M., Gimpel, K., Ziebart, B.~D., Yu, C.~T., 2015. A
  sense-topic model for word sense induction with unsupervised data enrichment.
  Transactions of the Association for Computational Linguistics 3, 59--71.
\newline\urlprefix\url{https://transacl.org/ojs/index.php/tacl/article/view/485}

\bibitem[{Wang et~al.(2017)Wang, Zhou, He, and
  Hopcroft}]{10.1007/978-981-10-6893-5_2}
Wang, W., Zhou, H., He, K., Hopcroft, J.~E., 2017. Learning latent topics from
  the word co-occurrence network. In: Du, D., Li, L., Zhu, E., He, K. (Eds.),
  Theoretical Computer Science. Springer Singapore, Singapore, pp. 18--30.

\bibitem[{Weng et~al.(2010)Weng, Lim, Jiang, and
  He}]{Weng:2010:TFT:1718487.1718520}
Weng, J., Lim, E.-P., Jiang, J., He, Q., 2010. Twitterrank: Finding
  topic-sensitive influential twitterers. In: Proceedings of the Third ACM
  International Conference on Web Search and Data Mining. WSDM '10. ACM, New
  York, NY, USA, pp. 261--270.
\newline\urlprefix\url{http://doi.acm.org/10.1145/1718487.1718520}

\bibitem[{{Wikidata}(2020)}]{wikidatadatasetsize}
{Wikidata}, 2020. {Wikidata:Statistics}. Accessed: 2020-05-15.
\newline\urlprefix\url{https://www.wikidata.org/wiki/Wikidata:Statistics}

\bibitem[{{Wikimedia Foundation}(2019)}]{WikidataOrg}
{Wikimedia Foundation}, 2019. Wikidata. Accessed: 2020-05-15.
\newline\urlprefix\url{https://www.wikidata.org/wiki/Wikidata:Main_Page}

\bibitem[{{Wikimedia Foundation}(2020)}]{WikidataSparql}
{Wikimedia Foundation}, 2020. Wikidata query service. Accessed: 2020-05-15.
\newline\urlprefix\url{https://query.wikidata.org/}

\bibitem[{{World {H}ealth {O}rganization}(2020)}]{icd-who-www}
{World {H}ealth {O}rganization}, 2020. {WHO} international classification of
  diseases. Accessed: 2020-05-28.
\newline\urlprefix\url{http://www.who.int/classifications/icd/en/}

\bibitem[{Yamada et~al.(2016)Yamada, Shindo, Takeda, and
  Takefuji}]{YamadaS0T16}
Yamada, I., Shindo, H., Takeda, H., Takefuji, Y., 2016. Joint learning of the
  embedding of words and entities for named entity disambiguation. CoRR
  abs/1601.01343.
\newline\urlprefix\url{http://arxiv.org/abs/1601.01343}

\bibitem[{Yan et~al.(2013)Yan, Guo, Lan, and Cheng}]{yan2013biterm}
Yan, X., Guo, J., Lan, Y., Cheng, X., 2013. A biterm topic model for short
  texts. In: Proceedings of the 22Nd International Conference on World Wide
  Web. WWW '13. ACM, New York, NY, USA, pp. 1445--1456, iSBN 978-1-4503-2035-1.

\bibitem[{{Yi} et~al.(2020){Yi}, {Jiang}, and {Wu}}]{8993771}
{Yi}, F., {Jiang}, B., {Wu}, J., 2020. Topic modeling for short texts via word
  embedding and document correlation. IEEE Access 8, 30692--30705.

\bibitem[{Yin et~al.(2018)Yin, Chao, Liu, Zhang, Yu, and
  Wang}]{Yin:2018:MCS:3219819.3220094}
Yin, J., Chao, D., Liu, Z., Zhang, W., Yu, X., Wang, J., 2018. Model-based
  clustering of short text streams. In: Proceedings of the 24th ACM SIGKDD
  International Conference on Knowledge Discovery \& Data Mining. KDD '18. ACM,
  New York, NY, USA, pp. 2634--2642.
\newline\urlprefix\url{http://doi.acm.org/10.1145/3219819.3220094}

\bibitem[{{Yin} et~al.(2019){Yin}, {Huang}, {Zhou}, {Li}, {Lan}, and
  {Jia}}]{bertEL2019}
{Yin}, X., {Huang}, Y., {Zhou}, B., {Li}, A., {Lan}, L., {Jia}, Y., 2019. Deep
  entity linking via eliminating semantic ambiguity with bert. IEEE Access 7,
  169434--169445.

\bibitem[{Yıldırım(2020)}]{EarlyPrototype}
Yıldırım, A., 2020. S-bounti: Semantic topic identification approach from
  microblog post sets. an application. Accessed: 2020-05-15.
\newline\urlprefix\url{https://doi.org/10.6084/m9.figshare.5943211}

\bibitem[{Yıldırım and Uskudarli(2018)}]{TopicDownload}
Yıldırım, A., Uskudarli, S., 2018. S-bounti: Semantic topic identification
  approach from microblog post sets using linked open data, published datasets.
  [cited 15 jun 2020]. database: figshare [internet].
\newline\urlprefix\url{https://doi.org/10.6084/m9.figshare.7527476}

\bibitem[{Yıldırım and Uskudarli(2019)}]{ontologyDetails}
Yıldırım, A., Uskudarli, S., 2019. About topico ontology. Accessed:
  2020-05-15.
\newline\urlprefix\url{http://soslab.cmpe.boun.edu.tr/sbounti/AboutTopico.pdf}

\bibitem[{Yıldırım et~al.(2020)Yıldırım, Uskudarli, and
  Ozgur}]{dataset2012}
Yıldırım, A., Uskudarli, S., Ozgur, A., 2020. Tf values, word frequency
  values for gathering idf values, and the evaluation data submitted to plos
  one, titled identifying topics in microblogs using wikipedia. Accessed:
  2020-05-15.
\newline\urlprefix\url{https://figshare.com/articles/data_tar_gz/2068665}

\bibitem[{Yıldırım et~al.(2016)Yıldırım, Uskudarli, and Özgür}]{bounti}
Yıldırım, A., Uskudarli, S., Özgür, A., 03 2016. Identifying topics in
  microblogs using wikipedia. {PLoS} ONE 11~(3), 1--20.

\bibitem[{{Zheng} et~al.(2018){Zheng}, {Han}, and
  {Sun}}]{LocationPredictionTwitterSurvey}
{Zheng}, X., {Han}, J., {Sun}, A., 2018. A survey of location prediction on
  twitter. IEEE Transactions on Knowledge and Data Engineering 30~(9),
  1652--1671.

\bibitem[{Çelebi and Özgür(2018)}]{celebi2017}
Çelebi, A., Özgür, A., 2018. Segmenting hashtags and analyzing their
  grammatical structure. Journal of the Association for Information Science and
  Technology 69~(5), 675--686.
\newline\urlprefix\url{https://onlinelibrary.wiley.com/doi/abs/10.1002/asi.23989}

\end{thebibliography}

\clearpage
\appendix

\section{The main object properties of topico:Topic}
\label{S1_Appendix}

This section specifies the main object properties for \topico. 
First the syntax is summarized in the following table.
Then, the object properties are provided.
The properties with inverse properties are indicated within their definition and not further defined in the table.

\begin{table*}[ht]
\label{tab:owl-dl}
\centering
\renewcommand{\tabcolsep}{0.4cm}
\begin{tabular}{lll}
\hline
 Descriptions  & Abstract Syntax & DL Syntax \\
\hline
\multirow{2}{*}{Concepts} 	& $intersection(C_1,C_2,\cdots,C_n)$ 	& $C_1 \sqcap C_2 \sqcap \cdots \sqcap C_n$\\
\multirow{2}{*}{}          	& $union(C_1,C_2,\cdots,C_n)$ 			& $C_1 \sqcup C_2 \sqcup \cdots \sqcup C_n$\\
\multirow{2}{*}{} 			& $partial(C_1,C_2,\cdots,C_n)$ 		& $A \sqsubseteq C_1 \sqcap C_2 \sqcap \cdots C_n$\\
\multirow{2}{*}{}           & $complete(C_1,C_2,\cdots,C_n)$ 		& $A \equiv C_1 \sqcap C_2 \sqcap \cdots C_n$\\  
\multirow{2}{*}{} 			& universal concept (top) & $\top$\\  \hline 
\multirow{3}{*}{Roles} & existential restriction  & $\exists R.C$\\
\multirow{3}{*}{}           & universal restriction   & $\forall R.C$\\
\multirow{3}{*}{}           & cardinality restriction   & $[=|\geq|\leq] n R.C$\\
\multirow{3}{*}{}           & Inverse of R & $R^-$\\
\hline
\end{tabular}
\caption{The basic syntax of OWL-DL.}
\end{table*}

\begin{longtable}{ll}
\label{tab:topic-object-properties}

{\bf Object Properties}  & {\bf Description Logic}  \\ \hline 
\property{}{hasAgent} & \ensuremath{\sqsubseteq}~\ensuremath{\forall}~\property{}{hasAgent}.\qnclass{foaf}{Agent} ~\ensuremath{\sqcap}~
 \ensuremath{\exists}~\property{}{hasAgent}.\ensuremath{\top}~\ensuremath{\sqsubseteq}~\qnclass{}{Topic}\\ 
& \ensuremath{\equiv}~\property{}{isAnAgentOf}\ensuremath{^-} \\ \hline

\property{}{hasPerson}& ~\ensuremath{\sqsubseteq}~\property{}{hasAgent}\\
& \ensuremath{\exists}~\property{}{hasPerson}.\ensuremath{\top}~\ensuremath{\sqsubseteq}~\qnclass{}{Topic} ~\ensuremath{\sqcap}~
 \ensuremath{\top}~\ensuremath{\sqsubseteq}~\ensuremath{\forall}~\property{}{hasPerson}.\qnclass{foaf}{Person}\\ 
& \ensuremath{\equiv}~\property{}{isAPersonf}\ensuremath{^-}\\ \hline 

\property{}{hasGroup}& \ensuremath{\sqsubseteq}~\property{}{hasAgent}\\
& \ensuremath{\exists}~\property{}{hasGroup}.\ensuremath{\top}~\ensuremath{\sqsubseteq}~\qnclass{}{Topic}~\ensuremath{\sqcap}~
\ensuremath{\forall}~\property{}{hasGroup}.\qnclass{foaf}{Group} \\ 
& \ensuremath{\equiv}~\property{}{isAGroupOf}\ensuremath{^-}\\ \hline 

\property{}{hasOrganization}& \ensuremath{\sqsubseteq}~\property{}{hasAgent}\\
& \ensuremath{\exists}~\property{}{hasOrganization}.\ensuremath{\top}~\ensuremath{\sqsubseteq}~\qnclass{}{Topic}~\ensuremath{\sqcap}~
\ensuremath{\forall}~\property{}{hasOrganization}.\qnclass{foaf}{Organization} \\ 
& \ensuremath{\equiv}~\property{}{isAnOrganizationf}\ensuremath{^-}\\ \hline

\property{}{hasLocation} & \ensuremath{\sqsubseteq}~\ensuremath{\forall}~\property{}{hasLocation}.\qnclass{}{Location}~\ensuremath{\sqcap}~ \ensuremath{\exists}~\property{}{hasLocation}.\ensuremath{\top}~\ensuremath{\sqsubseteq}~\qnclass{}{Topic} \\ 
& \ensuremath{\equiv}~\property{}{isLocationOf}\ensuremath{^-}\\ \hline

\property{}{hasTemporalExpression} & \ensuremath{\sqsubseteq}~\ensuremath{\forall}~\property{}{hasTemporalExpression}.\qnclass{}{TemporalExpression}~\ensuremath{\sqcap}\\ 
& \ \ \ \ \ensuremath{\exists}~\property{}{hasTemporalExpression}.\ensuremath{\top}~\ensuremath{\sqsubseteq}~\qnclass{}{Topic}\\ 
& \ensuremath{\equiv}~\property{}{isTemporalExpressionOf}\ensuremath{^-}\\ \hline

\property{}{hasTemporalEntity} & \ensuremath{\sqsubseteq}~\property{}{hasTemporalExpression} \\
& \ensuremath{\forall}~\property{}{hasTemporalEntity}.\qnclass{time}{TemporalEntity}~\ensuremath{\sqcap}\\
& \ \ \ \ \ensuremath{\exists}~\property{}{hasTemporalEntity}.\ensuremath{\top}~\ensuremath{\sqsubseteq}~\qnclass{}{Topic} \\ 
& \property{}{isTemporalEntityOf}\ensuremath{^-}\\ \hline

\property{}{hasTemporalTerm} & \ensuremath{\sqsubseteq}~\property{}{hasTemporalExpression} \\
& \ensuremath{\forall}~\property{}{hasTemporalTerm}.\qnclass{}{TemporalTerm}~\ensuremath{\sqcap}\\
& \ \ \ \ \ensuremath{\exists}~\property{}{hasTemporalTerm}.\ensuremath{\top}~\ensuremath{\sqsubseteq}~\qnclass{}{Topic}\\ 
& \ensuremath{\equiv}~\property{}{isTemporalTermOf}\ensuremath{^-}\\ \hline

\property{}{isAbout} & \ensuremath{\sqsubseteq}~\ensuremath{\forall}~\property{}{isAbout}.\ensuremath{\top}~\ensuremath{\sqcap}~ \ensuremath{\exists}~\property{}{isAbout}.\ensuremath{\top}~\ensuremath{\sqsubseteq}~\qnclass{}{Topic} \\ 
&\ensuremath{\equiv}~\property{}{inTopic}\ensuremath{^-} \\ \hline

\property{}{observationInterval} &  \ensuremath{\top}~\ensuremath{\sqsubseteq}~\ensuremath{=}~1~\property{}{observationInterval}.\ensuremath{\top} \ensuremath{\sqcap}\\
& \ \ \ \  \ensuremath{\exists}~\property{}{observationInterval}.\ensuremath{\top}~\ensuremath{\sqsubseteq}~\qnclass{}{Topic} \ensuremath{\sqcap}\\
& \ \ \ \  \ensuremath{\forall}~\property{}{observationInterval}.\qnclass{time}{Interval} \\
&\ensuremath{\equiv}~\property{}{isTopicOfObservationInterval}\ensuremath{^-} \\ \hline

\end{longtable}

\section{Supplementary Figures}

\subsection{Evaluation Tool}

The following shows theevaluation tool we developed to examine and rate the relevancy of topics generated by our approch.
This figure shows the \bounti\ and \sbounti\ topics extracted from  \vp\ [68-70).
The {\em See tweets} and {\em See word cloud} links show the related tweets and a word cloud generated from them.
The {\em Entity frequencies} link shows the list of linked entities and their frequencies.
All resources are reachable trough web-links for inspection.

\includegraphics[scale=0.2]{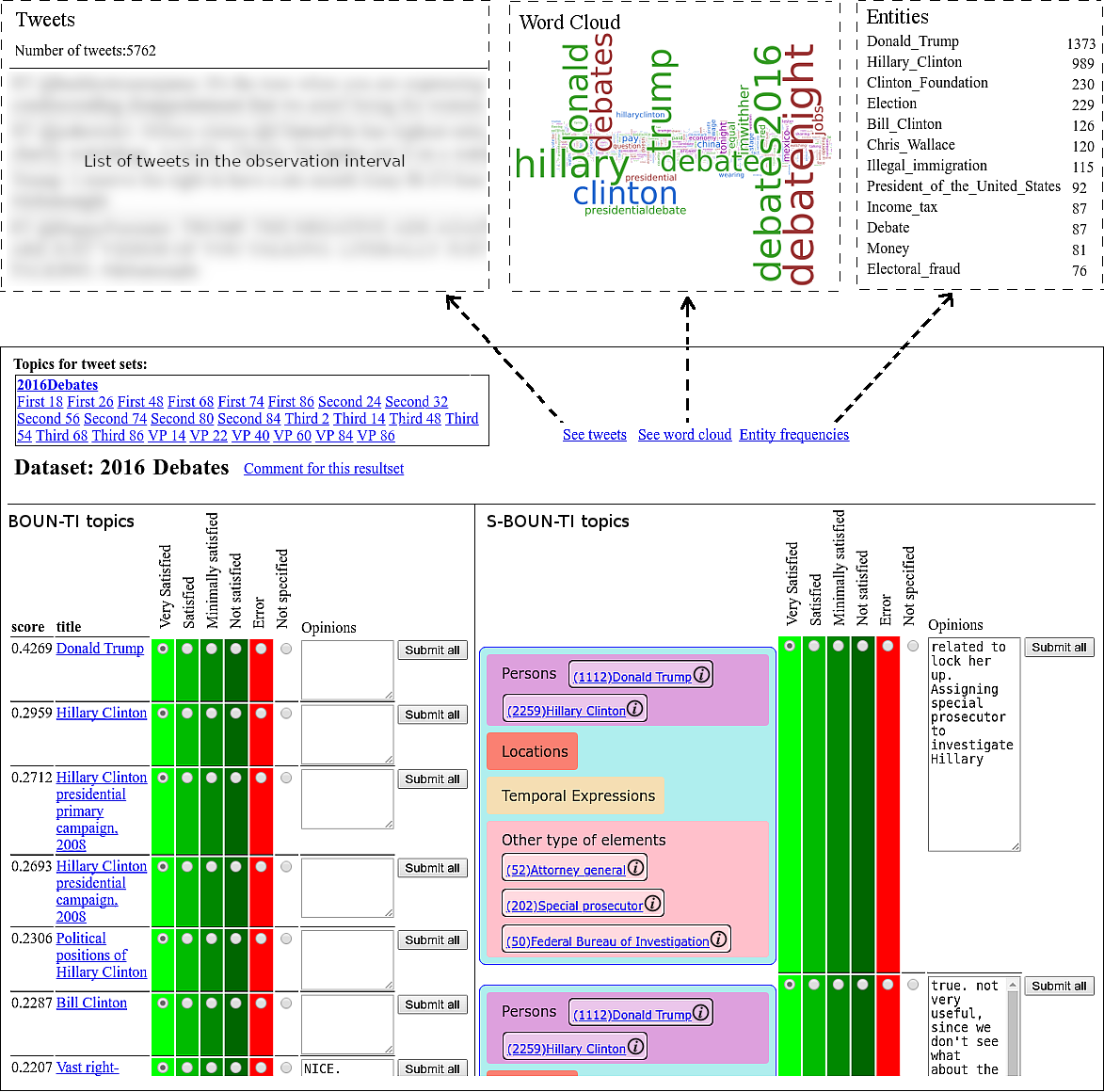}
\label{S1_Fig}

\subsection{Co-occurrence Graph Example}
\label{S2_Fig}

The following is a part of the sample co-occurrence graph of topic elements identified within  the first presidential debate.

\includegraphics[scale=0.35]{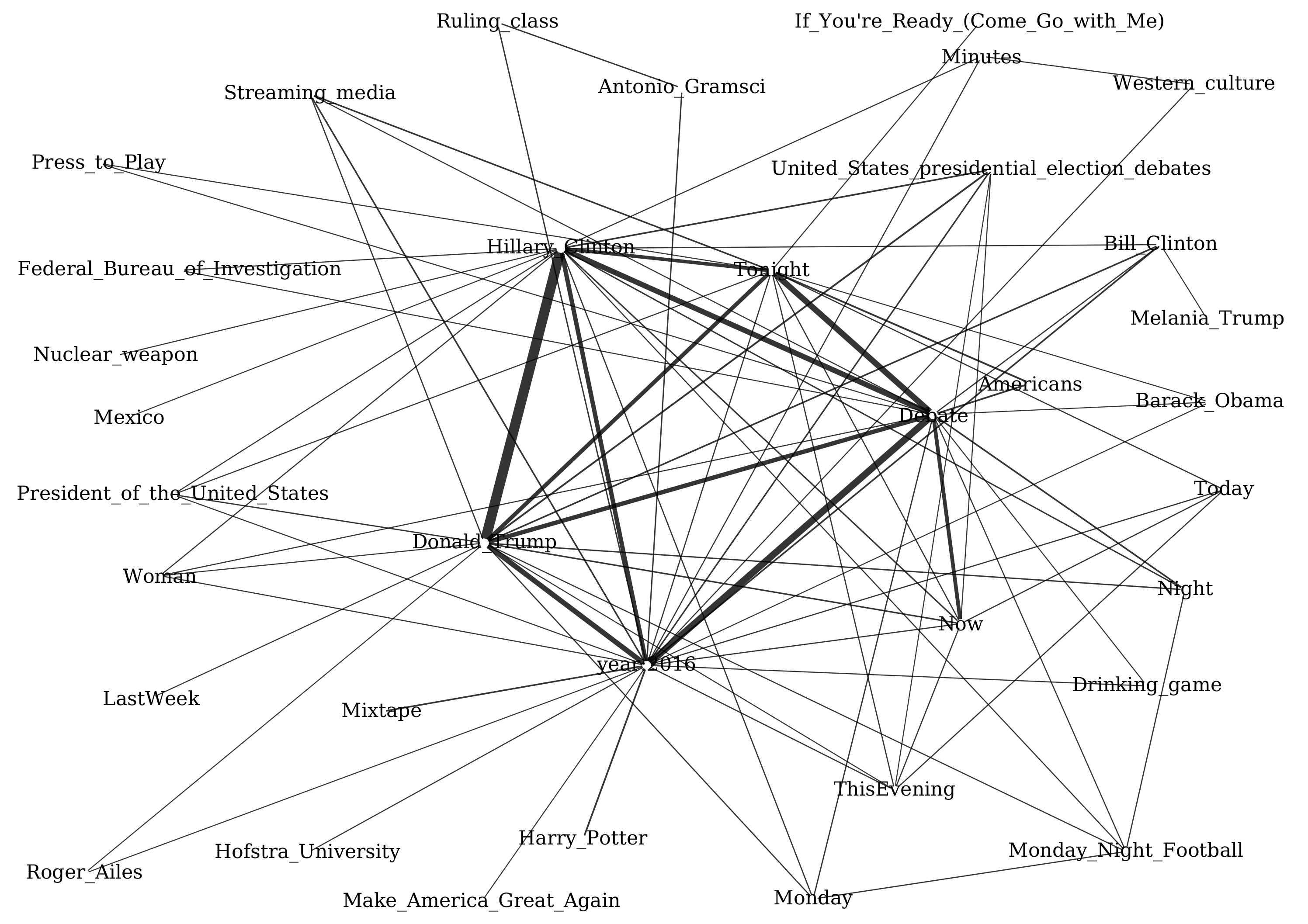}

\subsection{Namespaces}
\label{S1_Table}

The following table shows the namespace prefixes that are utilized in \topico\ and referred to in this paper.

\begin{table}[!ht]
\centering
\caption{{\bf The namespace prefixes that are utilized in \topico\ and referred to in this paper.}}
\begin{tabular}{|l|l|l|}
\hline
\textbf{Prefix} & \textbf{URI} & \textbf{Name}\\\thickhline
foaf & http://xmlns.com/foaf/0.1/ & \foaf\ ontology\\ \hline
schema & http://schema.org/ & Schema.org ontology \\ \hline
time & http://www.w3.org/2006/time\# &\wwwc\ time ontology \\ \hline
geo & http://www.w3.org/2003/01/geo/wgs84\_pos\# & \wwwc\ basic Geo vocabulary \\ \hline
geonames & http://www.geonames.org/ontology\# &Geonames ontology \\ \hline
dbo & http://dbpedia.org/ontology/ & \dbp\ ontology \\ \hline
dbr & http://dbpedia.org/resource/ & \dbp\ resource \\ \hline
dbc & http://dbpedia.org/resource/Category: & \dbp\ category \\ \hline
wikidata & http://www.wikidata.org/entity/ & Wikidata resource \\ \hline
\end{tabular}
\label{tab:namespaces}
\end{table}

\subsection{Tweet Collection Statistics}
\label{S2_Table}

The following table shows the percentage of tweets in the post sets that produce the vertices (topic elements), edges (co-occurring elements), and topics.
This table shows the percentage of tweets in the post sets that produce the vertices (topic elements), edges (co-occurring elements), and topics. 

\begin{table}[!ht]
\centering
\caption{{\bf Percentage of tweets in the post sets that produce the vertices (topic elements), edges (co-occurring elements), and topics.}}
\begin{tabular}{|c|S[table-format=2]|S[table-format=2]|S[table-format=2]|S[table-format=2]|S[table-format=2]|S[table-format=2]|}
\hline 
& \multicolumn {3}{c|}{\textbf{Vertices}} & \multicolumn {3}{c|}{\textbf{Edges}} \\ \hline
\textbf{Set} & \multicolumn {1}{c@{}}{\textbf{Before}} & \multicolumn {1}{c@{}}{\textbf{Pruned}} & \multicolumn {1}{c|}{\textbf{Topic}} & \multicolumn {1}{c}{\textbf{Before}} & \multicolumn {1}{c}{\textbf{Pruned}} & \multicolumn {1}{c|}{\textbf{Topic}} \\ \thickhline 
\pdone& 71 & 63 & 59 & 37 & 27 & 23 \\ \hline
\pdtwo& 71 & 65 & 61 & 38 & 30 & 24 \\ \hline
\pdthree& 67 & 57 & 51 & 34 & 23 & 16 \\ \hline
\vp& 71 & 61 & 55 & 37 & 26 & 20 \\ \hline
\brangelina& 42 & 30 & 26 & 17 & 13 & 8 \\ \hline
\carriefisher& 77 & 74 & 69 & 43 & 38 & 24 \\ \hline
\concert& 87 & 83 & 78 & 64 & 52 & 35 \\ \hline
\northdakota& 92 & 86 & 75 & 64 & 60 & 51 \\ \hline
\tonibraxton& 43 & 41 & 32 & 25 & 22 & 18 \\ \hline
\inauguration& 81 & 74 & 69 & 52 & 42 & 33 \\ \hline
\public& 47 & 10 & 0 & 20 & 2 & 0 \\ \hline
\end{tabular}
\label{tab:RelatedPostsBeforeAndAfter}
\end{table}

The columns labeled {\em Before} and {\em Pruned} show the impact of pruning the graph. 
The columns labeled {\em Topic} show how many were retained in the topic.

\pagebreak
\subsection{Dataset intervals used in Evaluation Collection}
\label{S3_Table}
The following table shows the  intervals within the datasets that were used for evaluating topics.

\begin{tabular}{|l|l|}
\hline
\multicolumn{1}{|c|}{\bf Set id}  & \multicolumn{1}{c|}{\bf Intervals}\\  \thickhline 
\pdone & [8-10), [18-20), [26-28), [38-40), [48-50), [68-70), [70-72), [74-76), [86-88) \\  \hline 
\pdtwo & [18-20), [24-26), [32-34), [36-38),  [56-58), [74-76), [76-78), [80-82), [84-86) \\  \hline 
\pdthree &  [0-2), [2-4), [14-16), [32-34), [48-50), [54-56), [62-64), [68-70), [86-88) \\  \hline 
\vp & [8-10), [14-16), [22-24), [36-38), [40-42), [60-62), [74-76), [84-86), [86-88) \\ \hline \end{tabular}

\end{document}